\documentclass[reprint,superscriptaddress,amsmath,amssymb,aps]{revtex4-1}
\usepackage{graphicx}
\usepackage{epstopdf}
\usepackage{color}
\usepackage{dcolumn}
\usepackage{bm}
\usepackage{hyperref}
\usepackage[mathlines]{lineno}
\usepackage{upgreek} 
\usepackage{threeparttable} 
\usepackage{booktabs}
\usepackage{siunitx}
\usepackage{textcomp} 
\usepackage{multirow}
\usepackage{tabularx} 
\usepackage{ragged2e} 
\definecolor{darkpurple}{RGB}{128,0,128}
\definecolor{darkgreen}{RGB}{0,150,0}

\newcommand{\GeVmass}{\ensuremath{\mathrm{GeV}/c^{2}}}
\newcommand{\dd}{$\mbox{D-D}$} 

\begin{document}

\title{Calibration, event reconstruction, data analysis and limits calculation for the LUX dark matter experiment}

\author{D.S.~Akerib} \affiliation{Case Western Reserve University, Department of Physics, 10900 Euclid Ave, Cleveland, OH 44106, USA} \affiliation{SLAC National Accelerator Laboratory, 2575 Sand Hill Road, Menlo Park, CA 94205, USA} \affiliation{Kavli Institute for Particle Astrophysics and Cosmology, Stanford University, 452 Lomita Mall, Stanford, CA 94309, USA}
\author{S.~Alsum} \affiliation{University of Wisconsin-Madison, Department of Physics, 1150 University Ave., Madison, WI 53706, USA}  
\author{H.M.~Ara\'{u}jo} \affiliation{Imperial College London, High Energy Physics, Blackett Laboratory, London SW7 2BZ, United Kingdom}  
\author{X.~Bai} \affiliation{South Dakota School of Mines and Technology, 501 East St Joseph St., Rapid City, SD 57701, USA}  
\author{A.J.~Bailey} \affiliation{Imperial College London, High Energy Physics, Blackett Laboratory, London SW7 2BZ, United Kingdom}  
\author{J.~Balajthy} \affiliation{University of Maryland, Department of Physics, College Park, MD 20742, USA}  
\author{P.~Beltrame} \affiliation{SUPA, School of Physics and Astronomy, University of Edinburgh, Edinburgh EH9 3FD, United Kingdom}  
\author{E.P.~Bernard} \affiliation{University of California Berkeley, Department of Physics, Berkeley, CA 94720, USA} \affiliation{Yale University, Department of Physics, 217 Prospect St., New Haven, CT 06511, USA} 
\author{A.~Bernstein} \affiliation{Lawrence Livermore National Laboratory, 7000 East Ave., Livermore, CA 94551, USA}  
\author{T.P.~Biesiadzinski} \affiliation{Case Western Reserve University, Department of Physics, 10900 Euclid Ave, Cleveland, OH 44106, USA} \affiliation{SLAC National Accelerator Laboratory, 2575 Sand Hill Road, Menlo Park, CA 94205, USA} \affiliation{Kavli Institute for Particle Astrophysics and Cosmology, Stanford University, 452 Lomita Mall, Stanford, CA 94309, USA}
\author{E.M.~Boulton} \affiliation{University of California Berkeley, Department of Physics, Berkeley, CA 94720, USA} \affiliation{Lawrence Berkeley National Laboratory, 1 Cyclotron Rd., Berkeley, CA 94720, USA} \affiliation{Yale University, Department of Physics, 217 Prospect St., New Haven, CT 06511, USA} 
\author{P.~Br\'as} \affiliation{LIP-Coimbra, Department of Physics, University of Coimbra, Rua Larga, 3004-516 Coimbra, Portugal}  
\author{D.~Byram} \affiliation{University of South Dakota, Department of Physics, 414E Clark St., Vermillion, SD 57069, USA} \affiliation{South Dakota Science and Technology Authority, Sanford Underground Research Facility, Lead, SD 57754, USA} 
\author{S.B.~Cahn} \affiliation{Yale University, Department of Physics, 217 Prospect St., New Haven, CT 06511, USA}  
\author{M.C.~Carmona-Benitez} 
\thanks{Corresponding author: carmona@psu.edu}
\affiliation{Pennsylvania State University, Department of Physics, 104 Davey Lab, University Park, PA  16802-6300, USA} \affiliation{University of California Santa Barbara, Department of Physics, Santa Barbara, CA 93106, USA} 
\author{C.~Chan} \affiliation{Brown University, Department of Physics, 182 Hope St., Providence, RI 02912, USA}  
\author{A.~Currie} \affiliation{Imperial College London, High Energy Physics, Blackett Laboratory, London SW7 2BZ, United Kingdom}  
\author{J.E.~Cutter} \affiliation{University of California Davis, Department of Physics, One Shields Ave., Davis, CA 95616, USA}  
\author{T.J.R.~Davison} \affiliation{SUPA, School of Physics and Astronomy, University of Edinburgh, Edinburgh EH9 3FD, United Kingdom}  
\author{A.~Dobi} \affiliation{Lawrence Berkeley National Laboratory, 1 Cyclotron Rd., Berkeley, CA 94720, USA}  
\author{J.E.Y.~Dobson} \affiliation{Department of Physics and Astronomy, University College London, Gower Street, London WC1E 6BT, United Kingdom}  
\author{E.~Druszkiewicz} \affiliation{University of Rochester, Department of Physics and Astronomy, Rochester, NY 14627, USA}  
\author{B.N.~Edwards} \affiliation{Yale University, Department of Physics, 217 Prospect St., New Haven, CT 06511, USA}  
\author{C.H.~Faham} \affiliation{Lawrence Berkeley National Laboratory, 1 Cyclotron Rd., Berkeley, CA 94720, USA}  
\author{S.R.~Fallon} \affiliation{University at Albany, State University of New York, Department of Physics, 1400 Washington Ave., Albany, NY 12222, USA}  
\author{A.~Fan} \affiliation{SLAC National Accelerator Laboratory, 2575 Sand Hill Road, Menlo Park, CA 94205, USA} \affiliation{Kavli Institute for Particle Astrophysics and Cosmology, Stanford University, 452 Lomita Mall, Stanford, CA 94309, USA} 
\author{S.~Fiorucci} \affiliation{Lawrence Berkeley National Laboratory, 1 Cyclotron Rd., Berkeley, CA 94720, USA} \affiliation{Brown University, Department of Physics, 182 Hope St., Providence, RI 02912, USA} 
\author{R.J.~Gaitskell} \affiliation{Brown University, Department of Physics, 182 Hope St., Providence, RI 02912, USA}  
\author{V.M.~Gehman} \affiliation{Lawrence Berkeley National Laboratory, 1 Cyclotron Rd., Berkeley, CA 94720, USA}  
\author{J.~Genovesi} \affiliation{University at Albany, State University of New York, Department of Physics, 1400 Washington Ave., Albany, NY 12222, USA}  
\author{C.~Ghag} \affiliation{Department of Physics and Astronomy, University College London, Gower Street, London WC1E 6BT, United Kingdom}  
\author{M.G.D.~Gilchriese} \affiliation{Lawrence Berkeley National Laboratory, 1 Cyclotron Rd., Berkeley, CA 94720, USA}  
\author{C.R.~Hall} \affiliation{University of Maryland, Department of Physics, College Park, MD 20742, USA}  
\author{M.~Hanhardt} \affiliation{South Dakota School of Mines and Technology, 501 East St Joseph St., Rapid City, SD 57701, USA} \affiliation{South Dakota Science and Technology Authority, Sanford Underground Research Facility, Lead, SD 57754, USA} 
\author{S.J.~Haselschwardt} \affiliation{University of California Santa Barbara, Department of Physics, Santa Barbara, CA 93106, USA}  
\author{S.A.~Hertel} \affiliation{University of Massachusetts, Department of Physics, Amherst, MA 01003-9337 USA} \affiliation{Lawrence Berkeley National Laboratory, 1 Cyclotron Rd., Berkeley, CA 94720, USA} \affiliation{Yale University, Department of Physics, 217 Prospect St., New Haven, CT 06511, USA}
\author{D.P.~Hogan} \affiliation{University of California Berkeley, Department of Physics, Berkeley, CA 94720, USA}  
\author{M.~Horn} \affiliation{South Dakota Science and Technology Authority, Sanford Underground Research Facility, Lead, SD 57754, USA} \affiliation{University of California Berkeley, Department of Physics, Berkeley, CA 94720, USA} \affiliation{Yale University, Department of Physics, 217 Prospect St., New Haven, CT 06511, USA}
\author{D.Q.~Huang} \affiliation{Brown University, Department of Physics, 182 Hope St., Providence, RI 02912, USA}  
\author{C.M.~Ignarra} \affiliation{SLAC National Accelerator Laboratory, 2575 Sand Hill Road, Menlo Park, CA 94205, USA} \affiliation{Kavli Institute for Particle Astrophysics and Cosmology, Stanford University, 452 Lomita Mall, Stanford, CA 94309, USA} 
\author{R.G.~Jacobsen} \affiliation{University of California Berkeley, Department of Physics, Berkeley, CA 94720, USA}  
\author{W.~Ji} \affiliation{Case Western Reserve University, Department of Physics, 10900 Euclid Ave, Cleveland, OH 44106, USA} \affiliation{SLAC National Accelerator Laboratory, 2575 Sand Hill Road, Menlo Park, CA 94205, USA} \affiliation{Kavli Institute for Particle Astrophysics and Cosmology, Stanford University, 452 Lomita Mall, Stanford, CA 94309, USA}
\author{K.~Kamdin} \affiliation{University of California Berkeley, Department of Physics, Berkeley, CA 94720, USA}  
\author{K.~Kazkaz} \affiliation{Lawrence Livermore National Laboratory, 7000 East Ave., Livermore, CA 94551, USA}  
\author{D.~Khaitan} \affiliation{University of Rochester, Department of Physics and Astronomy, Rochester, NY 14627, USA}  
\author{R.~Knoche} \affiliation{University of Maryland, Department of Physics, College Park, MD 20742, USA}  
\author{N.A.~Larsen} \affiliation{Yale University, Department of Physics, 217 Prospect St., New Haven, CT 06511, USA}  
\author{C.~Lee} \affiliation{Case Western Reserve University, Department of Physics, 10900 Euclid Ave, Cleveland, OH 44106, USA} \affiliation{SLAC National Accelerator Laboratory, 2575 Sand Hill Road, Menlo Park, CA 94205, USA} \affiliation{Kavli Institute for Particle Astrophysics and Cosmology, Stanford University, 452 Lomita Mall, Stanford, CA 94309, USA}
\author{B.G.~Lenardo} \affiliation{University of California Davis, Department of Physics, One Shields Ave., Davis, CA 95616, USA} \affiliation{Lawrence Livermore National Laboratory, 7000 East Ave., Livermore, CA 94551, USA} 
\author{K.T.~Lesko} \affiliation{Lawrence Berkeley National Laboratory, 1 Cyclotron Rd., Berkeley, CA 94720, USA}  
\author{A.~Lindote} \affiliation{LIP-Coimbra, Department of Physics, University of Coimbra, Rua Larga, 3004-516 Coimbra, Portugal}  
\author{M.I.~Lopes} \affiliation{LIP-Coimbra, Department of Physics, University of Coimbra, Rua Larga, 3004-516 Coimbra, Portugal}  
\author{A.~Manalaysay} \affiliation{University of California Davis, Department of Physics, One Shields Ave., Davis, CA 95616, USA}  
\author{R.L.~Mannino} \affiliation{Texas A \& M University, Department of Physics, College Station, TX 77843, USA} \affiliation{University of Wisconsin-Madison, Department of Physics, 1150 University Ave., Madison, WI 53706, USA} 
\author{M.F.~Marzioni} \affiliation{SUPA, School of Physics and Astronomy, University of Edinburgh, Edinburgh EH9 3FD, United Kingdom}  
\author{D.N.~McKinsey} \affiliation{University of California Berkeley, Department of Physics, Berkeley, CA 94720, USA} \affiliation{Lawrence Berkeley National Laboratory, 1 Cyclotron Rd., Berkeley, CA 94720, USA} \affiliation{Yale University, Department of Physics, 217 Prospect St., New Haven, CT 06511, USA}
\author{D.-M.~Mei} \affiliation{University of South Dakota, Department of Physics, 414E Clark St., Vermillion, SD 57069, USA}  
\author{J.~Mock} \affiliation{University at Albany, State University of New York, Department of Physics, 1400 Washington Ave., Albany, NY 12222, USA}  
\author{M.~Moongweluwan} \affiliation{University of Rochester, Department of Physics and Astronomy, Rochester, NY 14627, USA}  
\author{J.A.~Morad} \affiliation{University of California Davis, Department of Physics, One Shields Ave., Davis, CA 95616, USA}  
\author{A.St.J.~Murphy} \affiliation{SUPA, School of Physics and Astronomy, University of Edinburgh, Edinburgh EH9 3FD, United Kingdom}  
\author{C.~Nehrkorn} \affiliation{University of California Santa Barbara, Department of Physics, Santa Barbara, CA 93106, USA}  
\author{H.N.~Nelson} \affiliation{University of California Santa Barbara, Department of Physics, Santa Barbara, CA 93106, USA}  
\author{F.~Neves} \affiliation{LIP-Coimbra, Department of Physics, University of Coimbra, Rua Larga, 3004-516 Coimbra, Portugal}  
\author{K.~O'Sullivan} \affiliation{University of California Berkeley, Department of Physics, Berkeley, CA 94720, USA} \affiliation{Lawrence Berkeley National Laboratory, 1 Cyclotron Rd., Berkeley, CA 94720, USA} \affiliation{Yale University, Department of Physics, 217 Prospect St., New Haven, CT 06511, USA}
\author{K.C.~Oliver-Mallory} \affiliation{University of California Berkeley, Department of Physics, Berkeley, CA 94720, USA}  
\author{K.J.~Palladino} \affiliation{University of Wisconsin-Madison, Department of Physics, 1150 University Ave., Madison, WI 53706, USA} \affiliation{SLAC National Accelerator Laboratory, 2575 Sand Hill Road, Menlo Park, CA 94205, USA} \affiliation{Kavli Institute for Particle Astrophysics and Cosmology, Stanford University, 452 Lomita Mall, Stanford, CA 94309, USA}
\author{E.K.~Pease} \affiliation{University of California Berkeley, Department of Physics, Berkeley, CA 94720, USA} \affiliation{Lawrence Berkeley National Laboratory, 1 Cyclotron Rd., Berkeley, CA 94720, USA} \affiliation{Yale University, Department of Physics, 217 Prospect St., New Haven, CT 06511, USA} 
\author{L.~Reichhart} \affiliation{Department of Physics and Astronomy, University College London, Gower Street, London WC1E 6BT, United Kingdom}  
\author{C.~Rhyne} \affiliation{Brown University, Department of Physics, 182 Hope St., Providence, RI 02912, USA}  
\author{S.~Shaw} \affiliation{University of California Santa Barbara, Department of Physics, Santa Barbara, CA 93106, USA} \affiliation{Department of Physics and Astronomy, University College London, Gower Street, London WC1E 6BT, United Kingdom} 
\author{T.A.~Shutt} \affiliation{Case Western Reserve University, Department of Physics, 10900 Euclid Ave, Cleveland, OH 44106, USA} \affiliation{SLAC National Accelerator Laboratory, 2575 Sand Hill Road, Menlo Park, CA 94205, USA} \affiliation{Kavli Institute for Particle Astrophysics and Cosmology, Stanford University, 452 Lomita Mall, Stanford, CA 94309, USA}
\author{C.~Silva} \affiliation{LIP-Coimbra, Department of Physics, University of Coimbra, Rua Larga, 3004-516 Coimbra, Portugal}  
\author{M.~Solmaz} \affiliation{University of California Santa Barbara, Department of Physics, Santa Barbara, CA 93106, USA}  
\author{V.N.~Solovov} \affiliation{LIP-Coimbra, Department of Physics, University of Coimbra, Rua Larga, 3004-516 Coimbra, Portugal}  
\author{P.~Sorensen} \affiliation{Lawrence Berkeley National Laboratory, 1 Cyclotron Rd., Berkeley, CA 94720, USA}  
\author{T.J.~Sumner} \affiliation{Imperial College London, High Energy Physics, Blackett Laboratory, London SW7 2BZ, United Kingdom}  
\author{M.~Szydagis} \affiliation{University at Albany, State University of New York, Department of Physics, 1400 Washington Ave., Albany, NY 12222, USA}  
\author{D.J.~Taylor} \affiliation{South Dakota Science and Technology Authority, Sanford Underground Research Facility, Lead, SD 57754, USA}  
\author{W.C.~Taylor} \affiliation{Brown University, Department of Physics, 182 Hope St., Providence, RI 02912, USA}  
\author{B.P.~Tennyson} \affiliation{Yale University, Department of Physics, 217 Prospect St., New Haven, CT 06511, USA}  
\author{P.A.~Terman} \affiliation{Texas A \& M University, Department of Physics, College Station, TX 77843, USA}  
\author{D.R.~Tiedt} \affiliation{South Dakota School of Mines and Technology, 501 East St Joseph St., Rapid City, SD 57701, USA}  
\author{W.H.~To} \affiliation{California State University Stanislaus, Department of Physics, 1 University Circle, Turlock, CA 95382, USA} \affiliation{SLAC National Accelerator Laboratory, 2575 Sand Hill Road, Menlo Park, CA 94205, USA} \affiliation{Kavli Institute for Particle Astrophysics and Cosmology, Stanford University, 452 Lomita Mall, Stanford, CA 94309, USA}
\author{M.~Tripathi} \affiliation{University of California Davis, Department of Physics, One Shields Ave., Davis, CA 95616, USA}  
\author{L.~Tvrznikova} \affiliation{University of California Berkeley, Department of Physics, Berkeley, CA 94720, USA} \affiliation{Lawrence Berkeley National Laboratory, 1 Cyclotron Rd., Berkeley, CA 94720, USA} \affiliation{Yale University, Department of Physics, 217 Prospect St., New Haven, CT 06511, USA} 
\author{S.~Uvarov} \affiliation{University of California Davis, Department of Physics, One Shields Ave., Davis, CA 95616, USA}  
\author{V.~Velan} \affiliation{University of California Berkeley, Department of Physics, Berkeley, CA 94720, USA}  
\author{J.R.~Verbus} \affiliation{Brown University, Department of Physics, 182 Hope St., Providence, RI 02912, USA}  
\author{R.C.~Webb} \affiliation{Texas A \& M University, Department of Physics, College Station, TX 77843, USA}  
\author{J.T.~White} \thanks{deceased} \affiliation{Texas A \& M University, Department of Physics, College Station, TX 77843, USA}  
\author{T.J.~Whitis} \affiliation{Case Western Reserve University, Department of Physics, 10900 Euclid Ave, Cleveland, OH 44106, USA} \affiliation{SLAC National Accelerator Laboratory, 2575 Sand Hill Road, Menlo Park, CA 94205, USA} \affiliation{Kavli Institute for Particle Astrophysics and Cosmology, Stanford University, 452 Lomita Mall, Stanford, CA 94309, USA}
\author{M.S.~Witherell} \affiliation{Lawrence Berkeley National Laboratory, 1 Cyclotron Rd., Berkeley, CA 94720, USA}  
\author{F.L.H.~Wolfs} \affiliation{University of Rochester, Department of Physics and Astronomy, Rochester, NY 14627, USA}  
\author{J.~Xu} \affiliation{Lawrence Livermore National Laboratory, 7000 East Ave., Livermore, CA 94551, USA}  
\author{K.~Yazdani} \affiliation{Imperial College London, High Energy Physics, Blackett Laboratory, London SW7 2BZ, United Kingdom}  
\author{S.K.~Young} \affiliation{University at Albany, State University of New York, Department of Physics, 1400 Washington Ave., Albany, NY 12222, USA}  
\author{C.~Zhang} \affiliation{University of South Dakota, Department of Physics, 414E Clark St., Vermillion, SD 57069, USA}

\collaboration{LUX Collaboration}

\date{\today}

\begin{abstract}
The LUX experiment has performed searches for dark matter particles scattering elastically on xenon nuclei, leading to stringent upper limits on the nuclear scattering cross sections for dark matter. 
Here, for results derived from $\num{1.4}\times 10^{4}\;\mathrm{kg\,days}$ of target exposure in 2013, details of the calibration, event-reconstruction, modeling, and statistical tests that underlie the results are presented.
Detector performance is characterized, including measured efficiencies, stability of response, position resolution, and discrimination between electron- and nuclear-recoil populations. Models are developed for the drift field, optical properties, background populations, the electron- and nuclear-recoil responses, and the absolute rate of low-energy background events. Innovations in the analysis include \emph{in situ} measurement of the photomultipliers' response to xenon scintillation photons, verification of fiducial mass with a low-energy internal calibration source, and new empirical models for low-energy signal yield based on large-sample, \emph{in situ} calibrations.
\end{abstract}


\maketitle

\section{\label{sec:introduction}Introduction}

\noindent The Lambda Cold Dark Matter ($\rm \Lambda$CDM) model of the Universe is in excellent agreement with numerous different types of astrophysical observations \cite{Spergel:2007,Percival:2007}. According to the most recent results from Planck, the mass/energy content of the Universe comprises 26.8\% cold dark matter (CDM), which is nearly a factor of 5.5 more than the 4.9\% contribution of baryonic matter \cite{Ade:2015}. The identity of the CDM is unknown, but a number of theoretical models predict that dark matter particles interact very weakly with ordinary matter. Such models include weakly interacting massive particles (WIMPs) \cite{Goodman:1985,Lee:1977} and asymmetric dark matter \cite{Kaplan:2009,Zurek:2014}, and in both cases predict that the dark matter particles can elastically scatter with nuclei, producing nuclear recoils with energies of order 1--100 keV. Such elastic scattering events could be detectable using sufficiently sensitive instruments. A positive direct detection of dark matter interactions would open a window into physics beyond the Standard Model, give new insights to cosmology, and create a new field of direct dark matter observations.   

One of the most effective technologies for the direct detection of dark matter is the two-phase xenon time projection chamber (Xe TPC) \cite{Alner:2007,Akimov:2007,Aprile:2010bt,Akerib:2012:det,Cao:2014}. Xenon has low intrinsic radiological backgrounds since it has no long-lived isotopes other than the extremely long-lived $^{136}$Xe isotope, which undergoes double beta decay with a half-life of $2.1\times10^{21}$ years \cite{Albert:2014,Gando:2012} and, at the low energies relevant to WIMP detection, produces a background rate of events that is small in comparison to the electron scattering of solar neutrinos \cite{Baudis:2014}. Liquid xenon (LXe) is dense (2.9~g/cm$^{3}$), and produces large scintillation light and charge signals from both electron recoil (ER) and nuclear recoil (NR) events. The Xe TPC has the ability to discriminate ER from NR on an event-by-event basis using the charge-to-light ratio, and can be expanded to large homogeneous volumes. With excellent position resolution and short LXe gamma ray and neutron scattering lengths, this technology allows excellent self-shielding and extremely low backgrounds at low energies. 

The Large Underground Xenon (LUX) experiment~\cite{Akerib:2012:det} is a two-phase xenon-based dark-matter detector located at the 4850 ft. level of the Sanford Underground Research Facility (SURF) \cite{Heise:2015} in Lead, South Dakota, USA. LUX was assembled and first operated in a dedicated surface facility at SURF starting in 2009. A follow-up test of LUX on the surface (Run 2) started in October 2011 and ended in February 2012 \cite{Akerib:2012:surface}. The Davis Campus, at a depth of 4850 ft., was built to house LUX and other experiments, and completed at the end of May 2012. LUX began installation in the Davis Campus in June 2012 and the cooldown of the detector began in January 2013. 
The first WIMP search and calibration data (Run~3) were acquired during the period March-October 2013. First results from Run~3 were announced in October 2013 and subsequently published in Physical Review Letters \cite{Akerib:2013:run3}. 
The data were shown to be consistent with a background-only hypothesis, 
allowing 90\% confidence limits to be set on spin-independent WIMP-nucleon elastic scattering with a minimum upper limit on the cross section of 7.6$\times$10$^{-46}$ ~cm$^{2}$ at a WIMP mass of 33~GeV/c$^{2}$, the most stringent 
constraint on the dark matter scattering cross section at that time.  
Further extensive calibration studies, together with a large number of other improvements, enabled a more 
sensitive analysis of the Run~3 data to be performed, further improving the sensitivity, especially to lower mass WIMPs, and reducing the 90\% minimum upper limit exclusion to 6$\times$10$^{-46}$ ~cm$^{2}$, again for a WIMP mass of 33~GeV/c$^{2}$. This improved sensitivity was published as a second Physical Review Letter in early 2016~\cite{Akerib:2015:run3}, with a further Letter presenting results for spin-dependent interactions~\cite{Akerib:2016:SD}. Here, detailed and complete descriptions of the new calibrations are presented, together with the analysis methods that underpin these results, which were unable to be published in the short Letter format.

Recently, results from an extended 332-day exposure (Run 4) have been published~\cite{Akerib:2016:R4}, 
providing a further roughly four-fold improvement in sensitivity for high WIMP masses relative to previous results. With no evidence of WIMP nuclear recoils, WIMP-nucleon spin-independent cross sections above 2.2$\times$10$^{-46}$ cm$^{2}$ are now excluded at the 90\% confidence level. When combined with the previously reported LUX exposure, this exclusion strengthens to 1.1$\times$10$^{-46}$ cm$^{2}$ at 50~GeV/c$^{2}$. Many of the innovations outlined here persist in the analysis of these most recent data.

\section{\label{sec:LUX}The LUX Experiment}

\begin{figure}[htp]
\begin{center}
\includegraphics[width=0.38\textwidth,clip]{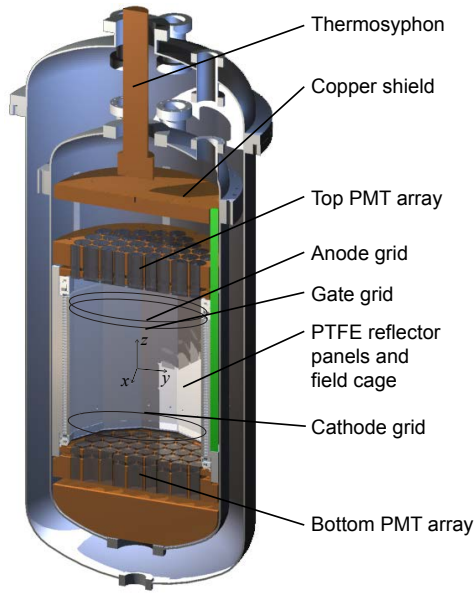}
\caption{Schematic of the LUX detector. LUX is a two-phase xenon time projection chamber containing an active LXe 
mass of 250~kg that is  viewed from above and below by PMT arrays.}
\label{fig:LUX_Schematic}
\end{center}
\end{figure}

\begin{figure}[htp]
\begin{center}
\includegraphics[width=0.48\textwidth,clip]{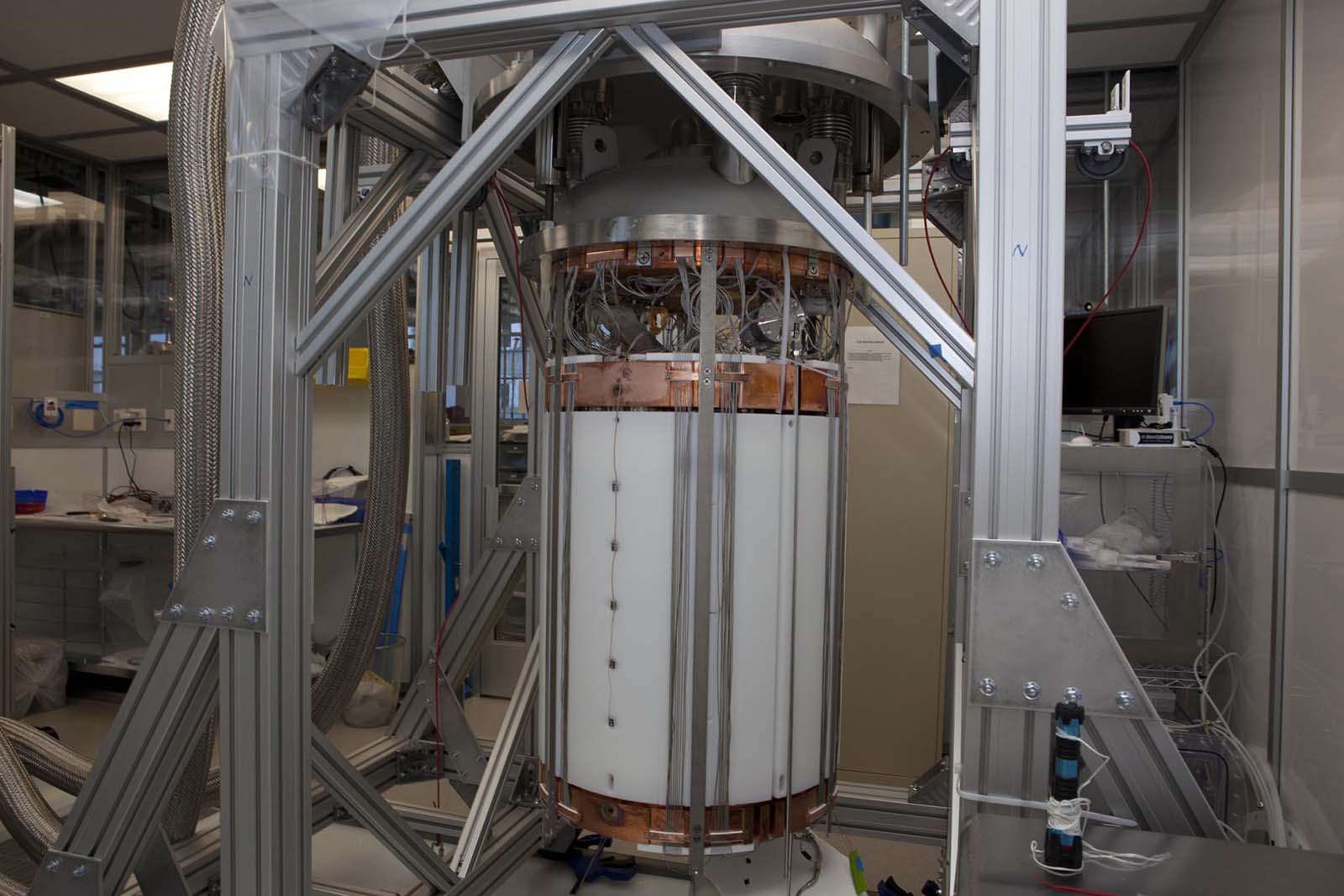}
\caption{Photograph of the LUX detector, with its inner vessel removed.}
\label{fig:LUX_no_can}
\end{center}
\end{figure}

\begin{figure}[htp]
\begin{center}
\includegraphics[width=0.48\textwidth,clip]{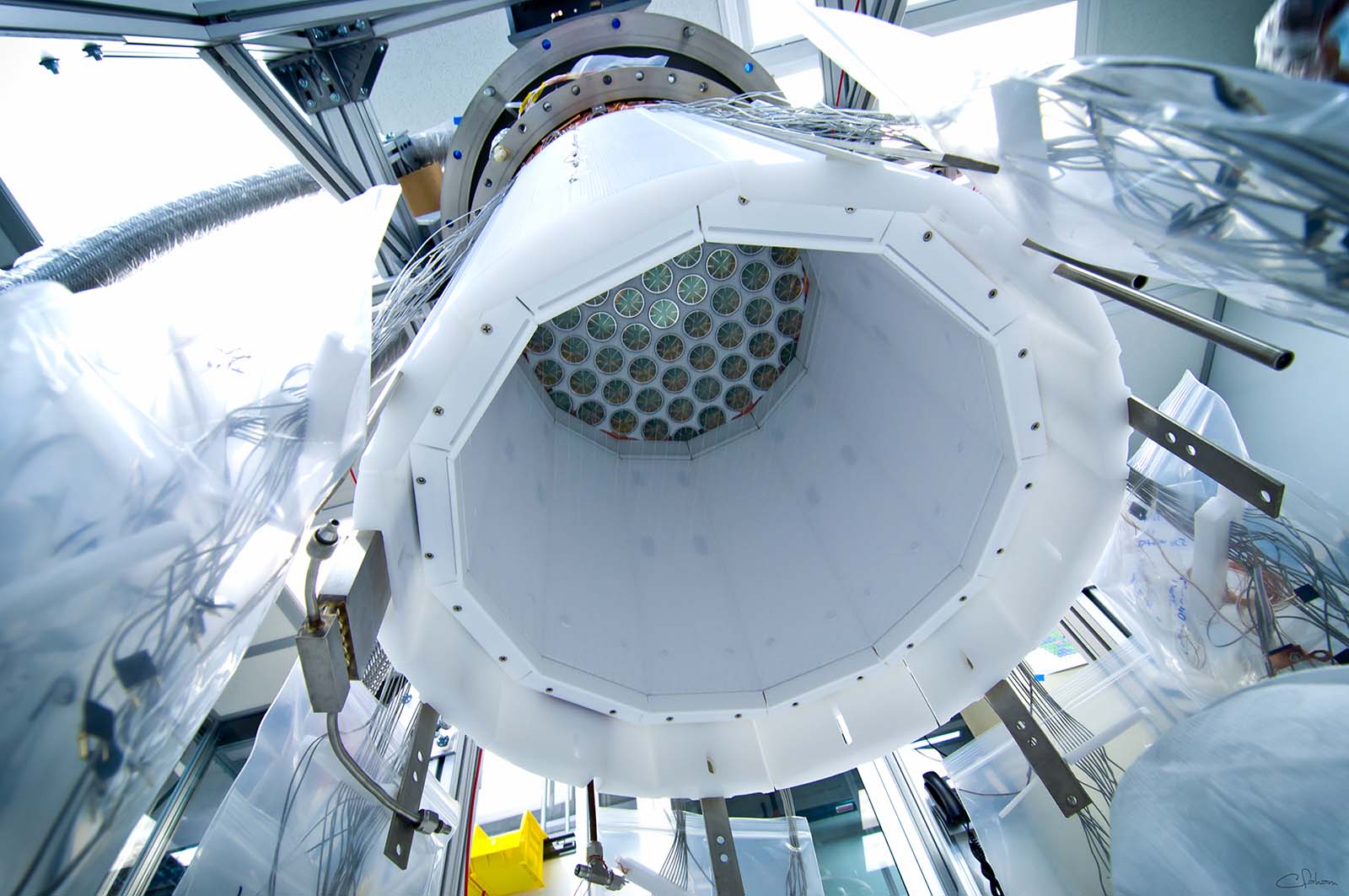}
\caption{Photograph of the inside of the LUX detector, with its lower PMT array removed.}
\label{fig:LUX_PTFE_photo}
\end{center}
\end{figure}

\begin{figure}[htp]
\begin{center}
\includegraphics[width=0.48\textwidth,clip]{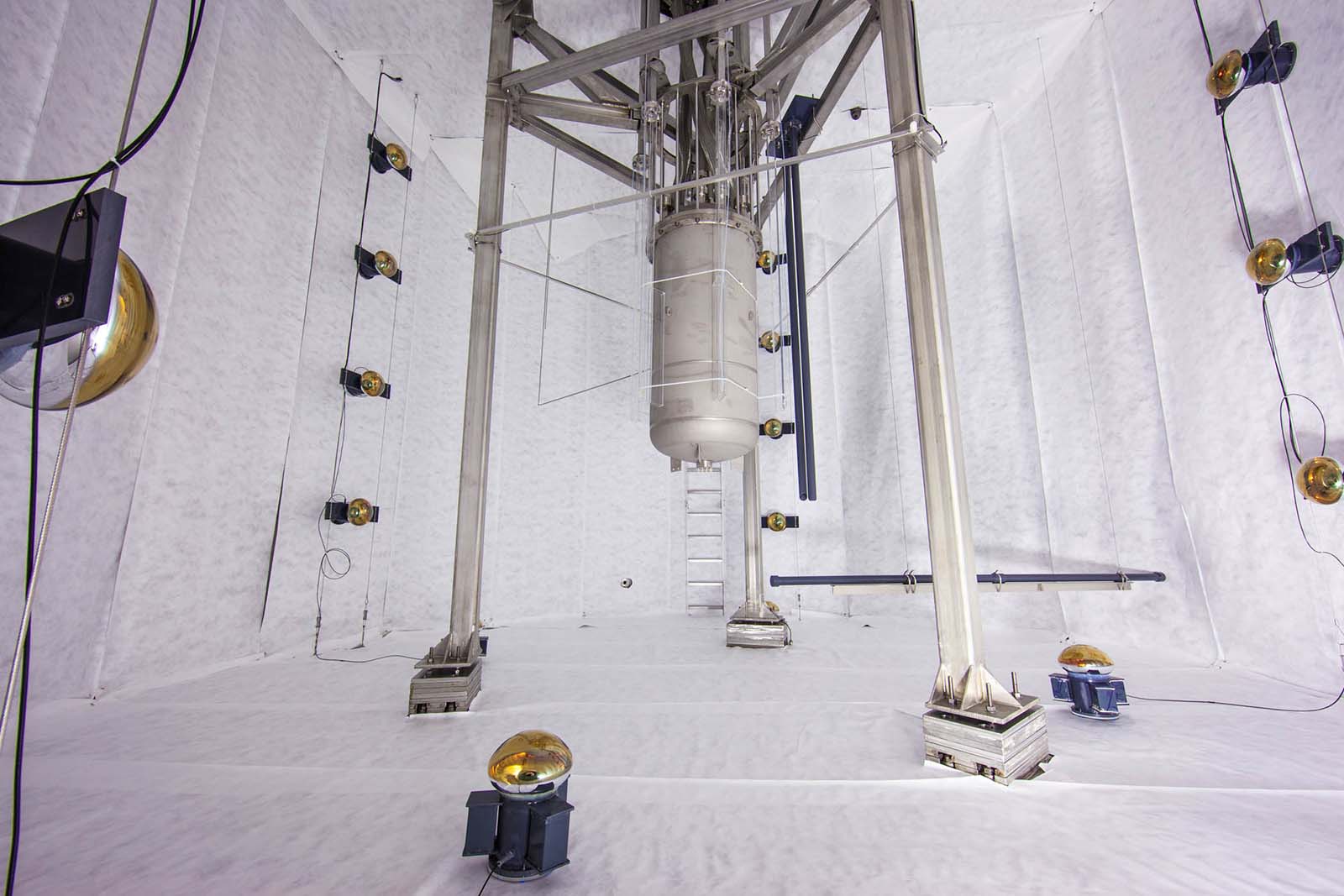}
\caption{Photograph of LUX installed in the Davis Cavern water shield at SURF.}
\label{fig:LUX_in_water_tank}
\end{center}
\end{figure}

\noindent LUX is a two-phase Xe TPC designed to detect both the prompt scintillation light and the delayed ionization electrons that result from ionizing radiation. A schematic of the LUX detector is shown in Fig.~\ref{fig:LUX_Schematic}, and photographs of LUX without its inner vessel, within the time projection chamber, and mounted in its water tank shielding, are shown in Fig.~\ref{fig:LUX_no_can}, \ref{fig:LUX_PTFE_photo}, and \ref{fig:LUX_in_water_tank}.

The central, fully-active Xe region is defined by 12 PTFE reflector panels, the gate grid located 0.5~cm below the Xe liquid surface, and the cathode grid located 48~cm below the gate grid. The diameter is 47~cm. Field shaping rings, spaced by 2~cm, are mounted on the outer PTFE vessel, and apply an electric field to drift the ionization electrons. An anode grid 0.5~cm above the surface, together with the gate grid, are used to generate high fields that extract the electrons and accelerate them, thus generating proportional scintillation light. Scintillation photons are produced almost uniformly throughout the gas gap, regardless of the electron extraction location. In Run~3, the LUX detector was operated with cathode, gate, and anode voltages of $-10.0$~kV, $-1.5$~kV, and $+3.5$~kV, respectively, corresponding to an average drift field of 180~V/cm and an electron extraction field of 5.53$\pm$0.30~kV/cm applied in the gas (2.84$\pm$0.16~kV/cm in the liquid).

An event in the LUX time projection chamber is characterized by two signals, corresponding to detection of direct scintillation light ($S1$) and proportional light from ionization electrons ($S2$). The two light pulses occur within a maximum drift time of 322~\textmu{}s, corresponding to a saturated electron drift velocity of 1.52~mm/\textmu{}s in LXe. Since electron diffusion in LXe is small, the $S2$ proportional scintillation pulse is produced in a small area at the gas-liquid interface. After corrections for any field-non-uniformities, the original event may be located accurately in the $xy$-plane, allowing 2D reconstruction. With precise $z$ information from the drift time measurement, the 3D event localization provides background discrimination via fiducial volume cuts.

Two arrays of 5.6~cm diameter Hamamatsu R8778 photomultiplier tubes (PMTs), 61~in each array, detect the $S1$ and $S2$ signals~\cite{Akerib:2013:pmt}. One PMT array above the liquid surface is primarily used to image the $xy$-position of the proportional light pulse. The second PMT array is in the liquid, below the cathode grid. Liquid xenon has a high refractive index, 1.69 at 170~K, causing internal reflection at the liquid surface that in turn leads to most prompt light being collected in this bottom array.  The high quantum efficiency of the R8778 PMTs, highly transparent grids, and use of PTFE reflectors between PMTs, means that a very high light yield is achieved, measured to be 8.8~photoelectrons/$\rm keV$ electron-equivalent energy (hereafter $\rm keV_{ee}$) for 122~keV $\rm \gamma$-rays at zero electric field.

The LUX collaboration has introduced a number of innovations, including a low-radioactivity titanium cryostat, nitrogen thermosyphons, high-flow Xe purification, two-phase Xe heat exchangers, internal calibration with gaseous sources of $^{83\mathrm{m}}$Kr and $^{3}$H, and nuclear recoil calibration using multiple scatters of monoenergetic neutrons produced with a deuterium-deuterium (\dd{}) neutron generator. 
The LUX cryostat vessels were fabricated from Ti with very low levels of radioactivity \cite{Akerib:2011:titanium}, rivaling the purities achieved in copper. The two-phase TPC technique requires precise control of the thermodynamic environment, and this was achieved in LUX through the development of a dedicated nitrogen thermosyphon system. This features precise, tunable, automated control delivering up to hundreds of kW of cooling power, plus reliable remote operation. 
A particularly important aspect of this system is that it has allowed highly controlled initial cooling of the LUX detector~\cite{Akerib:2012:det}, which is necessary to avoid warping of the large plastic structures of the TPC. A related development is the purification system that allowed rapid circulation of Xe through an external gas-phase getter. Flow rates exceeding 27 standard liters per minute (229 kg/day) were achieved, while a stable liquid surface was maintained through the use of a weir. 
Negligible overall heat load on the detector was then obtained through the use of a two-phase heat exchanger system \cite{Akerib:2013:heatexchanger} and very efficient heat transfer between evaporating and condensing Xe streams. Xe purification was further aided through the use of an innovative gas trapping and mass spectrometry system \cite{Dobi:2011:sampling,Dobi:2012:sampling},  sensitive to impurities at the sub-ppb concentration levels needed for good electron transport and light collection. This diagnostic capability allowed various portions of the gas system and detector to be monitored for contamination. Removal of Kr from the Xe, required to limit $^{85}$Kr and $^{81}$Kr beta-decay backgrounds, was performed before commencement of Run~3 using a charcoal column~\cite{Akerib:2016:kryptonRemoval}. These systems were demonstrated during the experiment's first science run, where cooldown was achieved in only nine days. Sufficient LXe purity to begin science operations was achieved only one month after the initial filling with LXe, at which time the electron drift lifetime (the mean time an electron remains in the LXe before being absorbed by an impurity) was over 500~\textmu{}s. Stable operation of the detector was maintained with mostly unattended operation over the five-month period, during which the pressure and liquid level had sufficient stability (1\%~and $<$ 500~\textmu{}m, respectively) to introduce no measurable correlations in the $S2$ or $S1$ signals.

\section{\label{sec:Data_acquisition}Data acquisition and reduction}

\subsection{\label{sec:DAQ}DAQ configuration and single photoelectron digitization acceptance}

The signal of each PMT output is amplified by a factor of 5 with a linear pre-amplifier located at the instrumentation breakout and is subsequently shaped by a post-amplifier that increases the pulse area by a further factor of 1.5. The post-amplifier boards have additional amplification outputs to feed the LUX trigger and discriminator boards. The output of the post-amplifier is digitized by 14-bit Struck SIS3301 ADC boards at a sampling rate of 100~MHz (10~ns data samples). The Struck board firmware was modified to use pulse-only digitization (POD), a zero-suppression mode that only digitizes signals above a specified threshold. Since signals in LUX are dominated by long periods of baseline with short bursts of $S1$ (width $<$ 100~ns) and $S2$ (width $<$ 5~\textmu{}s) signals, POD mode significantly reduces the number of recorded samples while increasing the maximum allowed acquisition rate. The signal threshold to begin digitizing the PMT signal was set to 1.5~mV at the Struck input; the signal threshold to end digitizing was set to 0.5~mV. An additional 24~samples before and 31~samples after the threshold crossings are also digitized. The LUX analog signal chain and DAQ maintain linearity in $S1$ and $S2$ signals at energies of $\sim$100~$\rm keV_{ee}$, well above the WIMP region of interest and comfortably above the $^{83\mathrm{m}}$Kr and tritium ER calibrations. For a more detailed description of the LUX DAQ system, refer to~\cite{Akerib:2011:daq}.

PMT signals are continuously recorded by the DAQ regardless of trigger conditions and the trigger pulse is digitized as an additional DAQ channel. Offline software, called the Event Builder, subsequently matches PMT signals within a specified time window around the trigger pulse for data processing and analysis. In this way, trigger changes can be made offline with no data loss.

\begin{figure}[t]
\begin{center}
\includegraphics[width=0.48\textwidth,clip]{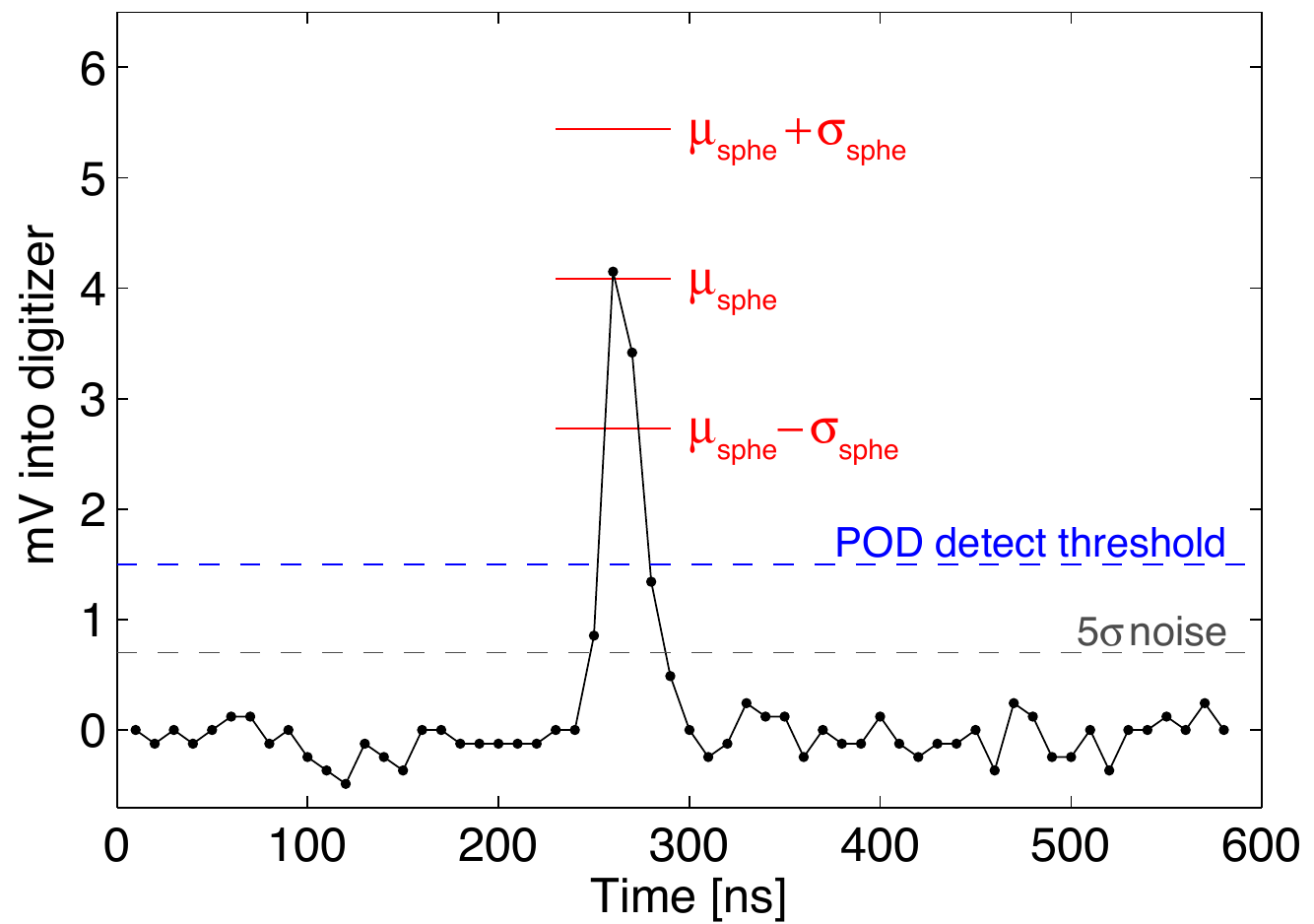}
\caption{A schematic diagram of the various thresholds applied to raw waveforms.}
\label{fig:sphe_threshold_diagram}
\end{center}
\end{figure}

For the nominal PMT gains of $4\times10^6$, a single photoelectron (sphe) generates a pulse with 4~mV amplitude and full width at half maximum (FWHM) of $\sim$20~ns. A typical sphe pulse is shown in Fig.~\ref{fig:sphe_threshold_diagram} with height markers for the mean sphe height ($\mu_\text{sphe}$) and its standard deviation ($\mu_\text{sphe}~\pm~\sigma_\text{sphe}$). The total noise from the electronics chain and the ADC, as measured at the Struck input, is 155 $\mu$V$_\text{rms}$. The 1.5~mV POD threshold is indicated in Fig.~\ref{fig:sphe_threshold_diagram}, as well as the height of a $5\sigma$ noise fluctuation.

\begin{figure}[t]
\begin{center}
\includegraphics[width=0.48\textwidth,clip]{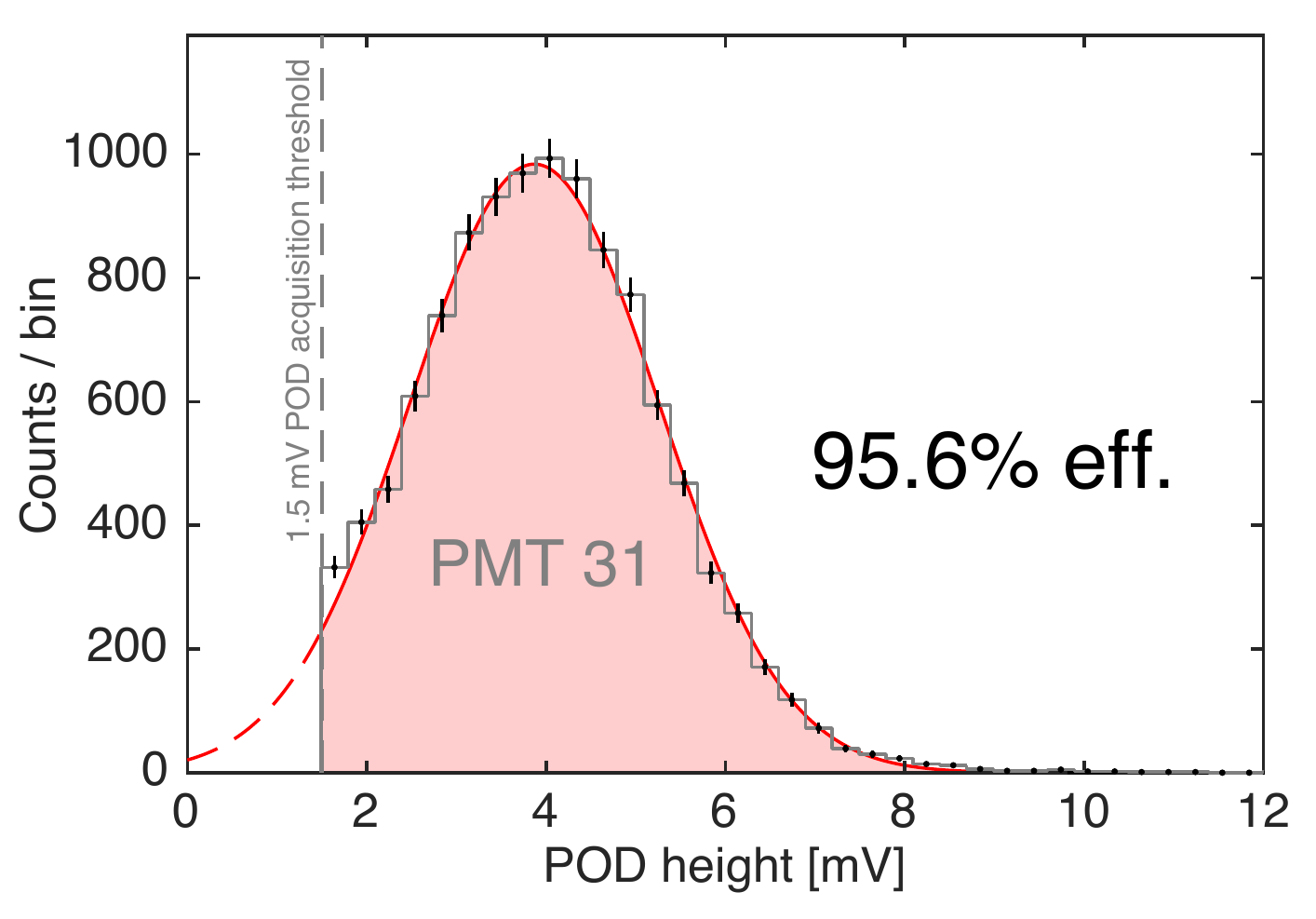}
\caption{Measured maximum sphe pulse height for PMT 31. The 1.5 mV POD threshold is indicated by the vertical dashed line. Based on a fit of the single photoelectron pulse height distribution, it is estimated that 95.6\% of the sphe in PMT 31 produce a pulse that exceeds the POD threshold.}
\label{fig:sphe_acceptance_example}
\end{center}
\end{figure}

\begin{figure}[t]
\begin{center}
\includegraphics[width=0.48\textwidth,clip]{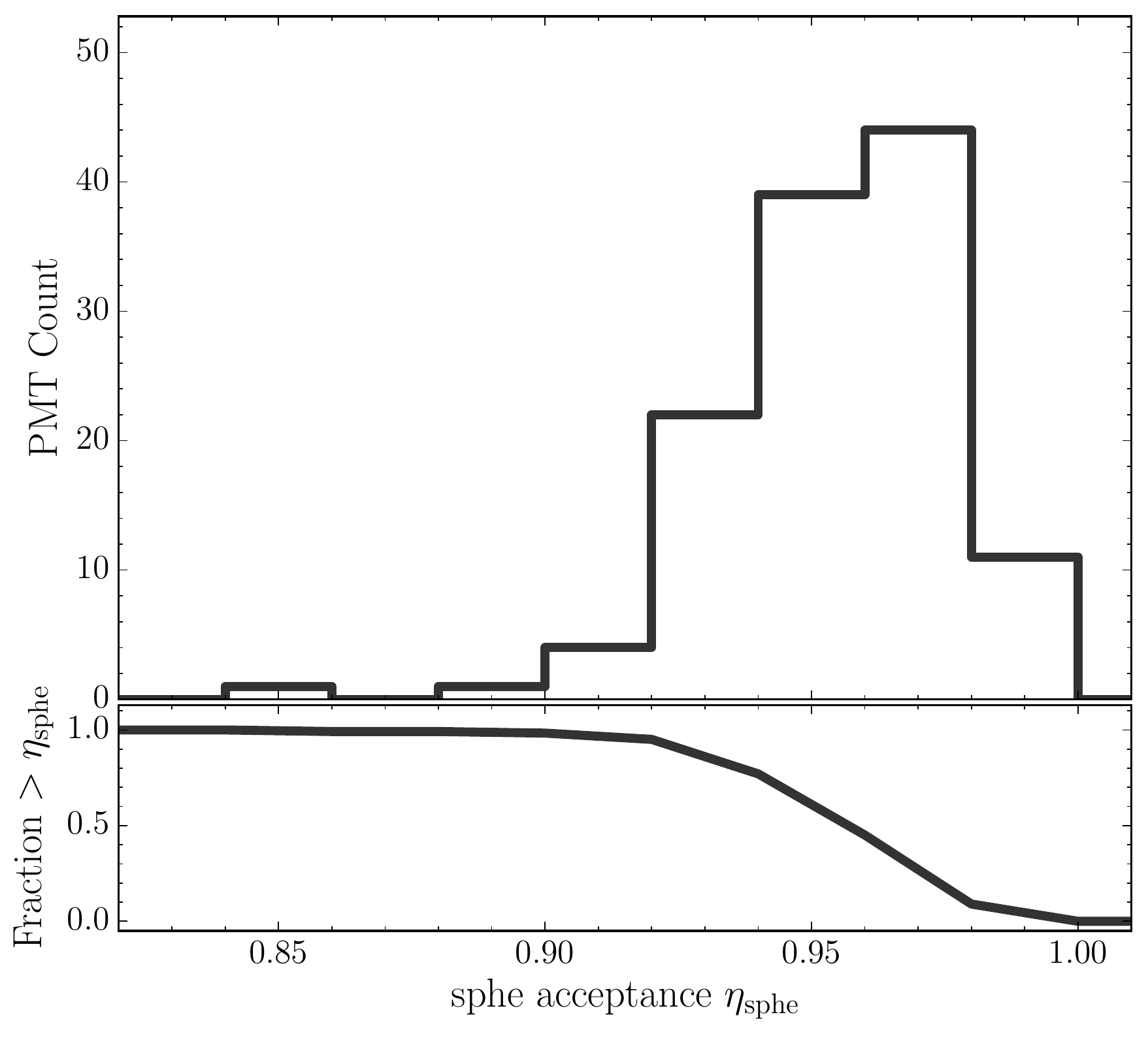}
\caption{Single photoelectron acceptance distribution for all 122 photomultiplier tubes}
\label{fig:sphe_acceptance_distribution}
\end{center}
\end{figure}

To estimate the sphe digitization acceptance for each PMT, $\eta_\text{sphe}$, a Gaussian distribution, truncated at the POD threshold of 1.5~mV, was fitted to each sphe spectrum. This sphe calibration was performed with 400~nm light pulses from the LED calibration system, which does not cause double photoelectron emission and therefore results in values for $\eta_\text{sphe}$ 
that are conservative, as double photoelectron emission would result in a very modest increase. 
Figure~\ref{fig:sphe_acceptance_example} shows an example distribution of sphe POD heights that has $\eta_\text{sphe}\sim0.95$. The distribution of $\eta_\text{sphe}$ values for the PMTs used during Run~3 is shown in Fig.~\ref{fig:sphe_acceptance_distribution}. All but two PMTs had a sphe digitization acceptance greater than 0.90; the mode of the distribution is above 0.95.

\subsection{\label{sec:Trigger}Trigger}

The LUX trigger system is described in detail in \cite{Akerib:2015:trigger}.
It is a digital FPGA-based system that flags events in the DAQ data stream for further analysis. 
All 122 PMTs are summed into 16 trigger channels, with no adjacent PMTs belonging to the same sum.
The sums are individually processed using two eight-channel DDC-8DSP digitizers/processors that communicate with a common Trigger Builder to make a final trigger decision. Internal digital filters perform baseline subtraction and signal integration.
In Run~3, the trigger required that at least two of these channels have a signal greater than 8~phe within a 2~\textmu{}s window. 
The overall trigger efficiency has a strong dependency on the required 2-fold coincidence and reaches $>$99\% efficiency for $S2$ signals with a total area of 100~photoelectrons (phe). Detailed study of the trigger efficiency is discussed in \cite{Akerib:LUX:TriggerEfficiency}. 
For the first third of Run~3 the hold-off time, which is a post-trigger time window during which we do not accept new triggers, was set to 4~ms. Analysis of the data being collected showed that the hold-off period did not have to be this long and subsequently was set to 1~ms for the remainder of the run. This improved the live-time slightly and was verified to have no negative impact on the overall results.

\subsection{\label{sec:DPF}The LUX data processing framework}

The LUX data processing framework (DPF) is a flexible, modular and multi-language framework developed by the LUX collaboration for extracting the relevant features from the raw digitized PMT data and returning a set of reduced quantities (RQs) that can be used for physics analyses. The LUX DPF employs interchangeable algorithm modules with standardized inputs and outputs that perform predefined tasks, such as calibration, pulse finding, classification and interaction vertex position reconstruction. These algorithm modules can be written in any supported programming language, which currently includes C++/ROOT, Python and MATLAB. A MySQL database (referred to as the ``LUG'') stores version-controlled calibration values and correction maps, data processing input settings and data processing logging, among other bookkeeping values. The LUX DPF was written entirely in Python and can be executed on computing clusters, and desktop and laptop computers. The list of modules to use, their order and specific configuration ({\it e.g.}, threshold value for a pulse finder algorithm) and required calibration constants must be provided to the framework in an input settings file in XML format ---which is stored in the LUG database and associated with a unique identifier (called the global setting number). This identifier allows the collaboration to easily establish the exact conditions that were used to process any dataset, and ensures that all the data used in a particular analysis campaign has been processed with the same settings.

\begin{figure}[t]
\begin{center}
\includegraphics[width=0.48\textwidth,clip]{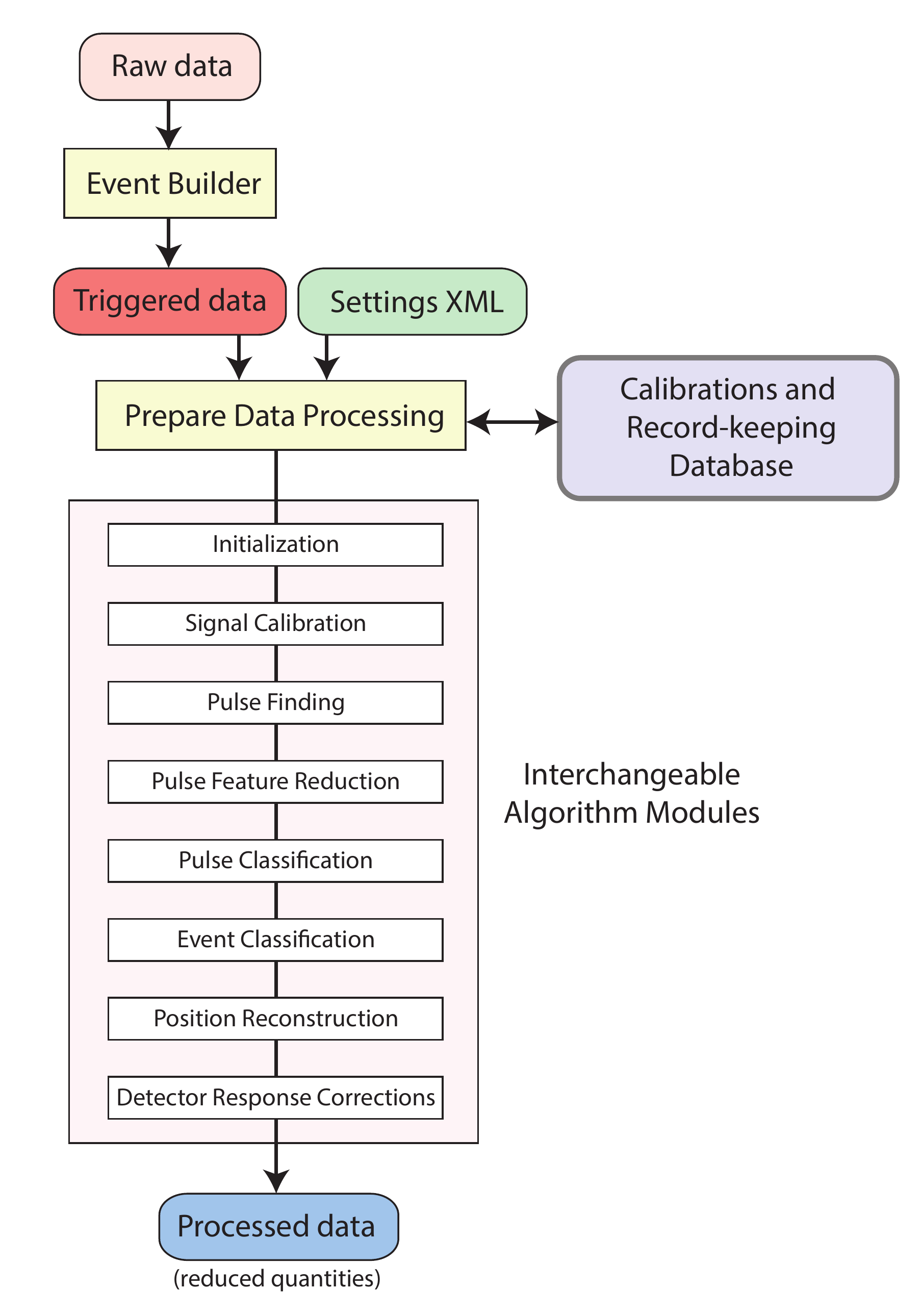}
\caption{Schematic view of the LUX Data Processing Framework}
\label{fig:data_processing_framework}
\end{center}
\end{figure}

Figure \ref{fig:data_processing_framework} shows a schematic representation of how raw data are processed through the LUX DPF to obtain the reduced quantities that are used in higher level analyses. The Event Builder reads the raw data as digitized by the DAQ and extracts the portions that are located in a time window before $\Delta t_\text{pretrigger}$ and after $\Delta t_\text{posttrigger}$. 
This window is referred to as an ``event''. For Run~3, both $\Delta t_\text{pretrigger}$ and $\Delta t_\text{posttrigger}$ = 0.5~ms. Given that the maximum Run~3 electron drift time was 322~\textmu{}s, these values ensure that both the $S1$ and the $S2$ pulses are contained in every event since either pulse type may induce a trigger.

The settings used in the Event Builder are stored in the LUG database, and the output file set is assigned a unique identifier that corresponds with the LUG record. The output of the Event Builder is read by the LUX DPF modules.

The LUG database, in addition to storing data processing bookkeeping values (such as the Event Builder settings, DPF global settings and details about each individual data processing run), also stores calibration constants for the detector. These include PMT gains, $x$,$y$,$z$ spatial calibration maps for $S1$ and $S2$ pulses, the electron lifetime, detector tilt measurements, light response functions for position reconstruction, and energy calibration parameters, among others. These are sent to each data processing run as specified in the XML settings file for access by the algorithm modules. The calibration constants stored in the LUG are stored with submission dates, version numbers, originating dataset name (from which the values were calculated) and algorithm names. The latter allows for different methods of obtaining a calibration parameter to be selected during a data processing run.

\subsection{\label{sec:algorithms}Data processing algorithms}

\subsubsection{\label{sec:PulseFinderClassifier}Pulse finder and classifier}

At the heart of a reduced-quantities-based analysis there is a pulse finding algorithm that searches for valid pulses in the acquired waveforms, and which then stores these relevant data for further analysis. Any detected pulses are subsequently classified according to their shape and properties.
The pulse finding algorithm module must fulfill the requirements of finding and separating all $S1$- and $S2$-type pulses, returning their start and end times accurately. It must do so without excluding a significant fraction of the pulse, as that would lead to under-representation of area. The pulse finder must also respect a limit imposed by the processing conditions of the DP framework of a maximum of 10 stored pulses, a limitation verified through extensive hand-scanning campaigns as not imposing significant efficiency loss.  

The pulse finder is based on a sliding boxcar filter (width of 4~\textmu{}s), applied to the full event waveform for each triggered event,  determining the region that maximizes the enclosed area. A moving average filter (30~ns width) smooths the regions before and after the maximum amplitude in the boxcar. The start and end times of the pulses are set at the point where the smoothed waveforms stay below the set baseline noise threshold of 0.1 detected photoelectrons (phd) per sample for a time of 0.5~\textmu{}s. In addition, valid pulses need to be at least 30~ns (3 samples) wide or their average amplitude must exceed the baseline noise threshold. If the pulse width exceeds 6~\textmu{}s, a moving average filter with a larger width parameter (250~ns) is used, moving forwards and subsequently backwards in time from the point of maximum amplitude in the waveform. If a falling edge is followed by a continuous rise in amplitude for a minimum of 0.5~\textmu{}s, this filter allows identification of possible additional signal clustered with the original pulse. If two or more individual signals are found, the algorithm splits the original waveform and stores only the start and end time of the largest of the two pulses. Finally, if a pulse has been found then the corresponding amplitude in the waveform is set to zero and all of the steps described above repeated until all pulses in the waveform, up to a maximum of 10, have been identified. 

\begin{figure}
\begin{center}
\includegraphics[width=0.48\textwidth,clip]{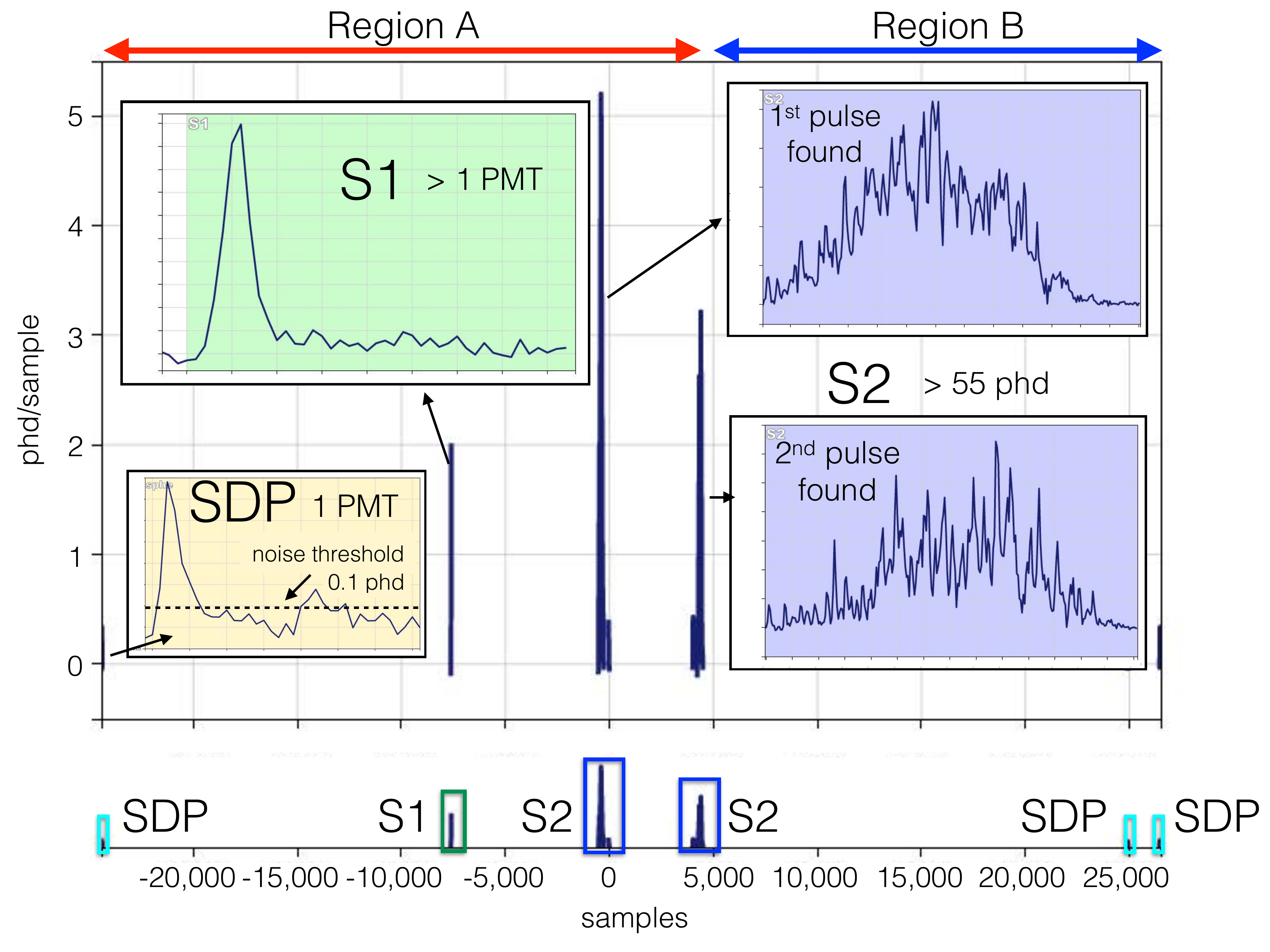}
\caption{An example of a double-scatter event waveform. The search regions are labeled A and B, with the first, A, prioritized. In addition to the main signal, this event contains three single detected photons (SDP), with one shown as an inset to indicate the noise threshold of the pulse finder. The bottom panel shows the location of pulses within the entire event record, with the $x$-axis as the number of 10~ns data samples either side of the trigger pulse, located at zero.}
\label{fig:waveform}
\end{center}
\end{figure}

The possibility of missing small $S1$ signals in the presence of single electrons, PMT afterpulsing, or other spurious signals, 
is minimized by using a complementary search logic that takes advantage of the preponderance for such pulses to dominate the region of the waveform following a large $S2$ signal. After the first two largest pulses have been found, the waveform is split into two search regions of differing priorities. The region before the pulse that occurred at the later time in the waveform is scanned first with the standard boxcar filter algorithm. If the maximum allowed number of pulses has not been reached after scanning this first search region, then the algorithm will continue to fill the empty slots with pulses from the second search region.
This methodology assists identification of events in which there are multiple interactions, {\it i.e.}, multiple scatters. Figure~\ref{fig:waveform} shows an example waveform of a multiple scatter event, indicating the regions of the search logic. Once pulses have been identified, independent modules subsequently parameterize the pulses for further analysis and classification of signal types.

\begin{figure}
\begin{center}
\includegraphics[width=3.2in,trim=0.5in 0.6in 0.5in 0.6in]{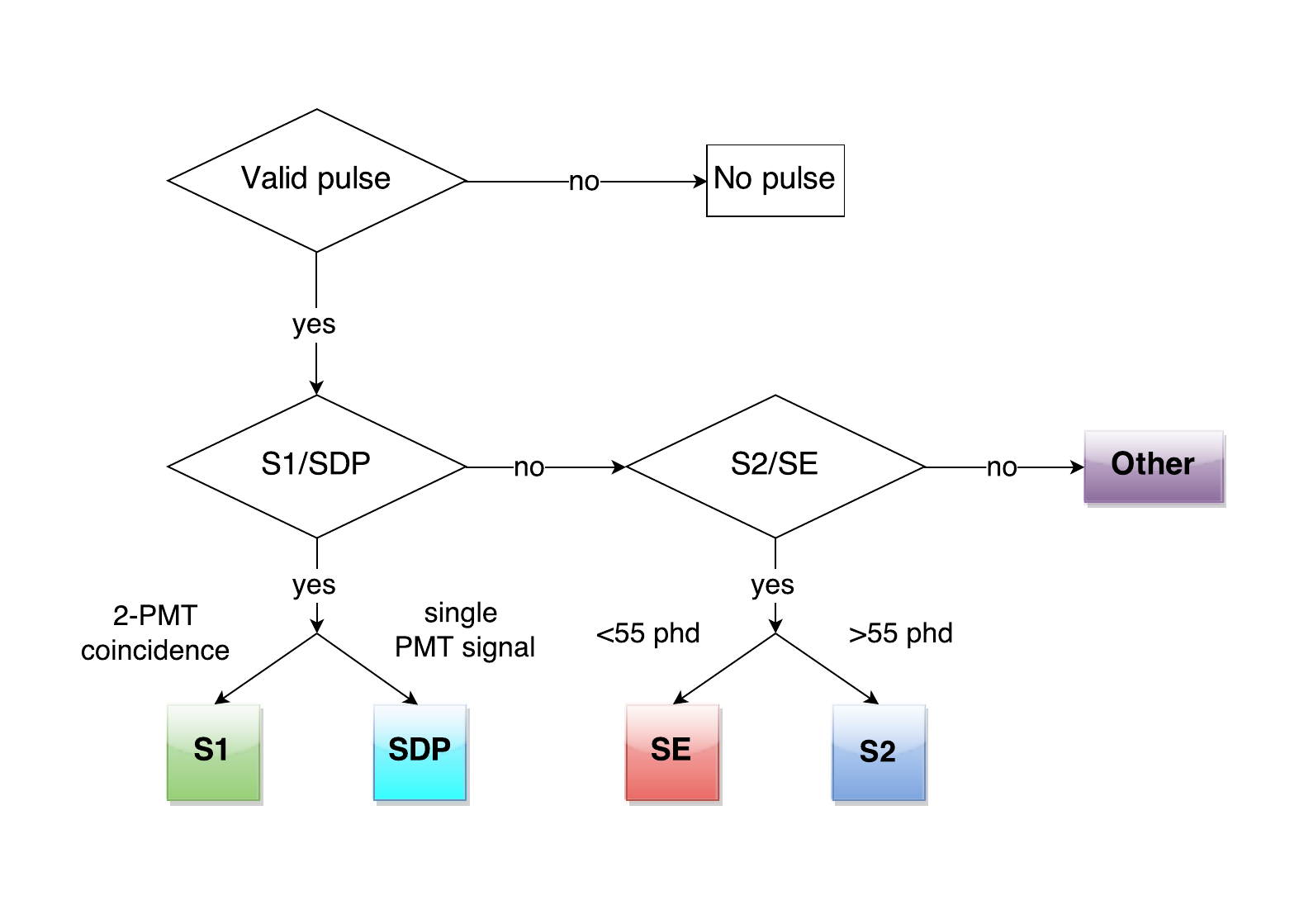}
\caption{Pulse classification decision tree.}
\label{fig:decisiontree}
\end{center}
\end{figure}

The pulse classification module assigns each identified pulse to one of the following five signal types, according to the extracted pulse parameters: $S1$, $S2$, single detected photon (SDP), single electron, and unclassified. The algorithm is represented by the decision tree diagram in Fig.~\ref{fig:decisiontree}. First, $S2$- and single electron-like signal types are assigned, followed by $S1$ and SDP type signals, ensuring that $S1$-like signals are not overwritten by $S2$-like ones. Events that fail all four previous categories are assigned an ``unclassified'' flag. The signal-type assignment is based on a number of pulse parameters as follows: a width selection, exploiting the much narrower width of an $S1$ in comparison to an $S2$-type signal, utilizing the ratio between two boxcar filters of length 0.5~\textmu{}s and 2~\textmu{}s; pulse rise time, based on width ratios and area fractions, using the distinct shape of an $S1$-like signal from scintillation that features a sharp rise at the pulse start in comparison to the rather symmetrical $S2$-like signals from electroluminescence in the gas; and PMT hit-distribution, where $S1$ signals are predominately recorded by the bottom PMT array due to total internal reflections off of the liquid-gas interface and the geometry of the TPC. The separation of SDPs and $S1$s, and likewise SEs and $S2$s, is an analysis classification dependent on signal size (rather than being physics-based) that enhances coincidental background rejection. For the assignment of an $S1$ signal two additional requirements have to be met. First, to reject pulses that are composed solely of baseline fluctuations, the maximum amplitude per sample within the full length of a pulse for at least 2 individual PMTs must exceed a 0.09~phd/sample amplitude threshold. Second, a 2-PMT coincidence requirement is imposed to reject designation of single detected photons from spontaneous PMT photocathode emission as $S1$-signal. The 2-PMT coincidence interval is set to 10 samples in width since it is expected that $\sim$80$\%$ of the $S1$ signal for over 90$\%$ of both ER and NR pulses will arrive within 100~ns. The 2-PMT coincidence requirement must be satisfied by at least two non-adjacent PMTs each exceeding a 0.3~phd integrated area threshold. A further discrimination between single electron-type and $S2$-type signals is gained by recognizing
their separation in energy, such that a valid $S2$ signal must exceed 33~phd in size (corresponding to $\sim$1.5 electrons) and classifying those pulses that fall below this threshold as single electrons (as long as their area is greater than 5~phd). Note that further (analysis-dependent) thresholds are usually applied on $S2$ size during event classification and selection.

\subsubsection{\label{sec:EventClassification}Event classification}

The event type of interest for the WIMP search analysis is a single scatter, or ``golden'', event. The definition and procedure for the selection of golden events is as follows: 
\begin{itemize}
\item There is only one valid $S2$ signal in the event after a valid $S1$ (golden events may include additional $S2$s that occur before the selected $S1$).
\item For the $S2$ to be valid, it must contain more than 55 ``spikes'' as counted by digital spike counting (see Sec.~\ref{sec:estimators}).
\item There is only one valid $S1$ signal before the selected $S2$ ($S1$s following the $S2$ signal are allowed).
\item The area of the $S2$ pulse must be larger than the area of the $S1$.
\end{itemize}

The purity of the golden events selected by the DPF, and the efficiency of the algorithms used, were evaluated through a detailed hand-scanning campaign. From an AmBe NR calibration dataset (live-time of 2.53~h), 4000 pre-selected events were categorized by eye using only the raw waveforms without any information from the reduced quantities. The pre-selection of events ($\sim$2\% of the dataset) was necessary to reduce the number to be scanned and, at the same time, to increase the sample size for single-scatter events in the region of interest. The criteria for pre-selection were predominately based on event information (number of non-empty samples in an event and the full event area), utilizing only very basic additional pulse parameters such as the largest pulse found and whether a clear sub-cathode event had been identified. Of the 4000 pre-selected events, approximately 200 single scatter events were identified. These were then compared to the result from the event classification DPF module. Applying all WIMP search analysis cuts (see Sec.~\ref{sec:EventSelec}), the purity of genuine golden events selected by the DPF was determined to be 98\%, and the efficiency of the DPF to select golden events from those identified in the data by the hand-scan was 98.8\%.

\subsubsection{\label{sec:PositionRec}Lateral position reconstruction}

The $xy$-position of an interaction in the LUX detector is recovered directly from the observed $S2$ signal, by considering the distribution of pulse areas in the upper photomultiplier array. The algorithm used is called ``Mercury'', and is based on the method developed for the ZEPLIN-III experiment~\cite{Solovov:2011}. The algorithm is a statistical search for the $xy$-position that matches the distribution of observed $S2$ pulse areas in the PMT array to those obtained using a pre-determined set of empirical functions, called light response functions (LRFs), which describe the average response of each individual PMT as a function of the interaction position. The major advantage of the Mercury method is that it needs only measured data, rather than simulations, to recover the position of interactions, and thus it can recover features from the data that are not well simulated.

The LRFs for each PMT are obtained through an iterative fit to experimental data, which in the case of Run~3 corresponds to calibration data obtained after the injection of $^{83\mathrm{m}}$Kr to the detector (see Sec.~\ref{sec:calibrations}).
In each iteration, new LRFs are obtained by fitting the response of the individual PMTs as functions of the event positions. These new functions are then input to the position reconstruction program to
derive improved estimates of the position of interactions, which are then used to find new improved LRFs. This process is iterated several times until the functions are stable. 
For the first fitting iteration only, the initial $xy$-positions are obtained using a simpler method of position reconstruction ({\it e.g.} a weighted mean, as used here). 

The simplest model for the LRFs, ${\mathcal H}_i$, consists of a radially symmetric function, $\eta$, that depends only on the distance  between the center of the PMT and the position of emission of light, $\rho$. It can be described as
\begin{equation}
 {\mathcal H}_i = {\mathcal A}_T {\mathcal C}_i \eta\left(\rho\right),
 \label{response_function}
\end{equation}
where ${\mathcal A}_T$ is the total pulse area of the event and the coefficients ${\mathcal C}_i$ normalize the response of each PMT in the top array~\cite{Akerib:positionreconstruction}. While this model was successfully used in ZEPLIN-III, it provides 
unsatisfactory results in the present work, due to the higher reflectivity of the inner walls. In LUX, the inner walls are covered with PTFE, which is a very good reflector for the xenon scintillation light, causing the amount of reflected light to increase for events near the walls. Consequently, a more sophisticated model that used 2-dimensional functions for the PMT response has been employed, which models the LRFs as sums of a radial component and a polar component, $\varepsilon_i$, defined for each PMT as
\begin{equation}
 {\mathcal H}_i = {\mathcal A}_T {\mathcal C}_i \left[\eta\left(\rho\right) + \varepsilon_i\left(r, \rho \right)\right].
 \label{response_function}
\end{equation}
The first component describes the light that propagates directly from the interaction to the PMT and depends only on $\rho$. The second component corresponds to the light that is reflected in the walls of the detector, and is described as a function of the event radial position and also the distance between the event and the center of the PMT~\cite{Akerib:positionreconstruction}. 

\begin{figure}[t]
\begin{center}
\includegraphics[width=0.48\textwidth,clip]{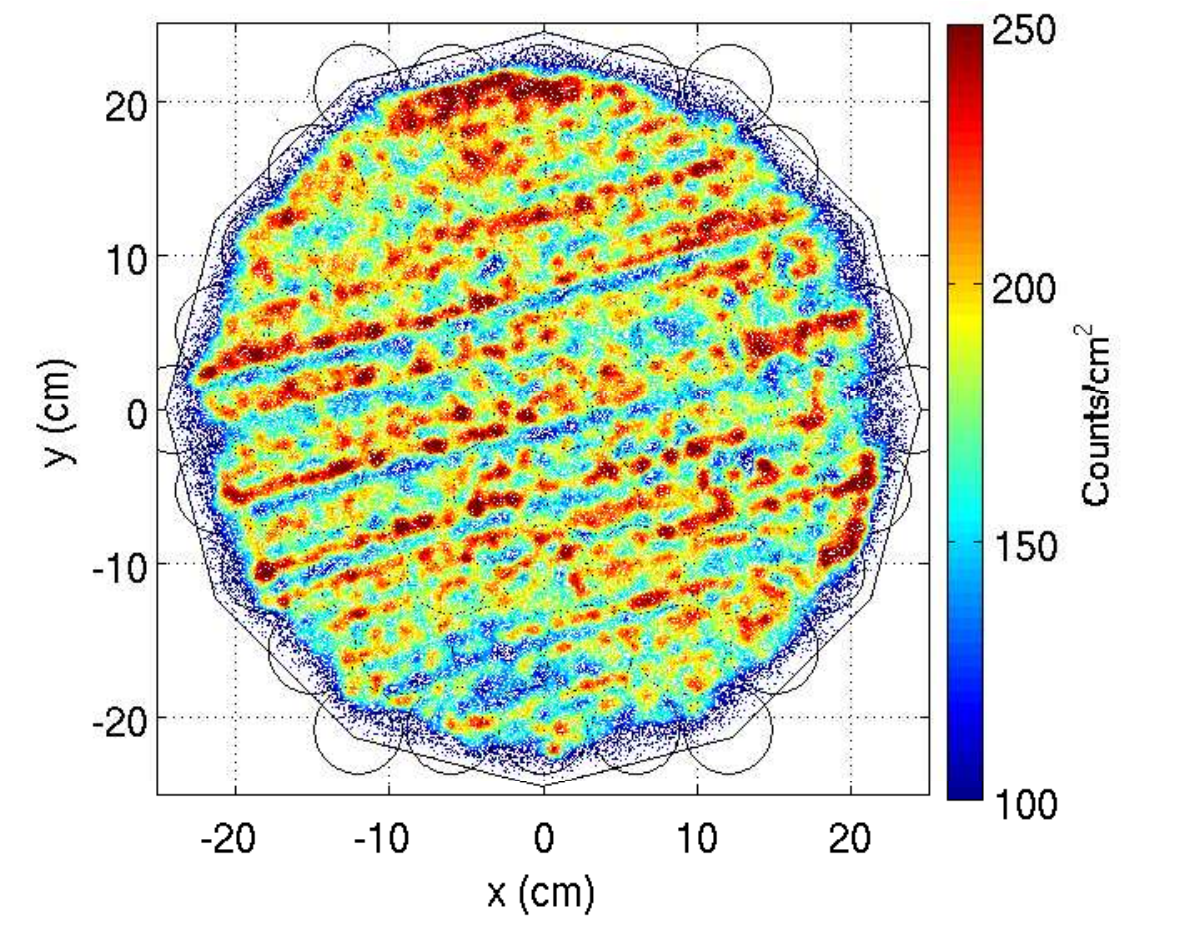}
\caption{Map of reconstructed positions in the $xy$-plane from data obtained after an injection of $^{83\mathrm{m}}$Kr in the detector. The observed striped pattern is the result of a focusing effect of the electrons near the gate wires, created by the difference between the fields in the drift and the extraction regions.}
\label{fig:Kr_XY_map}
\end{center}
\end{figure}

Figure~\ref{fig:Kr_XY_map} shows the final $xy$-distribution of events in the krypton calibration dataset using the Mercury algorithm. The observed striped pattern is a consequence of the geometry of the wires of the grid used to provide the electric field.  The pattern originates from a focusing effect of the electrons near the gate wire region, caused by the difference between the fields in the drift region (180$\pm$1~V/cm) and in the liquid extraction region (2.84$\pm$0.16~kV/cm). This effect is a limiting factor for the position resolution, and may be used to assess the quality of the reconstruction.
As discussed later, a non-uniform field within the xenon volume results in events at greater depth having their positions reconstructed at less than their true radius: the correction of this effect is discussed separately in Sec.~\ref{sec:PosRec}. 

\subsubsection{\label{sec:estimators}Estimators of light and charge}
The $S1$ signals generated by WIMP recoils are expected to have a mean number of detected photons per PMT that is significantly less than one. This small signal may be reconstructed with improved precision, as compared to 
simple pulse area, by counting of the candidate single-photon pulses in individual PMT waveforms, termed here `spikes'. Such digital counting avoids the variance that is inherent in the width of PMTs' single-detected-photon area distributions. The raw number of spikes is obtained in LUX by counting maxima in the regions of waveform that are above the POD-start threshold. A simple Monte Carlo model of spike overlap in time, based on average $S1$ and single-photon pulse shapes, was used to generate a look-up table of the most likely number of true photons as a function of raw area and as a function of counted spikes. Data from tritiated methane calibrations (see Sec.~\ref{sec:calibrations}) were used to demonstrate that the average of the simple area estimator to this combined area estimator agreed systematically to within 5\% everywhere, and to within 1\% from 16~phd to 80~phd, above which areas alone were used as pileup becomes significant.

The last step of event reconstruction is to account for spatial and temporal variation in the detector response.
The dominant sources of non-uniformity are the geometry-dependent probability for signal photons to be detected by the PMTs and, additionally for $S2$, the time-varying concentration of impurities in the xenon which capture ionized charge as it drifts towards the liquid surface. Using the krypton calibration data, maps of the relative position-dependent response were generated for all events. New calibrated $S1$ and $S2$ signal areas were defined such that they would equal the raw pulse area for these events if they had occurred at the detector center and there been  no signal loss due to impurities. The symbols $S1_c$ and $S2_c$ were used to denote these final, flat-fielded estimators of light and charge: $S2$ is always measured via pulse areas, with $S1$ also utilizing digital spike counting for small pulses (up to 80~phd).

The absolute energy scales of scintillation photons and ionization electrons were obtained from a set of responses to monoenergetic ER sources using the Platzmann model (see Sec.~\ref{sec:Doke}); however, the WIMP search does not rely on these scales as the detector's NR and ER responses in $S1_c$ and $S2_c$ are calibrated \emph{in situ}.

Collectively, the above procedures result in high-level output of the data processing framework that is a set of observables measuring position, light and charge for each reconstructed event above threshold in the active region of LUX: $x$, $y$, $z$, $S1_c$ and  $S2_c$. Subsequent analyses apply appropriate event selection and make inferences about physics models by comparing observed and predicted distributions in these observables.

\section{\label{sec:LUXSim}Simulations: LUXS\lowercase{im}}
\subsection{\label{sec:NEST}Infrastructure}
\noindent A detailed understanding of the physics capability of advanced instrumentation is frequently achieved through use
of a sophisticated Monte Carlo computer simulation of the apparatus. This is useful in both design and exploitation. 
Here, the LUX simulation package is presented, known as LUXSim~\cite{Akerib:2012:luxsim}. Its structure can be divided into 5 overarching, mostly serial functions:
\begin{itemize}
\item Recording of simulated energy deposits. This may be for several different particles in multiple volumes at different times, and uses the Geant4 toolkit~\cite{Agostinelli:2003,Allison:2006}, specifically 
Geant version 4.9.4 patch 4.
\item The production of both VUV scintillation photons and thermal ionization electrons.
This is modeled with the NEST (Noble Element Simulation Technique)~\cite{Szydagis:2011,Szydagis:2013} formalism, a frequently updated semi-empirical collection of models based on past detectors' calibration data. 
The number of such photons and electrons that are generated is in general anti-correlated and depends 
on the interacting particle type, the magnitude of the electric drift field and the energies of the recoils.
\item Propagation of photons using the Geant4 optical model (default libraries within the version specified above). Photons from the initial primary scintillation in the liquid are propagated until reaching PMTs or becoming absorbed by impurities. This is simulated directly by means of either an exponential mean free path for photon propagation through the xenon, or by imperfectly reflective surfaces.
\item Drifting of ionization electrons. Low-energy electrons liberated by an interaction are drifted up through the xenon using NEST, diffusing in three dimensions and being absorbed in a similar manner to the photons, but now using the empirically-determined (with $^{83\mathrm{m}}$Kr) electron absorption length, and the field-dependent, binomial electron extraction efficiency at the liquid-gas interface. Once the electrons reach the gas, then NEST produces the secondary scintillation as a function of field, density, and gas region length. The electroluminescence photons are again simply propagated with Geant4 optical processes. The drift and extraction fields are both sufficiently uniform in LUX that modeling them as scalar constants serves as an accurate representation for most purposes. See Sec.~\ref{sec:field}.
\item Pseudodata generation. Quantum efficiencies (QEs) are simulated as a function of incoming photon wavelength, angle, temperature, and PMT variation. A full custom simulation for the unique DAQ takes numbers and arrival times of primary and secondary photons and generates output files from LUXSim that are in an identical format to empirical data. These may then be processed using the same data processing framework, enabling direct comparisons.
\end{itemize}
LUXSim provides a component-centric approach: this makes it possible to define any parts of the detector, not just the PMT arrays, as sensitive volumes, making testing and validation studies easier to perform. Although the NEST framework, based on earlier experiments' results, is used for the absolute photon and electron yields, slight modifications are made to the values of various free parameters, fine-tuning them to more closely match LUX calibration data at its particular electric field, for both ER and NR, as discussed in Sec.~\ref{sec:calibrations}.

Full ER and NR simulations were performed and passed through the data-processing chain. 
Initially, the simulation was validated in terms of ability to reproduce raw waveform shapes, followed by 
reproduction of calibration results. With this achieved, LUXSim was then used to produce ER and NR energy spectra
for non-calibration scenarios. Examples included generating samples of ``pure'' single-scatter events, 
NR events with no contribution from ER contamination, misidentified multiple-scatter fiducial events, 
or multiple scatters with vertices outside of the drift region. These are all populations of events that can occur, 
especially with AmBe or $^{252}$Cf sources, demonstrating that the simulation allows exploration of specific (and/or rare)
event topologies. The simulation was then also used to provide estimates of the expected WIMP search 
spectra overlaid with the dominant background contributions from (measured) radioactive impurities in the detector
components, contributing ER-type signals overwhelmingly. Predicted WIMP NR signal models could then also be
generated, for various candidate WIMP masses and, together with the predicted background spectra, used as 
probability density functions (PDFs) for use in the Profile Likelihood Ratio (PLR) analysis, see Sec.~\ref{sec:limit}. 
LUXSim combined with the $^{83\mathrm{m}}$Kr calibration of the position-dependent light collection is critical in simulating WIMP interactions; none of the NR calibrations are uniformly distributed, so they were not direct WIMP analogues.

\begin{figure}
\begin{center}
\includegraphics[width=0.38\textwidth,clip] {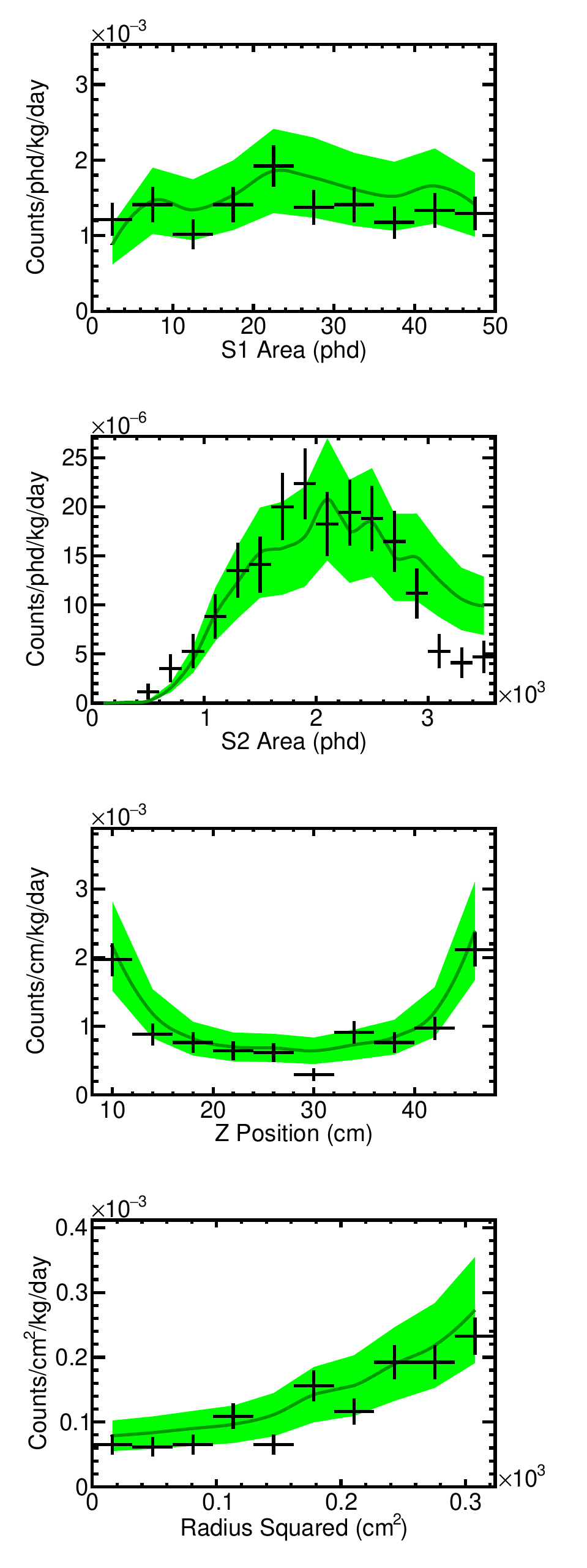}
\caption{From top to bottom: Profile of the absolute number of background events as function of $S1$ size, $S2$ size, depth, and radius, combining all known sources of background. S1 is cut off at 50~phd (photons detected), the WIMP search cut, causing drop-off in $S2$ spectrum. The black points and crosses are data, while the smooth gray center-lines come from a simulation based on a data-informed full background model. The green bands represent the 1-sigma uncertainties in the simulation: they are $\sim$30\% of the value of the background model, aggregating uncertainties across different background components~\cite{Akerib:2014:bg}. The disagreement at high $S2$ comes from NEST overestimating high-energy charge yield for electron recoils. Because this was at high energy, the impact on the WIMP search sensitivity was negligible. (This was corrected in the following run, 332 live-days, with new calibrations.)}
\label{fig:BGeSpec}
\end{center}
\end{figure}

Extensive campaigns of radioactivity screening with high-purity germanium detectors, inductively coupled plasma mass spectrometry, and neutron activation, provided detailed knowledge of the radioactive contamination of many of the detector components, crucial to the high fidelity of any simulation. The results of these studies formed the LUX background model~\cite{Akerib:2014:bg} and provided input into the Monte Carlo in terms of radioisotope levels in each component. 
Simulated $S1$ and $S2$ spectra from these contributions were then compared with the actual background observed during the WIMP search, in terms of not only the absolute count rate but also the position-dependent profile, crucial for the PLR analysis (Fig.~\ref{fig:BGeSpec}). 
The primary background constituents are ERs from the PMTs' uranium and thorium decay-chain gamma-rays, with further 
contributions from internal sources (such as $^{127}$Xe) and $^{40}$K.
To precisely match data, the initial contamination values required fine-tuning, though well within errors, from the predictions based upon material screening values.

\begin{figure}
\begin{center}
\includegraphics[width=0.499\textwidth,clip] {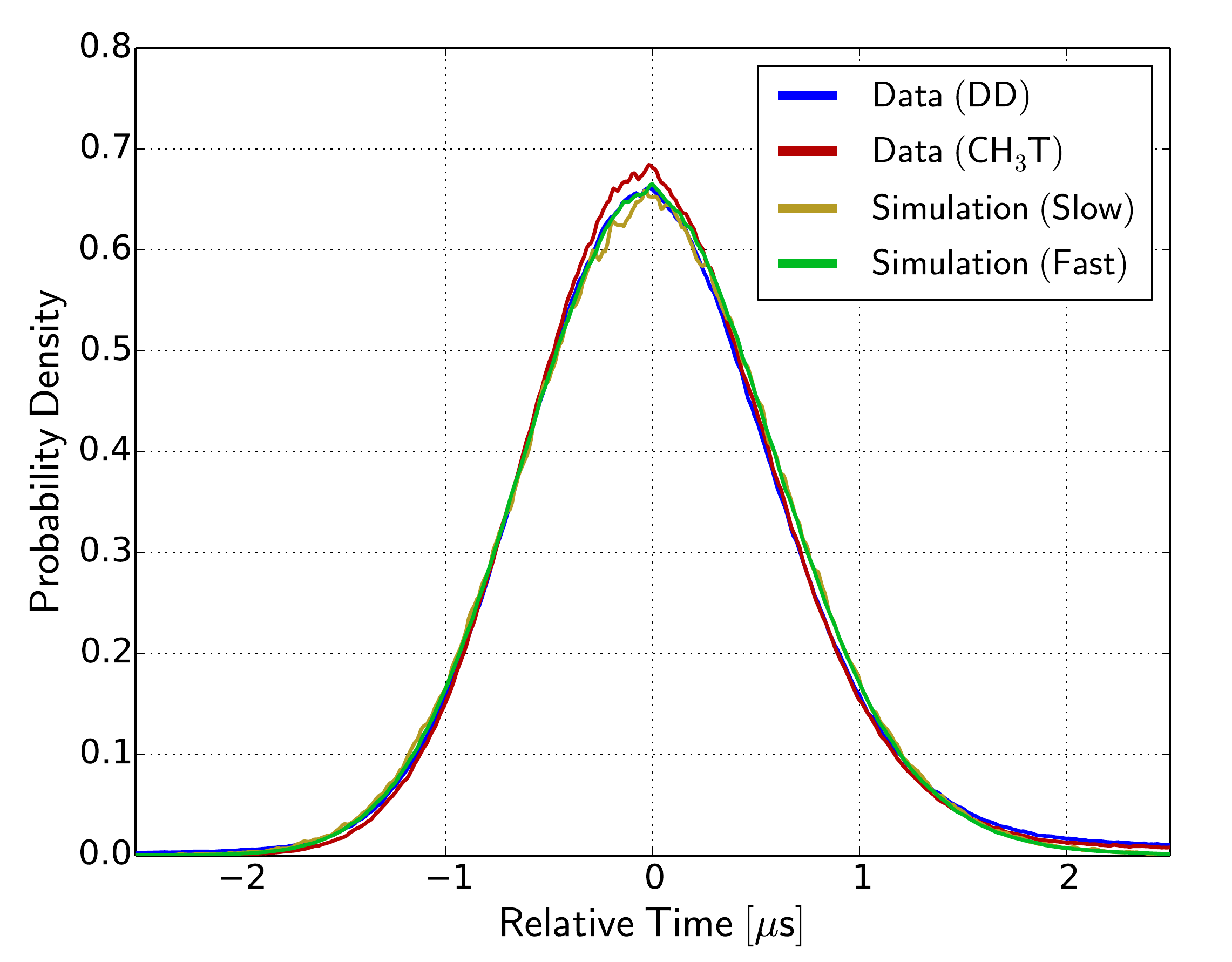}
\caption{The averaged $S2$ shapes from NR (\dd{} neutron calibration data, in blue) and ER (CH$_{3}$T, in red). Slow simulation (yellow) refers to a G4/NEST prediction from first principles based on the electrical and geometric properties of the detector. Fast simulation (green) refers to the parametrized PMT hit-map library mode of LUXSim that is forced to agree with data as closely as possible. A drift time cut of 100-120~\textmu{}s is applied in all cases so that \dd{} and tritium can be compared directly. In contrast to $S1$ signals, no significant differences are expected for different interaction types.}
\label{fig:S2_spec}
\end{center}
\end{figure}

The simulated sphe area (in voltage) and shape (roughly Gaussian) were calibrated to sphe events occurring in LED calibration and WIMP search data sets. In principle, the size of single electrons, in terms of number of phe, could 
have been predicted by calculating the absolute $S2$ photon yield~\cite{Monteiro:2007,Fonseca:2004} and knowing the light collection efficiency for the gas region from optical simulations, and the measured QEs. This method 
in fact agrees within 10\% of the observed single electron pulse area~\cite{Malling:2013,Chapman:2014}. The width and general shape are matched using a Gaussian distribution for the absolute yield in the gas gap and binomial light collection and QE, yielding a slightly non-Gaussian shape that is indeed observed in data. A comparison of the single electron area between simulations and data is shown in Fig.~\ref{fig:SE_spec} and discussed further in Sec.~\ref{sec:calibrations}. The single electron width in phe number is about twice the value expected from a Poisson process ($\sqrt{N}$), though this is understood as being due to field non-uniformities between the anode and liquid-gas border~\cite{Oliveira:2011}. The temporal profile of a single electron, or of an $S2$ of any size in general (see Fig.~\ref{fig:S2_spec}), matched well that expected given the known electron diffusion constants in liquid xenon, the electron drift speeds in liquid and gas, the electron extraction delay at the liquid surface, the (small) light travel time to the PMTs, and the singlet and triplet time constants characteristic of the excited molecular (excimer) states of xenon dimers. The last two contributions, plus the time it takes for ionization electrons to recombine with ions, also allows reproduction of the $S1$ pulse shape~\cite{Mock:2014}.

\subsection{\label{sec:OpticalModel}Optical model}
In contrast to NEST, the model for photon propagation has been specifically tailored to the present detector conditions. The parameters within the Geant4 model that were tuned, such that the simulation accurately replicated the data, were:
\begin{itemize}
\item the bi-hemispherical reflectance of the stainless steel field-generating grids assumed to be constant with the angle of incidence and the bi-hemispherical  reflectance of PTFE in liquid and in gas assumed to be only diffuse (Lambertian);
\item the photon mean free path, separately tuned for liquid and gas phases;
\item the reflectivity of the aluminum flashing on the PMT quartz faces;
\item the average QE for the top PMT array relative to the bottom, an absolute normalization systematic used to provide better agreement with data for the ratio of top/bottom signal size. This correction, around 2\%, was within the expected uncertainty.
\end{itemize}
The indices of refraction of xenon and quartz were based on previous data~\cite{Silva:2010,Hitachi:2005} while the Rayleigh scattering length came from a theoretical calculation~\cite{Seidel:2002}. Of these, the parameters with the most impact were found to be the reflectivity of the PTFE walls in the liquid and the photon absorption length in the liquid. For both, an increase in parameter value implies better light collection, and thus they are partially degenerate, although some discrimination was afforded through comparisons of $S1$; $S2$; the position dependence of each; and the ratio of light in the top or bottom PMTs to the total light collected (Figs.~\ref{fig:OpticalModel1} and \ref{fig:OpticalModel2}). The strongest optimization of the optical model was found to come from consideration of the $S1$ pulse area as a function of depth from the main mono-energetic peak (32.1~keV) in high-statistics $^{83\mathrm{m}}$Kr calibration data. The radial and angular dependencies, due to the orientation of the grids and the variation of PMT QEs, were found to be of secondary importance. The vertical symmetry is significantly broken 
by total internal reflection at the liquid surface.

\begin{figure}
\begin{center}
\includegraphics[width=0.48\textwidth,clip] {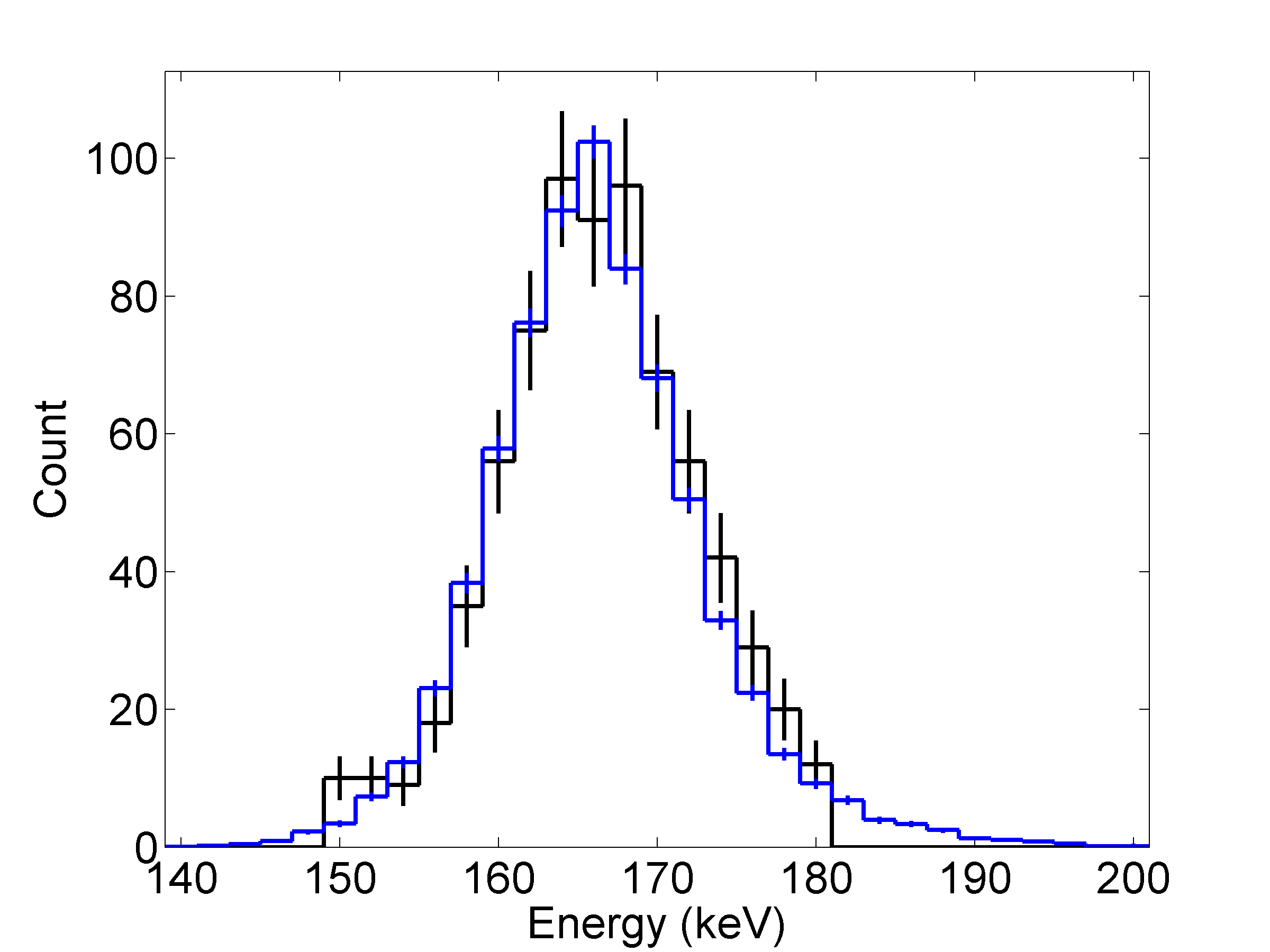}
\caption{Comparison of simulation results (blue) to data (black, with error bars representing the statistical uncertainties in each bin) for the $^{131\mathrm{m}}$Xe cosmogenic-neutron activation peak, which is the only mono-energetic ER in the background that has sufficient statistics to allow such a comparison to be made.
Presented here is the reconstructed ``combined'' energy scale, calculated from both $S1$ and $S2$ and which therefore has improved resolution as compared to the use of a single channel alone, for the data, while for the simulation the $x$-axis represents the Monte Carlo truth energy. The width of the peak is determined by the binomial statistics of $S1$ light collection, sphe resolution, electron lifetime, extraction efficiency, $S2$ collection, single electron pulse size variance, and recombination fluctuations. To reproduce the mean precisely, the values of $S1$ and $S2$ collection efficiencies had to be varied by 5\% (upwards), well within their uncertainties at the 1--2 sigma level, and the recombination fluctuations are increased by 1-sigma compared to a best fit at low energies. The default NEST mean yields from V0.98, 2013, are assumed without modification. The source of the slight asymmetry is unknown, but minor.}
\label{fig:SimDataMonoEPeak}
\end{center}
\end{figure}

\begin{figure}
\begin{center}
\includegraphics[width=0.49\textwidth,clip]{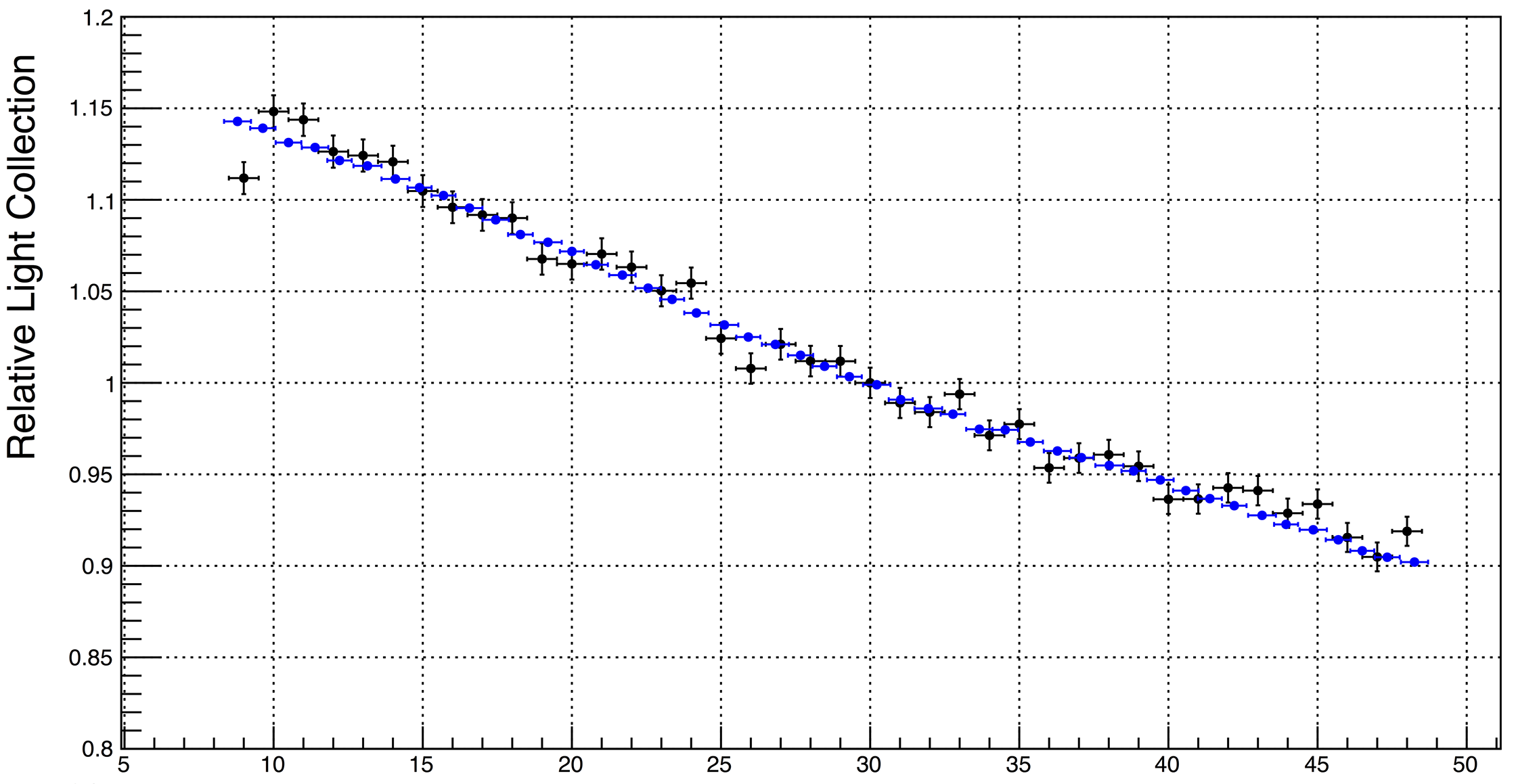}
\includegraphics[width=0.49\textwidth,clip]{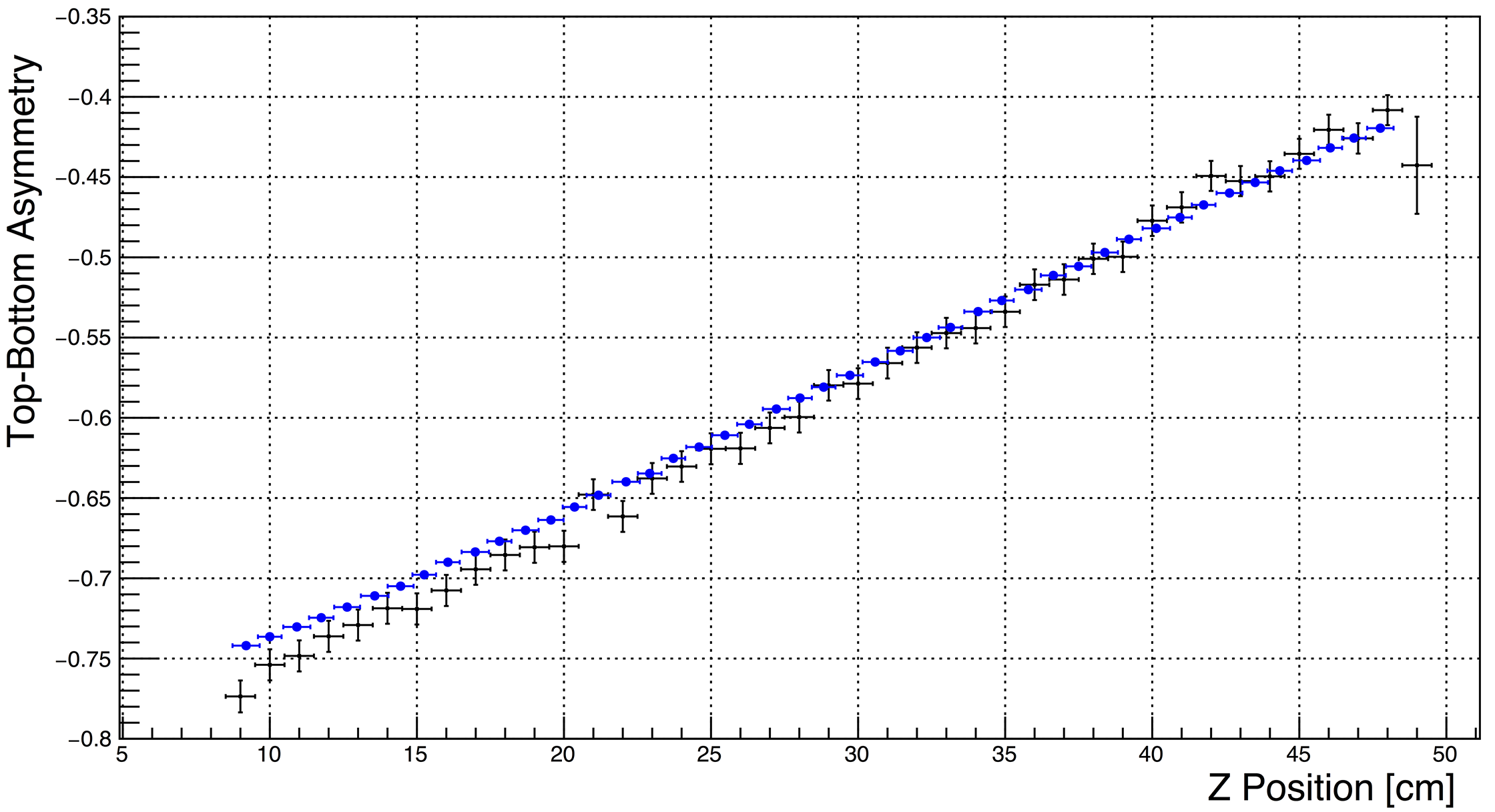}
\caption{Top: the relative light collection of $S1$ signals as a function of the depth in the detector. Bottom: The asymmetry in the 
$S1$ light collection efficiencies between the top and bottom PMT arrays, defined as (T-B)/(T+B), where T and B are the values of $S1$ recorded in the top and bottom PMT arrays, respectively (simultaneous fit to both quantities). In both plots, data are presented in black with statistics-only error bars while the results of the simulation are shown in blue.}
\label{fig:OpticalModel1}
\end{center}
\end{figure}

\begin{figure}
\begin{center}
\includegraphics[width=0.49\textwidth,clip] {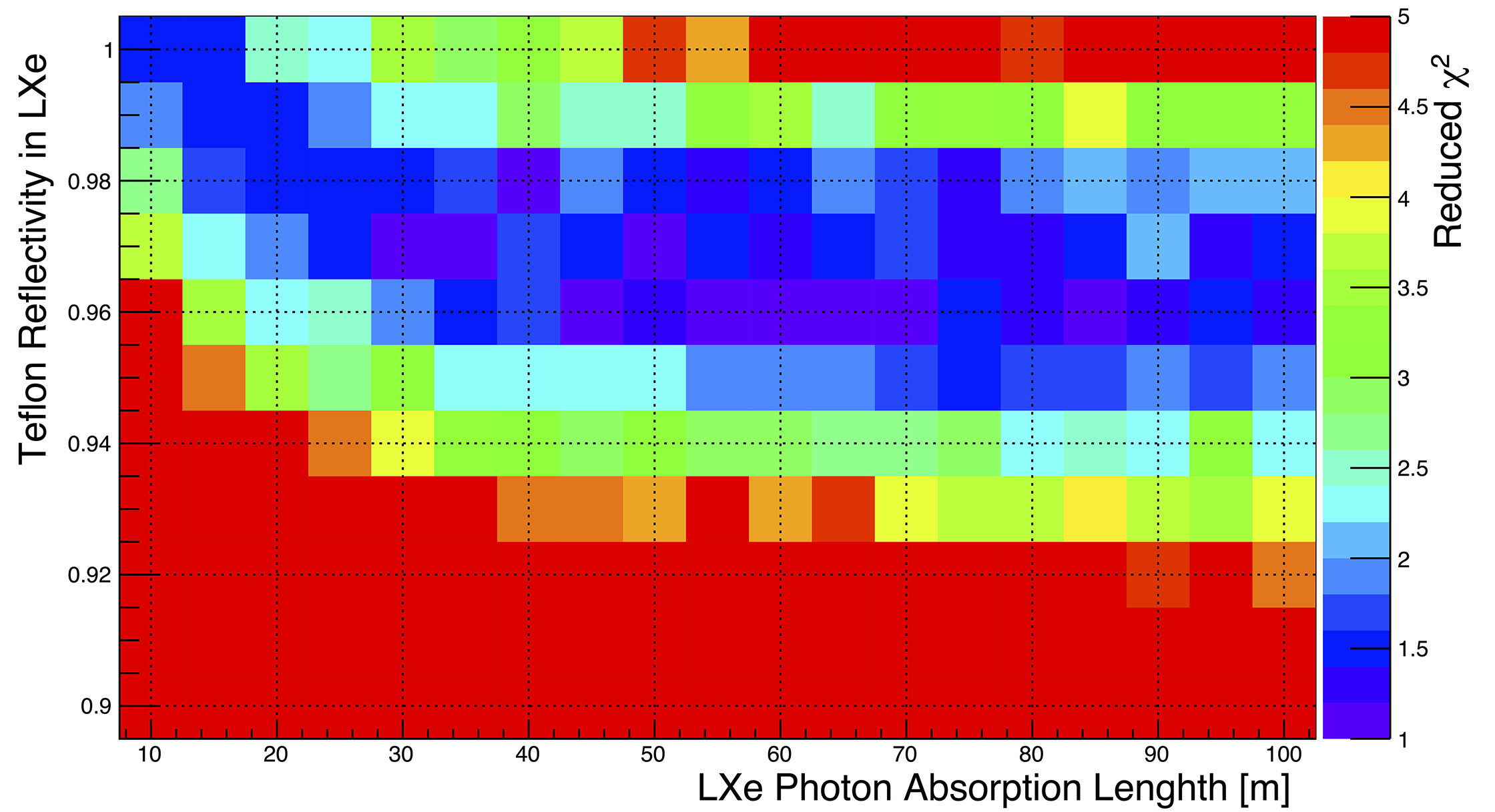}
\caption{$\chi^{2}$ map for Teflon reflectivity and photon absorption length in liquid xenon, the two free parameters that have the greatest impact on both relative and absolute light collection. The degeneracy between the reflectivity of the Teflon and the absorption length for photons in the liquid can be seen, with each having similar effects on the reduction in the light collected. As the former asymptotes to $\sim$96\%, a reasonable value~\cite{Silva:2009}, the latter approaches infinity.}
\label{fig:OpticalModel2}
\end{center}
\end{figure}

After the relative light collection efficiency, as a function of depth and radius, had been matched for the $S1$ signals with LUXSim, it was validated through analysis of single electron signal size and position dependence of $S2$ signals. The use of a $^{83\mathrm{m}}$Kr calibration source leads to a model that is valid for all particles because, once light is generated, the nature of the original incident particle is irrelevant. 
The absolute photon detection efficiency, which is a combination of light collection and QE, was then simulated for the events that occurred at the center of the detector, and this was then used to surpass the relative efficiency comparison. The resulting simulated $S1$ photon detection efficiency, $g_{1}$ (as defined in Sec.~\ref{sec:ER} as the dimensionless ratio of photons detected to all those emitted), of 11.9~$\pm$~0.1\% is in excellent agreement with the value of 11.7~$\pm$~0.3\% determined from analysis
of the data presented in Fig.~\ref{fig:dokeplot}. The simulation result of 11.9\% includes a factor of 1.122 arising from the increase in QE occurring at cryogenic temperatures~\cite{Araujo:2004,Sorensen:2008}. This is consistent with the previously reported result of 14~$\pm$~1\%~\cite{Akerib:2013:run3}, which was also based on simulation, but which did not include the recently measured average 17.3\% probability for the R8778 PMTs to generate 2 phe from a single detected photon~\cite{Akerib:2015:run3, Faham:2015}.
Lastly, comparison of mono-energetic $S1$ peak mean positions to NEST predictions agreed within known uncertainties (Fig.~\ref{fig:SimDataMonoEPeak} is one example). The revised calculation of the extraction efficiency (49\% compared to the previous value of 65\%) is now in better agreement with \cite{Gushchin:1979} and \cite{Gushchin:1982}, but differs from \cite{AprileDoke:2010}, thus leading to a lower extraction electric field than originally estimated.

Deviations between data and simulation in Fig.~\ref{fig:OpticalModel1} demonstrate that a purely physics-motivated approach, even one which includes tuned optical properties such as reflectivity, is still imprecise. It cannot, for instance, account for all the potential microscopic surface deformations that create position-dependent reflectivities on a surface, or possible exotic non-exponential mean free paths for photon absorption by impurities in the xenon. Thus, to ensure the most precise possible background and signal models for the sensitivity calculation, the position corrections based on the $^{83\mathrm{m}}$Kr calibrations (see Secs.~\ref{sec:estimators} and \ref{sec:calibrations}, and the data within Fig.~\ref{fig:OpticalModel1}) applied to the real data sets were also used to generate empirical $S1$ and $S2$ three dimensional look-up libraries for every PMT. These libraries avoided the need for photon propagation in every simulation, and thus allowed two to three orders of magnitude larger
simulated data sets to be generated.

In summary, the LUXSim simulation described here provides near-perfect replication of numerous
empirical results, providing exceedingly reliable support for further data analysis. 

\begin{table}[htp]
\setlength{\extrarowheight}{3pt}
\caption{Optical model parameters used in LUXSim optimized for replication of the experimental data.}
\begin{center}
\begin{threeparttable}
\begin{tabularx}{\linewidth}{X p{1.5cm} p{1.cm} }
\toprule    \hline\hline
Quantity  & Liquid & Gas \\
\midrule \hline
PTFE diffuse reflectivity (\%)                 &  97$^{+3}_{-2}$ & 75$^{+10}_{-5}$\\
Stainless steel grid reflectivity (\%) & 5$\pm$5 & 20 $\pm$5 \\
PMT aluminum reflectivity\tnote{$\ast$} (\%)  &{100${}^{+0}_{-10}$} & {100${}^{+0}_{-10}$}\\
Photon absorption (m) & 30${}^{+40}_{-20}$ & 6 $\pm$3 \\
PMT array QE/predicted &1.024 & 1.000\\
\bottomrule \hline\hline	
\end{tabularx}
\begin{tablenotes}
\item[$\ast$] The aluminum is in contact with the PMT quartz window. 
\end{tablenotes}
\end{threeparttable}
\end{center}
\label{tab:optparam}
\end{table}

\section{\label{sec:calibrations}Calibrations}
\subsection{\label{sec:SPE_SE}Responses to single quanta}

Energy depositions in experiments such as LUX arise from scintillation and ionization, both  of which result in photons detected by PMTs. The variance in such signals is dominated by the number of phe emitted at the photocathodes, as this is where the signal quanta is at a minimum (after the QE and before dynode amplification). 
Consequently, the basic unit of measurement in LUX is the number of phe.
Calibration begins with consideration of single-phe (sphe). To stimulate the emission of sphe for calibration purposes, six blue light-emitting diodes (LEDs), located in the LXe but outside of the TPC, were individually pulsed at a rate of 1\,kHz and a pulse width of 100\,ns.  The pulse amplitude was set so that a given PMT sees no signal for most LED pulses, in which case the number of phe observed per LED pulse in that PMT obeys a Poisson distribution.  If the amplitude is small enough, LED pulses that show non-zero signal will be a nearly pure sample of sphe, and may be used to extract the average amplification ({\it i.e.},~the `gain') of the PMT.

\begin{figure}[hb!]
\begin{center}
\includegraphics[width=.48\textwidth]{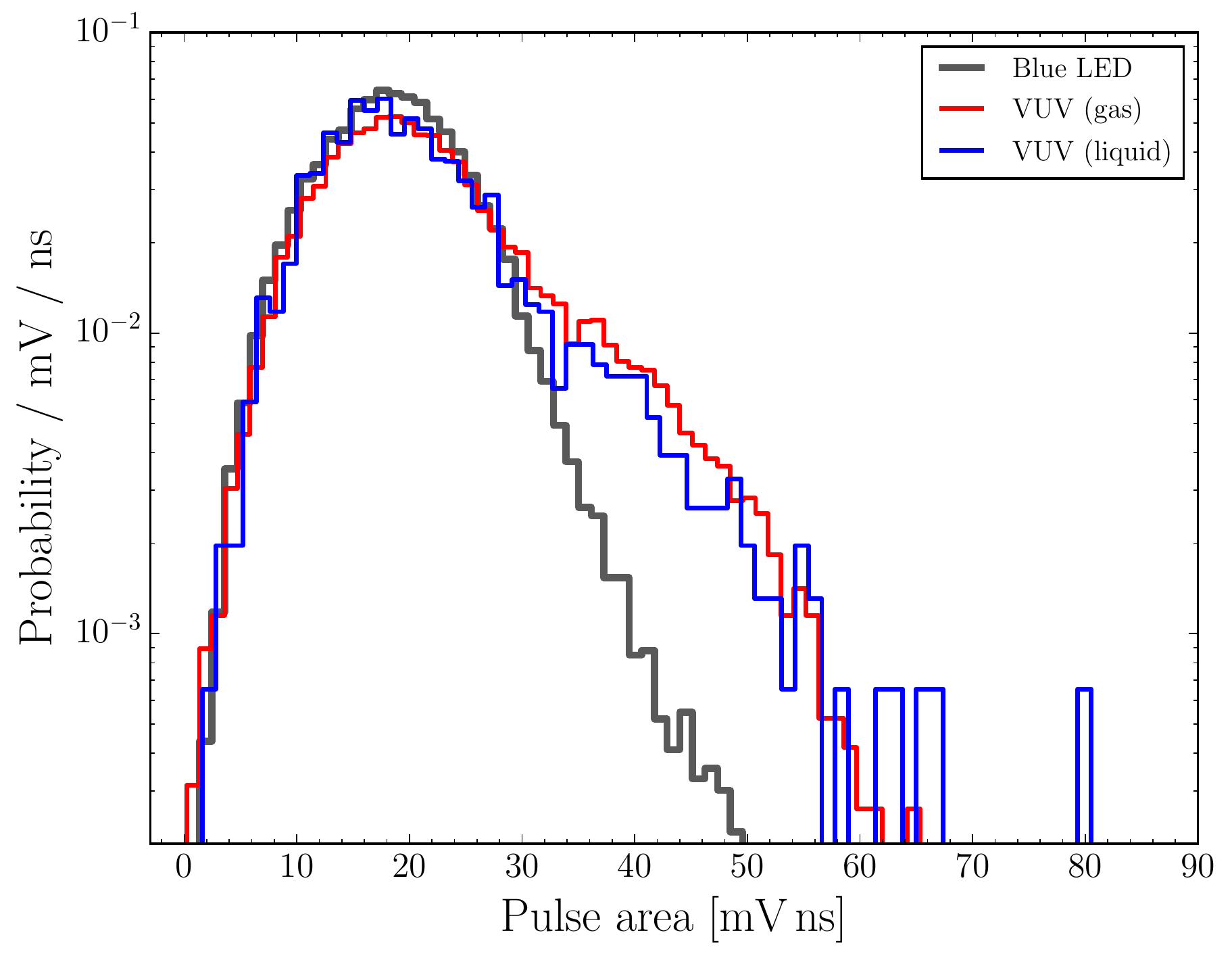}
\end{center}
\caption{\label{fig:singlePHE_DPH} The pulse area spectrum for one of the LUX PMTs under a range of conditions. The black histogram shows single photoelectrons stimulated by exposure to a pulsed blue LED, the red histogram shows the signals recorded as a consequence of VUV photons emitted from gaseous xenon, and the blue histogram shows the signals recorded as a consequence of VUV photons emitted from liquid xenon. Evidence of 2~phe emission from the VUV photon exposures is clearly evident.}
\end{figure}

The single-phe response of one LUX PMT is shown in Fig.\,\ref{fig:singlePHE_DPH} (gray histogram).  The gain of the PMT, defined as the average number of electrons collected at the anode from a single electron emitted from the photocathode, is on average $(3.9\pm0.4)\times10^6$, with a resolution ($\sigma/\mu$) of 35\%. LED calibrations were carried out weekly throughout the duration of the experiment.

Typically in the use of PMTs, a sphe is synonymous with a single detected photon. However, as shown in Fig.~\ref{fig:singlePHE_DPH}, the LUX experiment has discovered that the detection of a single VUV scintillation photon often causes the emission of two phe.  Because the QE reported by the PMT manufacturer is defined as the number of phe emitted from the photocathode per incident photon, this effect leads to a difference between the QE and the photon detection efficiency, a fact that has important implications in accurately reconstructing the energy of an event based on absolute yields \cite{Faham:2015}.

The rate of double-phe emission-per-VUV-photon has been measured in two ways: firstly using scintillation photons emitted by the liquid, and secondly by using scintillation photons emitted from the gas.  For the former, a sample of events from a CH$_3$T calibration were selected, for which the total $S1$ light collected for each event is on average $\sim$5\,phe.  With this selection, the average number of phe collected in a single PMT is less than 0.05 per event.  The number of detected photons---\textit{not} the number of phe---is Poisson distributed in this PMT (similar to the strategy utilized in LED calibrations).
If $\mu$ is the average number of detected photons per event for a given PMT, the fraction of non-zero events which are contaminated by multiple detected photons is $1-(\mu e^{-\mu})/(1-e^{-\mu})$.  Therefore, taking $\mu<0.05$, the set of non-zero hits (for that PMT) is a nearly pure sample of single detected photons, with multiple detected photons contributing less than 2.5\%.  Figure\,\ref{fig:singlePHE_DPH} shows, for an example PMT, a comparison between the spectrum obtained from the (optical) LED calibration and the (VUV) CH$_3$T calibration (``VUV (liquid)'').  The shoulder on the tail of the single-phe peak is readily visible, which indicates the presence of this double-phe emission process.  Plots such as this were used to construct a ``VUV gain'' for each PMT, which indicated the average number of electrons collected at the PMT anode for a single detected VUV photon.  Consequently, the basic unit of measurement is no longer the number of phe, but the number of photons detected, phd.  

The second method for measuring VUV-photon response uses electroluminescence light from the gas region, in the form of single electron ionization pulses from calibration data. Photoionization of impurities in the bulk liquid following $^{83\mathrm{m}}$Kr $S1$s provide a large and pure sample of single electrons. Light from each extracted electron is approximately uniform in time over the 1~$\upmu$s drift from the liquid surface to the anode, and sums to an average of 25 detected photons across the 122-PMT array; the signal therefore appears in individual PMT traces predominantly as single photons or two clearly separable photons (single photons having FWHM around 30~ns). The mean area of the single photon response in a given PMT is obtained in three steps. First, the mean area of those single electron responses with one identified maximum above a 1.4~mV threshold is calculated. Second, the number of unresolved pileup events contributing to that mean is estimated from those responses with two photons resolved in time: the interval between first and last DAQ samples above threshold for the two-spike responses has the expected linear distribution above 7 samples, which is extrapolated and integrated over the region of smaller intervals, where the two photons may not be resolved. Third, the mean area of single spike events is corrected for the pileup, with the area of the contaminating 2-photon responses taken to be the same as the resolved 2-photon responses. This correction is small, on average 3\% for top array PMTs and 1\% for bottom array PMTs. The resulting gain estimates are systematically 2.5\% higher than the liquid-scintillation estimates, which may be due to the difference in scintillation wavelength. The liquid-scintillation values are adopted since it is for $S1$ light that the number of detected photons implies a detection efficiency for the fundamental signal quanta. It is worth noting that for the case of ionization, any pulse-area unit used for both $S2$s and single electrons cancels out when one divides to estimate the signal size in units of electrons.

\begin{figure}[hb!]
\begin{center}
\includegraphics[width=.45\textwidth]{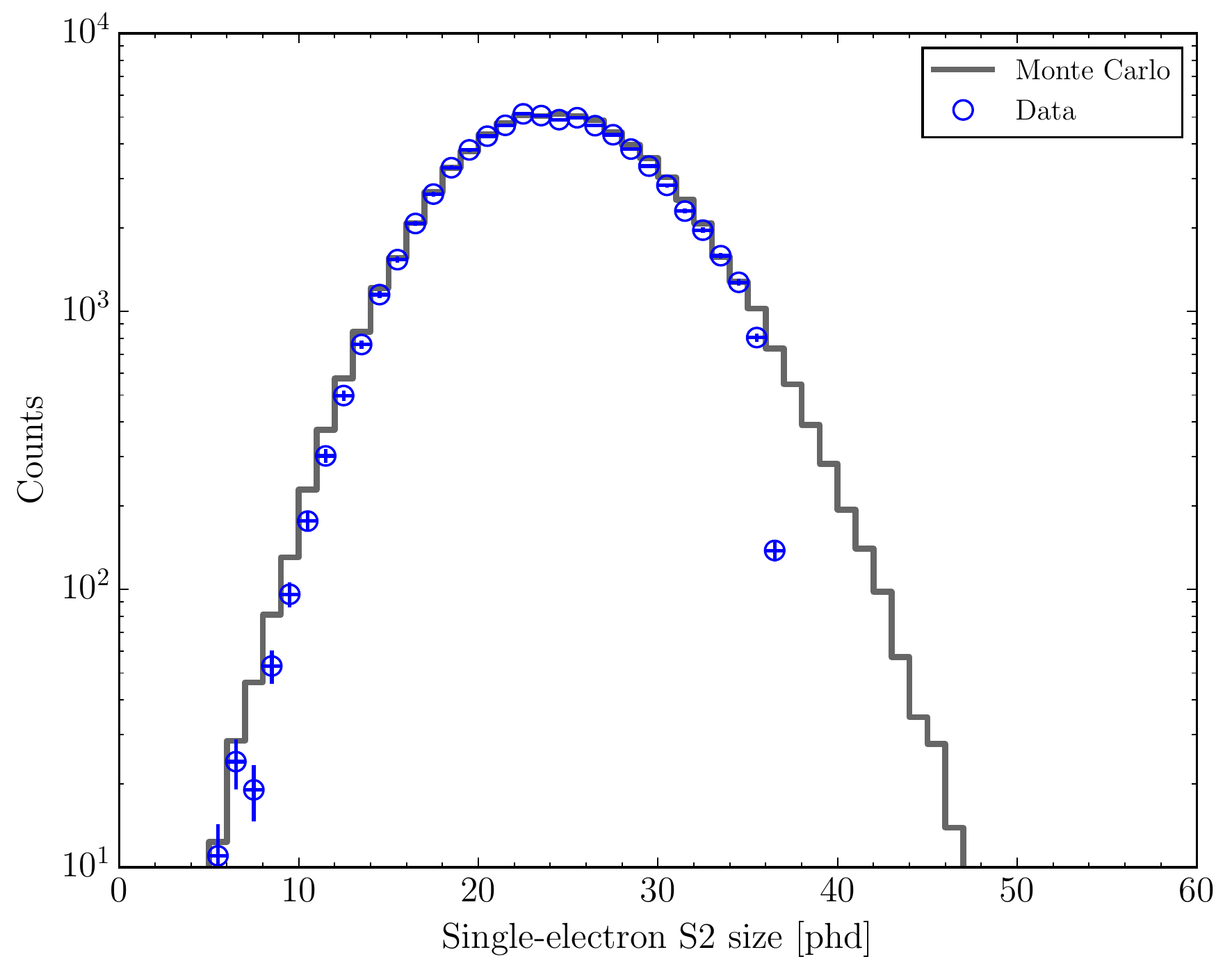}
\end{center}
\caption{\label{fig:SE_spec}The spectrum of ionization signals from single electrons, shown as the blue data points. The spectrum, fitted with a skew-normal distribution, gives a mean of $24.66\pm0.02$(stat)~phd and a 1-$\sigma$ width of $5.95\pm0.02$(stat)~phd. The black line shows a simulation of this spectrum produced from first principles, based on a 5.8~kV/cm field in a 1.6-bar, 5-mm ``gas gap'', with an $S2$ light collection efficiency of 10\%. The fall-off in the data for large phd values is caused by the switch from single electron to $S2$ pulse definition.}
\end{figure}

\begin{figure}[hb!]
\begin{center}
\includegraphics[width=.45\textwidth]{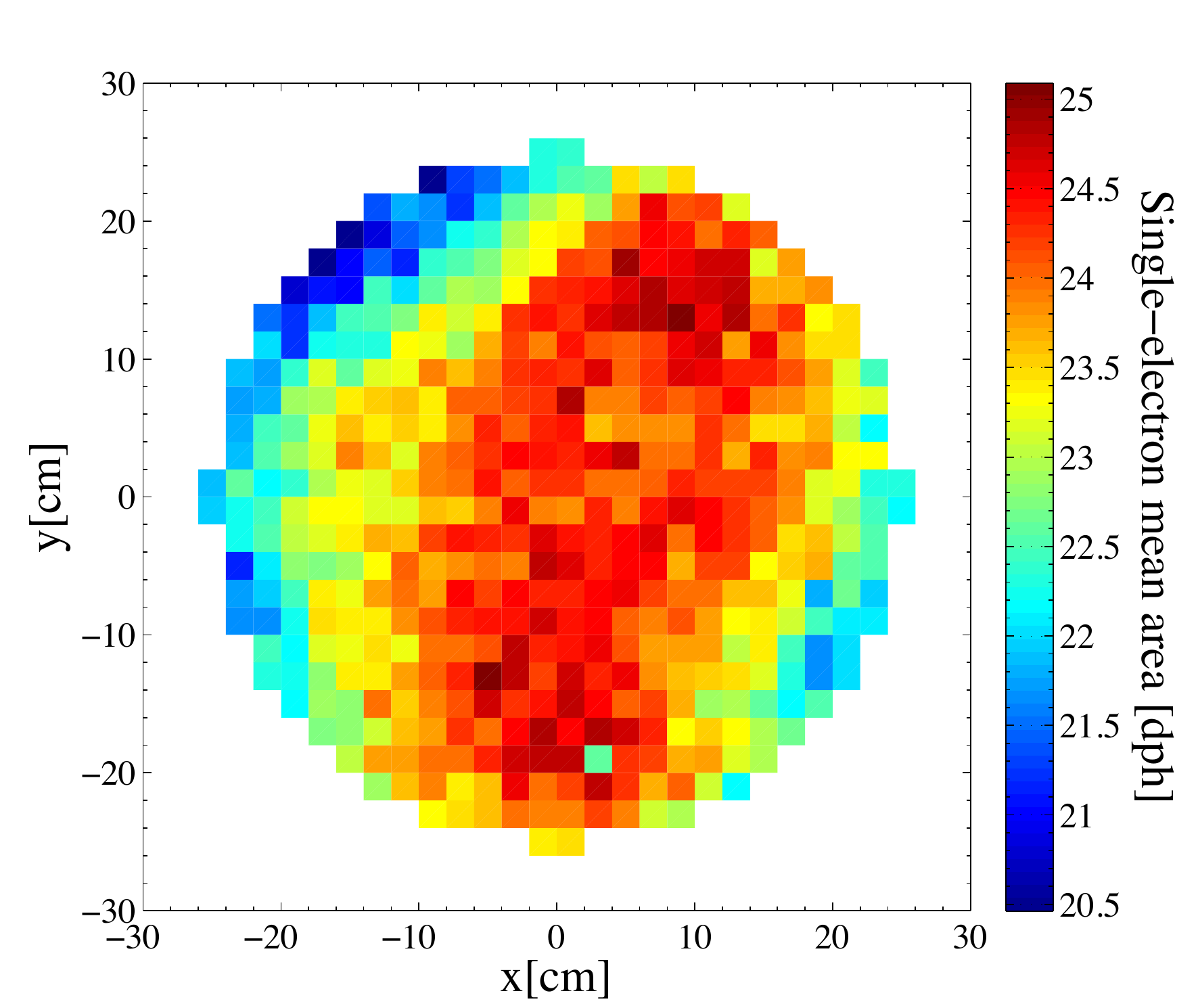}
\end{center}
\caption{\label{fig:SE_xy}Variation of the mean of the single electron distribution in the plane of the liquid surface.}
\end{figure}

In a two-phase Xe TPC, ionization signals generated by a particle (or gamma-ray) interaction result in a burst of VUV photons.  The size of these pulses can be understood more usefully by reporting the absolute number of electrons creating the signal.  Doing so requires a calibration of the detector to single electrons.  Fortunately, single electrons are periodically emitted from the liquid surface, a phenomenon that has been described in the ZEPLIN-II\,\cite{Edwards:2007nj}, ZEPLIN-III\,\cite{Santos:2011ju}, XENON10\,\cite{Aprile:2010bt}, and XENON100\,\cite{Aprile:2013blg} experiments.  A sample of pure single electrons is selected by searching the event record in $^{83\mathrm{m}}$Kr events between the $S1$ and $S2$ signals.  Since it is known that these events are essentially single-site, any $S2$ features in the event record between $S1$ and $S2$ are likely to be single electrons.  The rate of single electrons is low enough such that the probability for two electrons to randomly overlap in time is negligible.  Figure\,\ref{fig:SE_spec} shows the spectrum of the ionization signal from these single electrons, in units of number of detected photons.  As the proportional scintillation process depends on the extraction field and gas gap\,\cite{bolozdynya:1999}, variations of these parameters in time and across the plane of the liquid surface can lead to variations of the average single electron size.  Figure\,\ref{fig:SE_xy} shows the average of the single-electron distribution in the $x$-$y$ plane, while Fig.\,\ref{fig:SE_Stability} shows the average single-electron size over the duration of the science run.

\subsection{\label{sec:Stability} Stability}
\subsubsection{\label{sec:PMTGainStability} PMT gain stability}

\begin{figure}[hb!]
\begin{center}
\includegraphics[width=0.45\textwidth,clip]{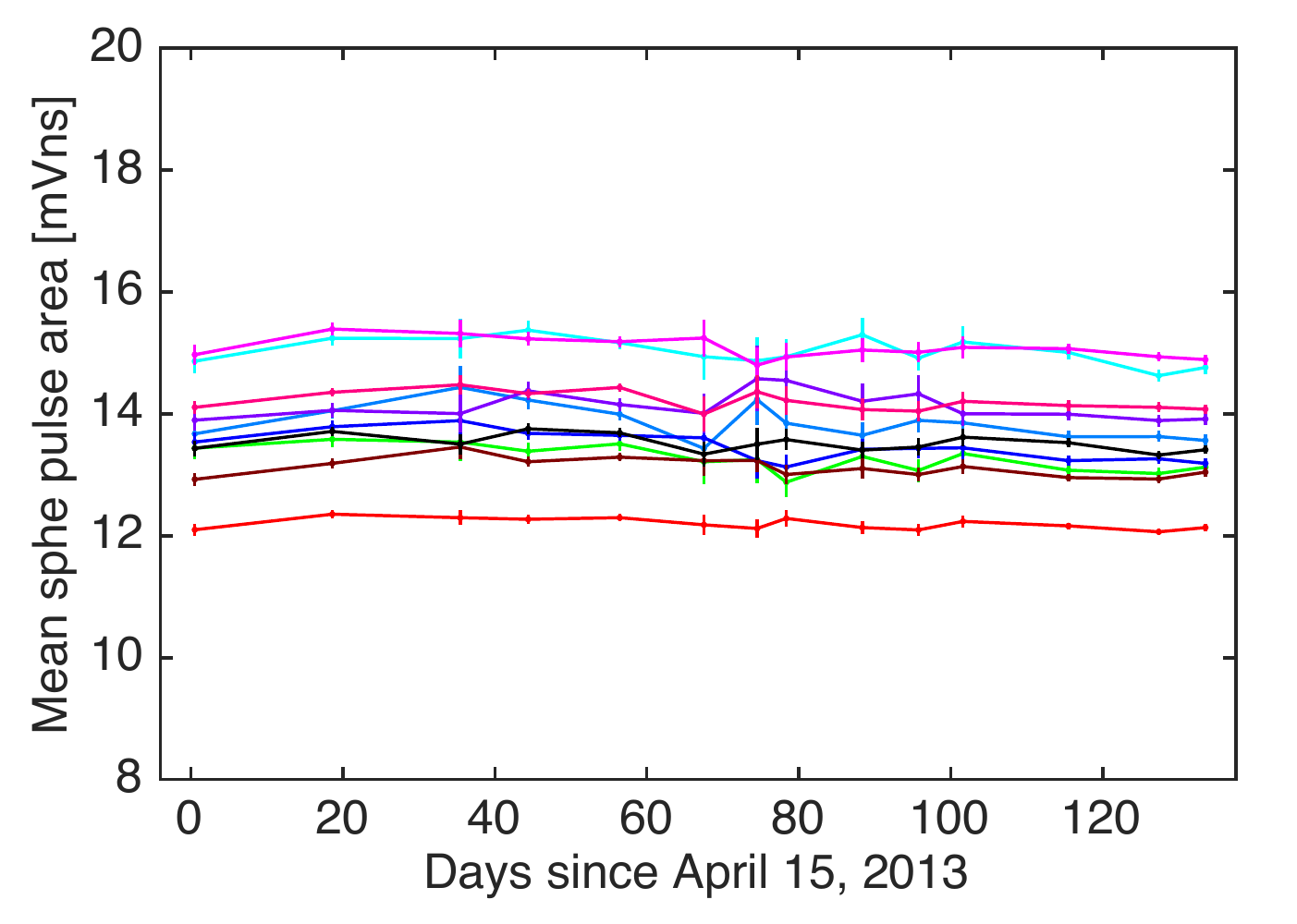}
\caption{The gain as a function of time for 10 representative PMTs}
\label{fig:gain_time_trend_example}
\end{center}
\end{figure}

\begin{figure}[hb!]
\begin{center}
\includegraphics[width=0.45\textwidth,clip]{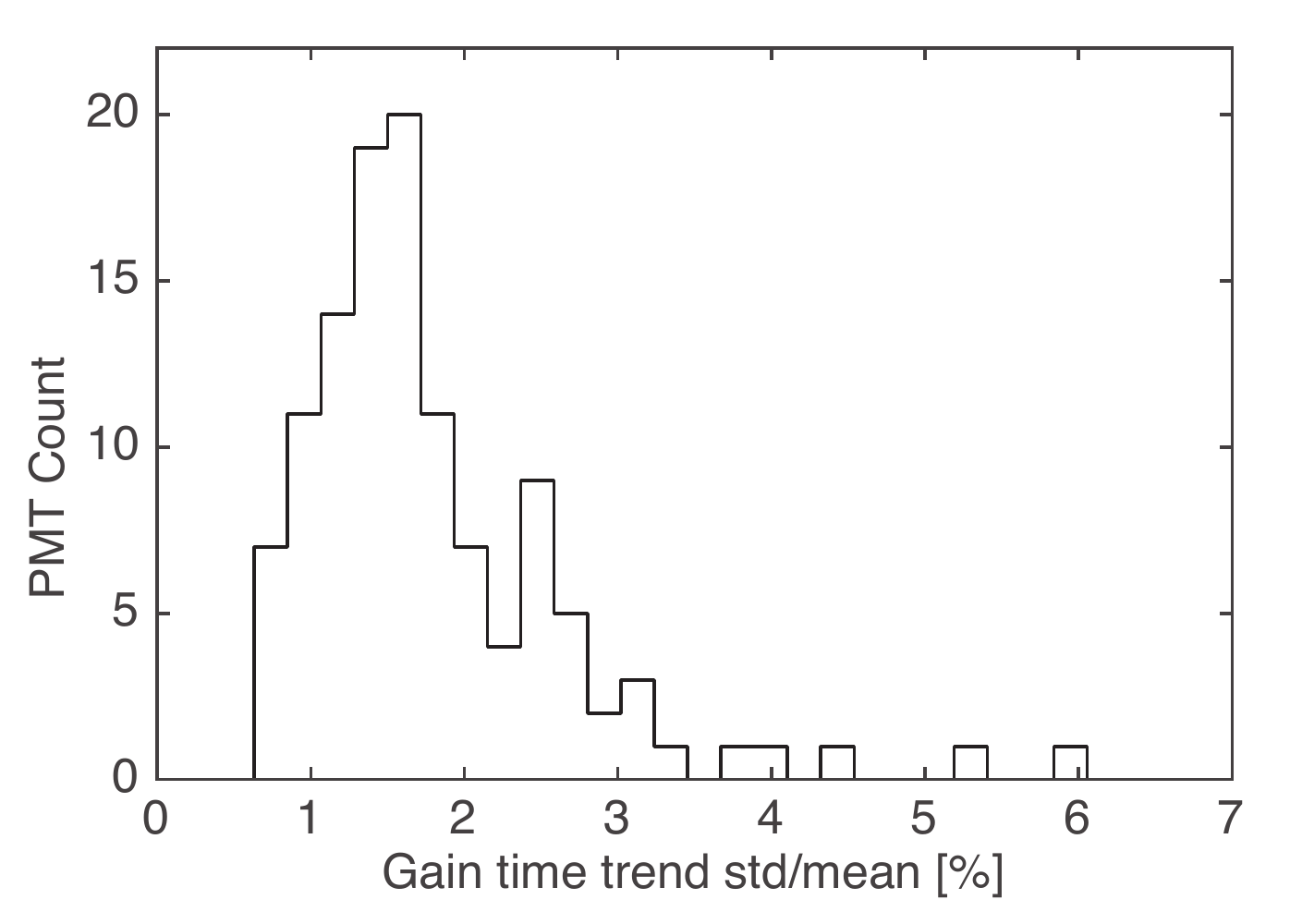}
\caption{The gain of each PMT was measured each week throughout the run with an LED calibration to stimulate sphe emission. Shown is a histogram of the standard deviations of these gains, each normalized to the PMT's mean gain. Most gains are stable to better than 2\%.}
\label{fig:gain_time_deviation_distribution}
\end{center}
\end{figure}

The LED calibrations, described above, were performed periodically over the course of the WIMP search.  Figure\,\ref{fig:gain_time_trend_example} shows the deduced gains for ten PMTs over most of the duration of the run, illustrating the stability achieved.  For all PMTs, the relative level of fluctuations are presented in Fig.\,\ref{fig:gain_time_deviation_distribution}; most gains are stable to better than 2\% (standard deviation/mean).

\subsubsection{\label{sec:SEStability} Single electron size stability}

\begin{figure}[htp!]
\begin{center}
\includegraphics[width=.45\textwidth]{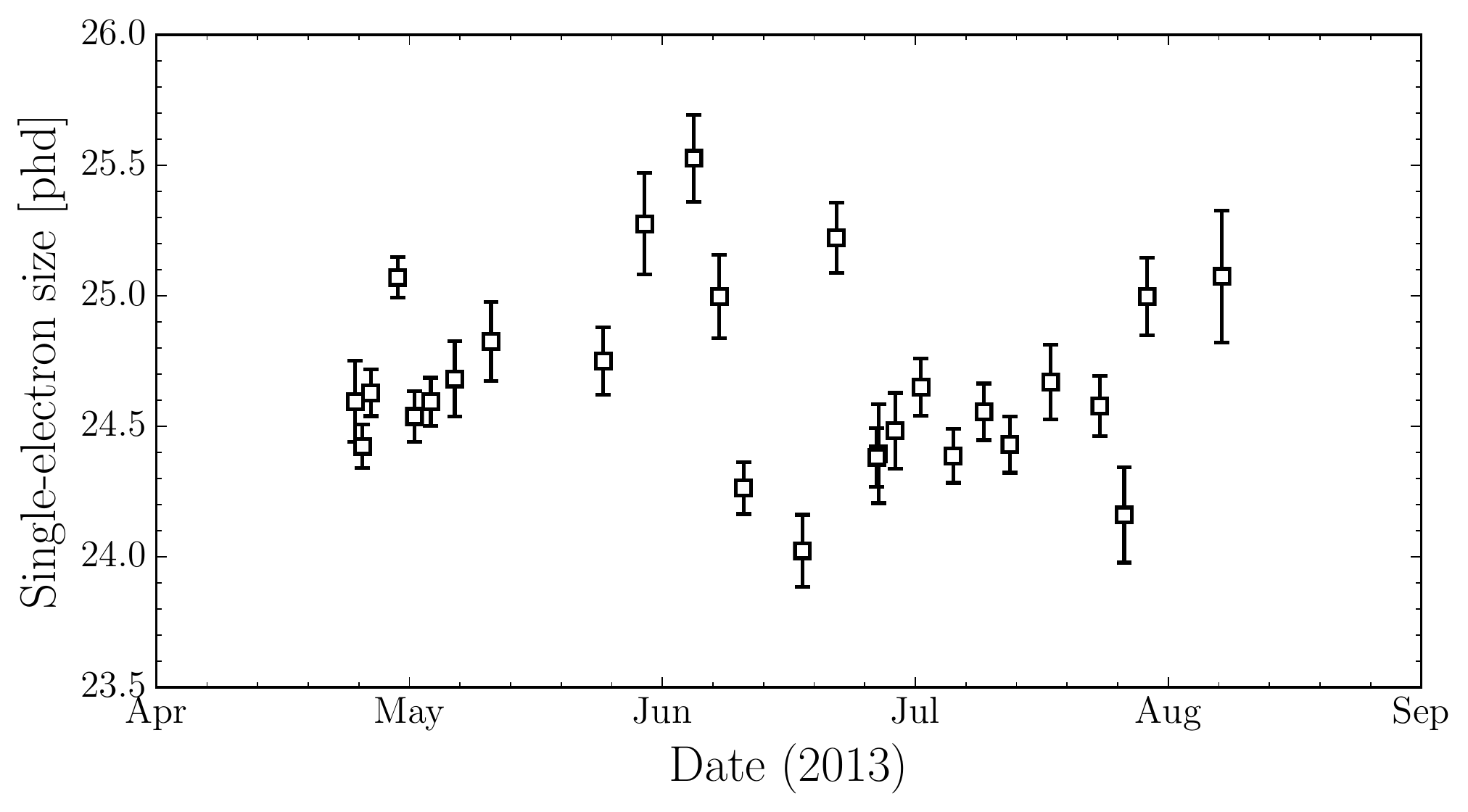}
\end{center}
\caption{\label{fig:SE_Stability}The variation in the size of a single electron $S2$ pulses over the duration of the WIMP search.}
\end{figure}

The gain for $S2$ signals, generated by electroluminescence in the gas, is determined by a number of factors. These include the pressure of the gas, the level of the liquid and the extraction electric field.  While each of these parameters was monitored by the slow-control system, an absolute measurement of the $S2$ gain was determined using $^{83\mathrm{m}}$Kr calibration data. As mentioned previously, these data contain single electrons emitted from the liquid surface, allowing the average size of a single electron $S2$ to be determined periodically over the course of the WIMP search. The resulting gains are
shown in Fig.\,\ref{fig:SE_Stability}.  These data indicate that the single electron size was stable at the level of 1.4\% over the duration of the run.

\subsubsection{\label{sec:ElifetimeStability} Electron lifetime stability}

\begin{figure}[htp!]
\begin{center}
\includegraphics[width=.45\textwidth]{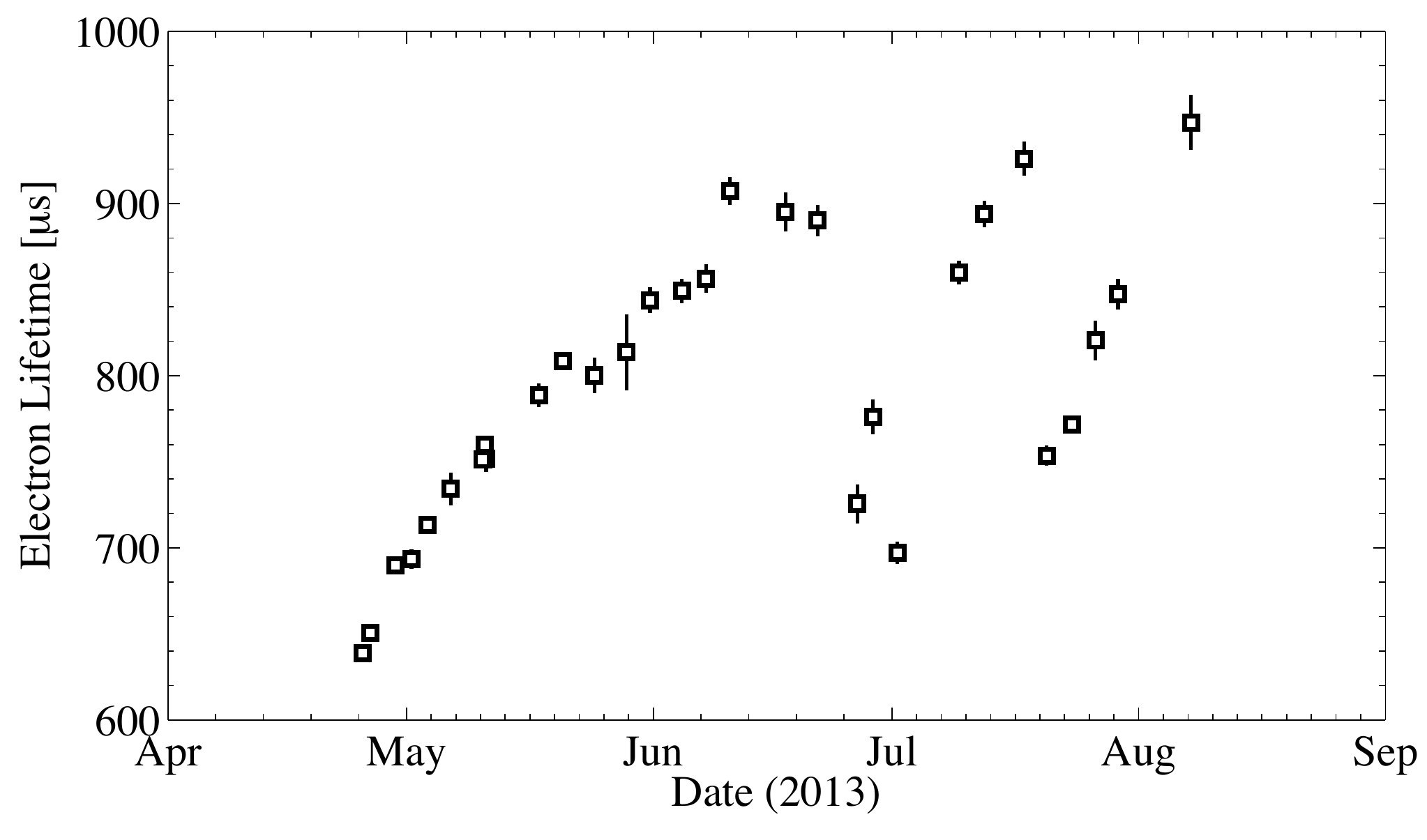}
\end{center}
\caption{\label{fig:ElifetimeVsTime}The measured electron lifetime over the duration of the experiment. Sporadic dips in electron lifetime are a result of temporary stoppages in the LXe recirculation.}
\end{figure}

As electrons are drifted through the liquid, they may be captured by residual electronegative impurities, such as O$_2$. This produces a depth-dependent attenuation of the ionization signal and allows the concentration of O$_2$ to be expressed as the average time that an electron will propagate before capture; this quantity is known as the electron lifetime.  As this quantity varies over the course of the experiment (and is used to correct observed ionization signals), it is measured periodically during the WIMP search.  Periodic calibrations with $^{83\mathrm{m}}$Kr, in a technique similar to those described in \cite{Kastens:2009,Manalaysay:2009yq}, were carried out.  This source decays via two sequential transitions depositing 32.1~keV and 9.4~keV, each of which consists mostly of conversion and Auger electrons, with small contributions of gamma- and X-rays. The half-life of the 9.4\,keV state is 154\,ns, sufficiently short 
such that the sequential decays occur at essentially the same physical location, and also such that that they often occur too close in time to be resolved as separate pulses.  The depth-dependence of the combined 41.5\,keV ionization signal is measured, from which the electron lifetime is calculated.  Figure\,\ref{fig:ElifetimeVsTime} shows the measured electron lifetime during the WIMP search.

\subsubsection{\label{sec:Kr83Stability} $^{83\mathrm{m}}$Kr light yield stability}

\begin{figure}[htp!]
\begin{center}
\includegraphics[width=.45\textwidth]{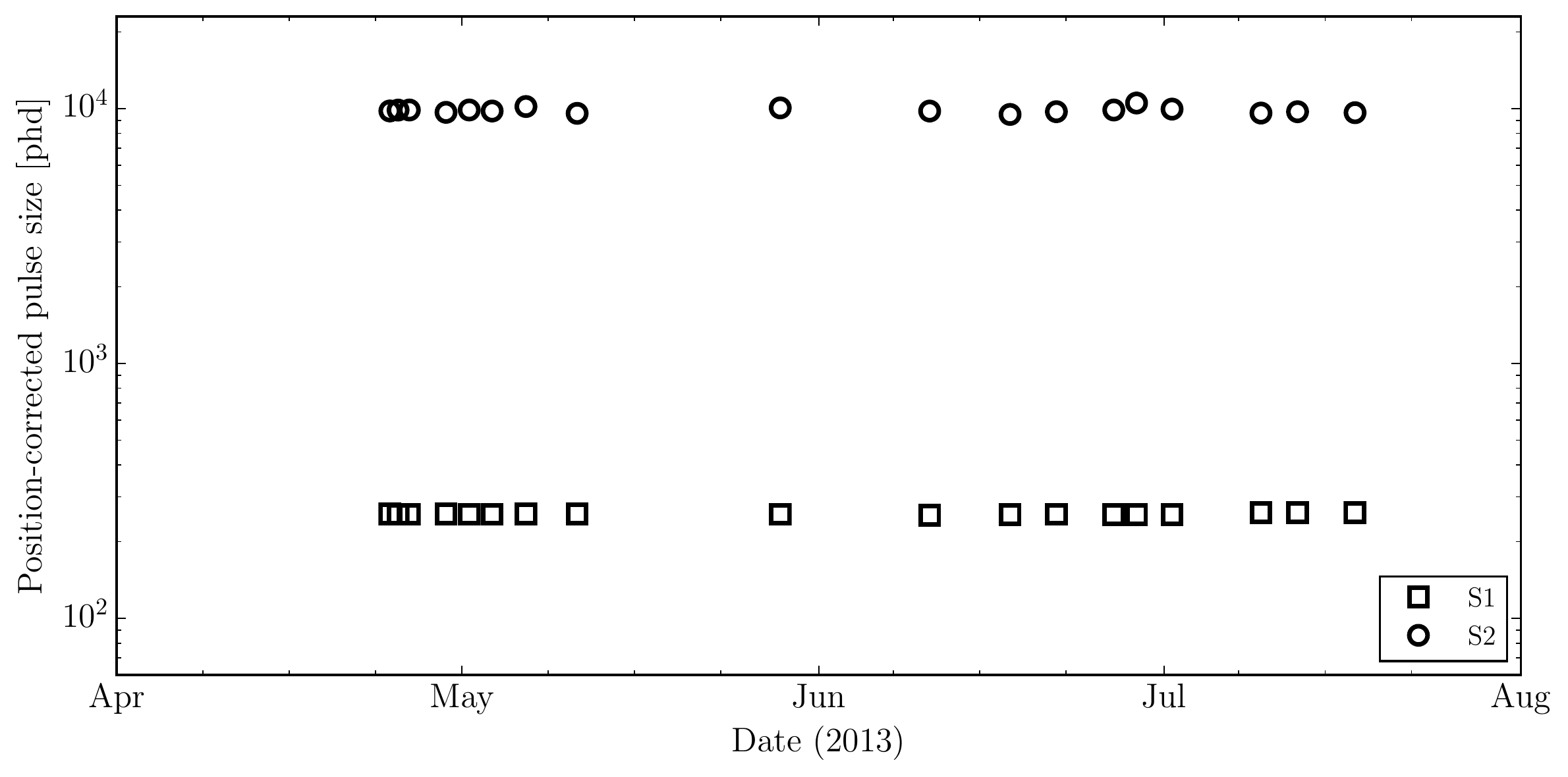}
\end{center}
\caption{\label{fig:KrStability}The scintillation (squares) and ionization (circles) response of the LUX detector over time to the decays of $^{83\mathrm{m}}$Kr.
}
\end{figure}

The stability of the scintillation and ionization was monitored with periodic $^{83\mathrm{m}}$Kr calibrations~\cite{Akerib:2017:Kr83m}. Figure\,\ref{fig:KrStability} shows these responses over the time period of the WIMP search.  The relative time variation (standard deviation/mean) of the scintillation response from these calibrations is 0.6\%, while that of the ionization signal is 2.4\%.

\subsection{\label{sec:PosRec}Position reconstruction}
\subsubsection{\label{sec:FV}Measuring the fiducial volume}

The total xenon mass in the active volume between the cathode and the gate grids is determined as:
\begin{equation}
\label{eq:active_xe_mass}
M_{\textrm{Xenon}} = 3R^2 L\rho_{\textrm{Xenon}} = 250.9 \pm 2.1\,\mathrm{kg},
\end{equation}
where R=24.48$\pm$0.05~cm is the distance between the center of the detector and the corners of the PTFE panels, and $L=48.32\pm0.34$~cm is the distance between the cathode and the gate grid (dimensions when the detector is cold). $\rho$=2.887$\pm$0.005~g/mL is the average xenon density during the run, corresponding to an average temperature of 173.19$\pm$0.07~K.

The fiducial region, that is the volume within which candidate events must occur, is defined as a cylinder with radius 20~cm and a vertical extent corresponding to drift times between 38 and 305~\textmu{}s. The conversion between drift time and $z$ position, measured relative to the surface of the PMTs in the bottom array, is given by:
\begin{equation}
z = s_c + L - v_d\left(\tau - \tau_g\right) \qquad \mathrm{with} \qquad v_d = \frac{L}{\left(\tau_c-\tau_g \right)},
\end{equation}
where $v_d$ is the drift velocity, measured at 0.1518$\pm$0.0011~cm/\textmu{}s; $s_c=5.6$~cm is the distance between cathode and the surface of the PMTs; and $\tau_g$ and $\tau_c$ correspond to the drift time of events at the gate and at the cathode, respectively. These values are estimated using the krypton calibration data, from which $\tau_g=3.4\pm0.4$~\textmu{}s and $\tau_c=322.3\pm0.4$~\textmu{}s are obtained.  The resulting fiducial volume mass is found to be 147$\pm$1~kg. 

The fiducial mass of the detector may also be calculated using the distribution of the tritium events in the detector. This method takes the ratio of the observed number of tritium events in the fiducial volume to the number of events between the cathode and the gate, and then multiplies it by the total xenon mass between the cathode and the gate. This method relies on the tritiated methane being isotropically distributed throughout the detector, but has the advantage of not requiring physical dimensions to be defined, such that possible systematic uncertainty from the drift time conversion and position reconstruction are eliminated.

\begin{figure}[t]
\begin{center}
\includegraphics[width=0.48\textwidth,clip]{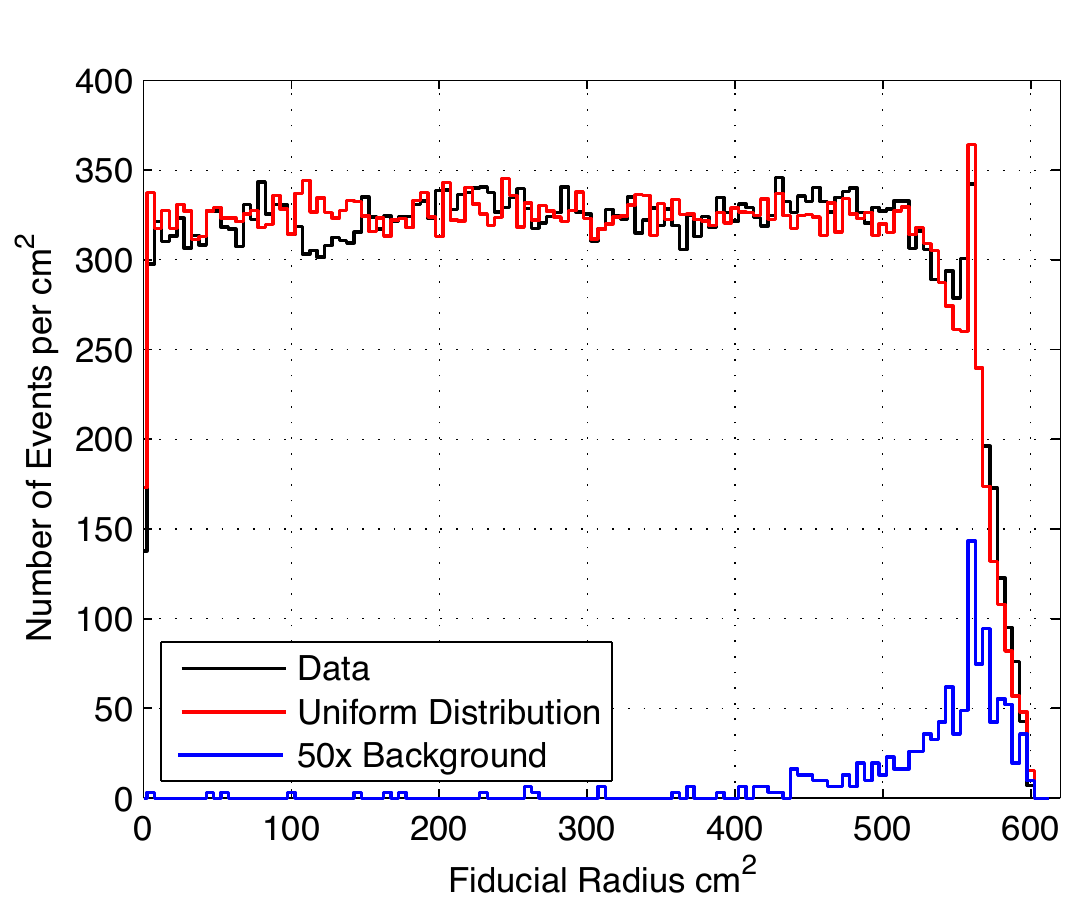}
\caption{Histogram of the number of the observed tritium events as function of the corrected radius and for a drift time between 38 \textmu{}s and 305 \textmu{}s (solid black line). The red line represents a simulation of a perfectly uniform distribution of events. The blue line represents an estimate of the background observed during the lifetime of the tritium calibration, multiplied by a factor of 50.}
\label{fig:tritium_vs_squared_radius_forFiducialMass}
\end{center}
\end{figure}

For this calculation, the tritium events are selected using the same quality cuts as applied in the WIMP search data (described in Sec.~\ref{sec:EventSelec}). Additionally, the events from the first hour after the tritium injection are not included to ensure uniform distribution of tritium. The mixing time observed for $^{83\mathrm{m}}$Kr injections was observed to be less than 10 minutes, and and a similar mixing time is expected for tritium injections.  After the mixing period, the distribution of tritium events was found to be highly uniform along the radius, with only a small accumulation of events near the walls of the detector, as shown in Fig.~\ref{fig:tritium_vs_squared_radius_forFiducialMass}.

\begin{figure}[t]
\begin{center}
\includegraphics[width=0.48\textwidth,clip]{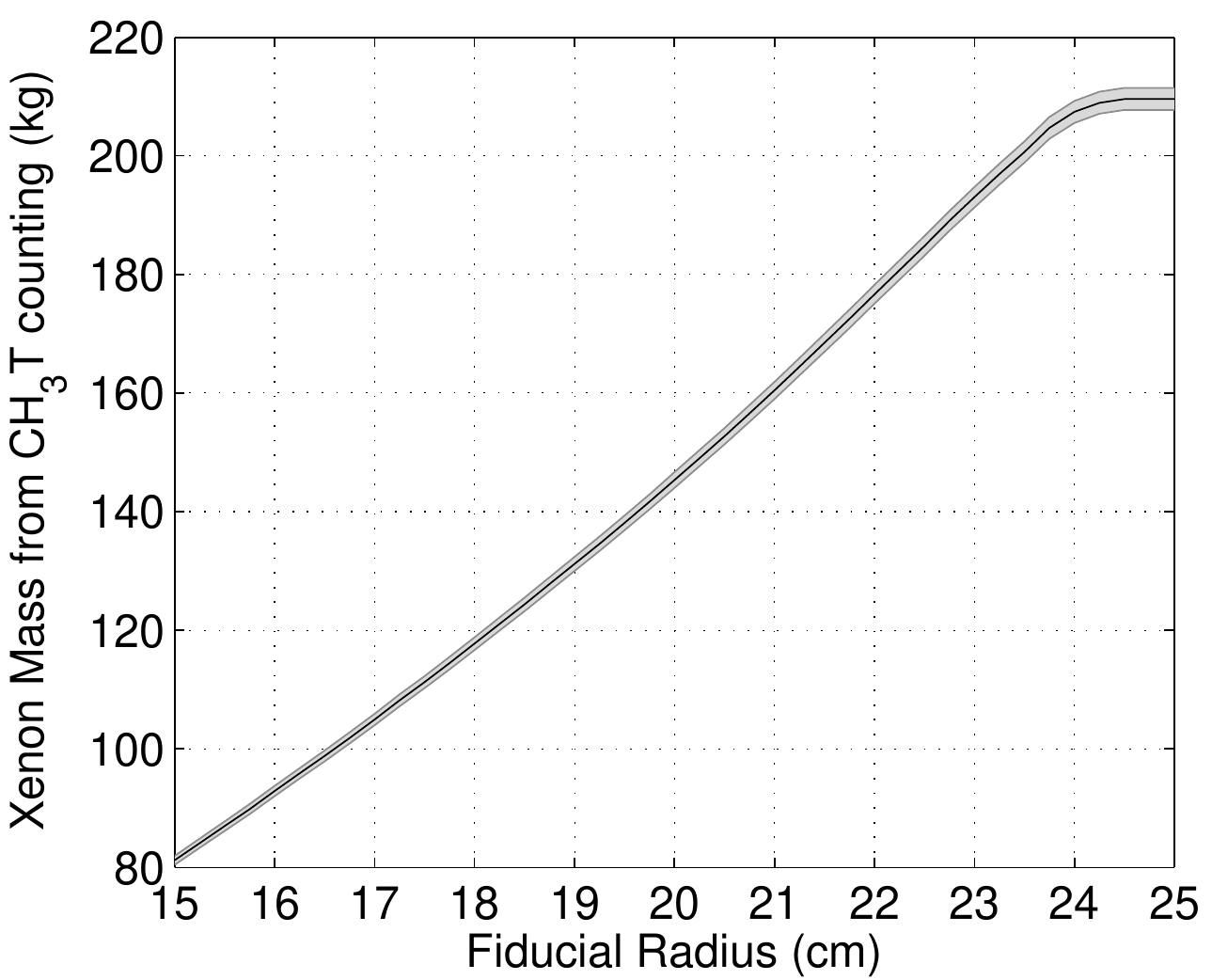}
\caption{Xenon mass calculated directly from tritium counting as function of the radius of the fiducial volume. The $\pm$ 1 sigma band is represented by the gray band. The drift time considered is between 38~\textmu{}s and 305~\textmu{}s and the mass between the gate and anode is 250.9 $\pm$ 2.1 $\mathrm{kg}$.}
\label{fig:xenon_mass}
\end{center}
\end{figure}

Using a drift time cut to select events in the active volume between the cathode and the gate grid, $(223.1\pm0.5)\times10^3$ tritium events were observed. A fiducial volume cut selected events with a drift time between 38~\textmu{}s and 305~\textmu{}s and a radius less than $r$. The ratio between these two event counts is multiplied by the xenon mass between the gate and cathode, calculated in Eq. \ref{eq:active_xe_mass}.  Figure~\ref{fig:xenon_mass} shows the fiducial mass as function of the radius $r$.  For a radius of  $r=20$~cm, the number of events observed is $(129.1\pm0.4)\times10^3$, which yields a fiducial mass of 145$\pm$1~kg.

\begin{figure}[t]
\begin{center}
\includegraphics[width=0.48\textwidth,clip]{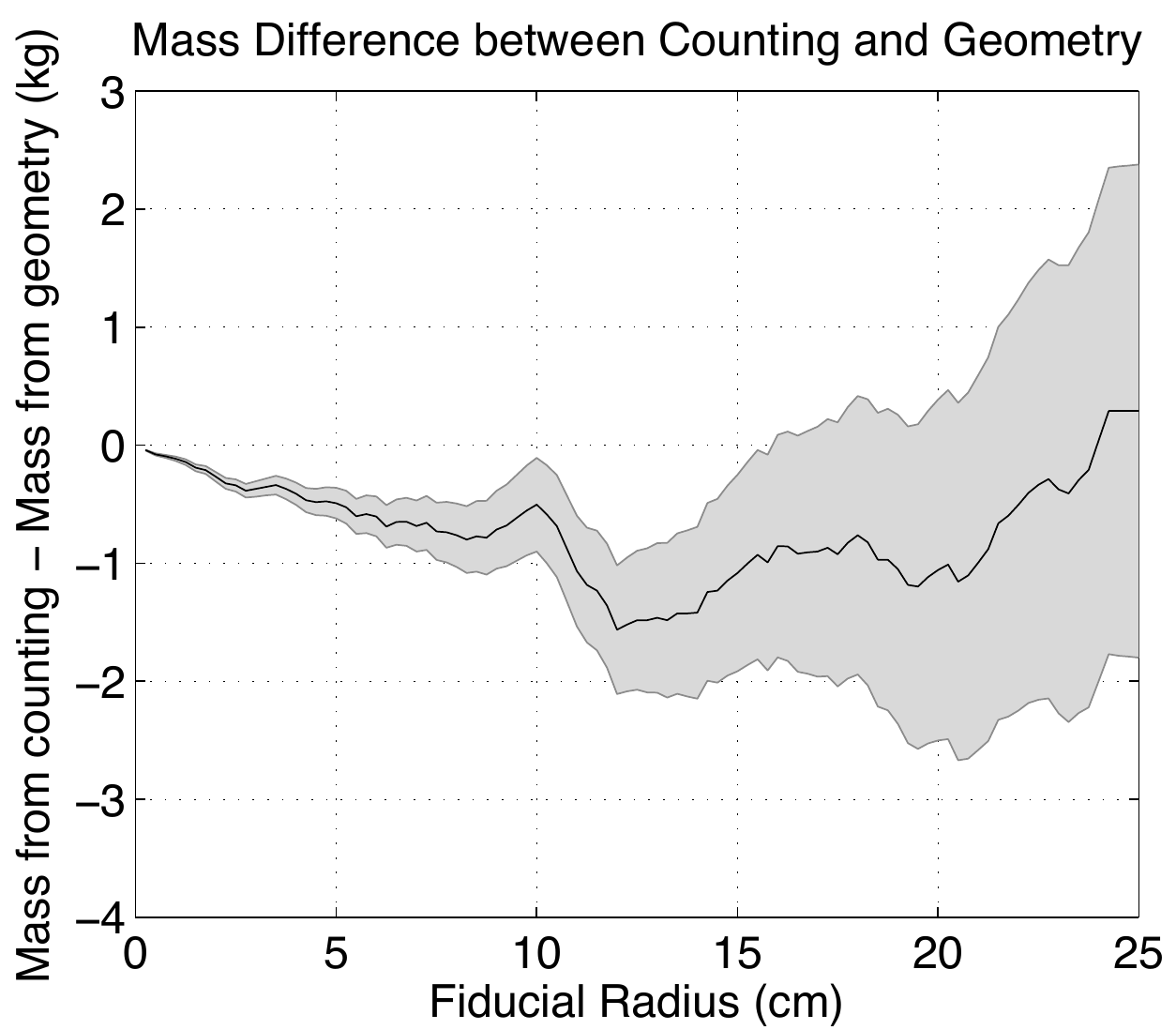}
\caption{The difference between the xenon mass calculated from the counting of tritium events and the xenon mass obtained from the geometry of the fiducial volume as function of the radius of the fiducial volume. The $\pm~1\sigma$ region is represented by the gray band.}
\label{fig:difference_xenon_mass}
\end{center}
\end{figure}

Figure~\ref{fig:difference_xenon_mass} shows the difference between the mass calculated using the event counting method, and the mass calculated directly from the geometry, as function of the fiducial volume radius. The difference between these two methods is smaller than 1.4 kg for any fiducial radius.

\subsubsection{\label{sec:PosResolution}Accuracy of position reconstruction}

Given the geometry of the LUX detector, the statistical uncertainty in $xy$-position reconstruction is most naturally handled in polar coordinates. In this case the uncertainty breaks down into three categories:

\begin{enumerate}
\item The centripetal uncertainty: the radial uncertainty in the direction towards zero radius
\item The centrifugal uncertainty: the radial uncertainty in the direction towards the wall
\item The polar uncertainty: the uncertainty in the polar angle, treated as symmetric
\end{enumerate}

\begin{figure}
\begin{center}
\includegraphics[width=0.48\textwidth,clip] {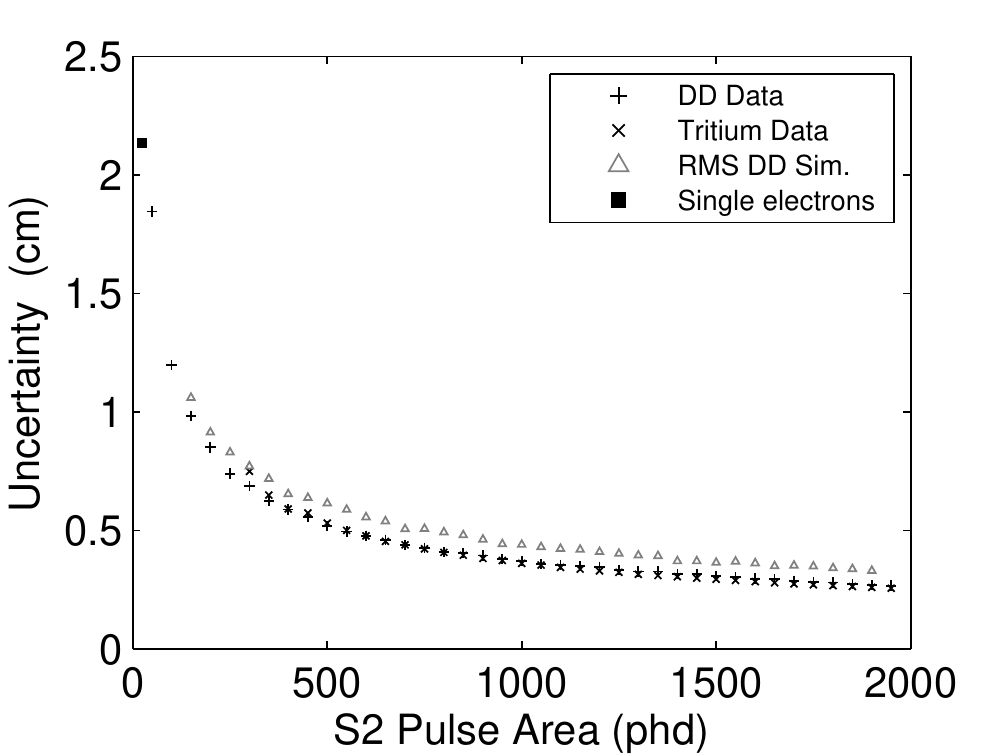}
\caption{Comparison of centripetal uncertainties in Mercury in \dd{} data (vertical black crosses), tritium data (diagonal black crosses), single electrons (black squares), and the root mean square between the reconstructed radius and the original radius, obtained in \dd{} simulations (triangles).}
\label{fig:Merc_Uncer_Comp}
\end{center}
\end{figure}

This parameterization is convenient as the radial position is paramount in determining whether an event looks more signal-like (in the bulk, towards the center of the detector) or background-like (towards the wall of the detector). These three uncertainties are calculated from the Mercury error oval based on the goodness of fit. Figure~\ref{fig:Merc_Uncer_Comp} shows the centripetal uncertainty versus $S2$ size for \dd{} data and tritium data,
and the root mean square of the difference between the reconstructed radius and true radius, from \dd{} simulations. Here the uncertainties, binned by $S2$ size, are the average over the entire fiducial volume. Unsurprisingly, the uncertainty decreases as $S2$ size increases, as a result of there being more information carriers. The uncertainties calculated from the \dd{} data and tritium data agree between each other while the results obtained from the \dd{} simulations are~13\% larger.

\begin{figure}
\begin{center}
\includegraphics[width=0.48\textwidth,clip] {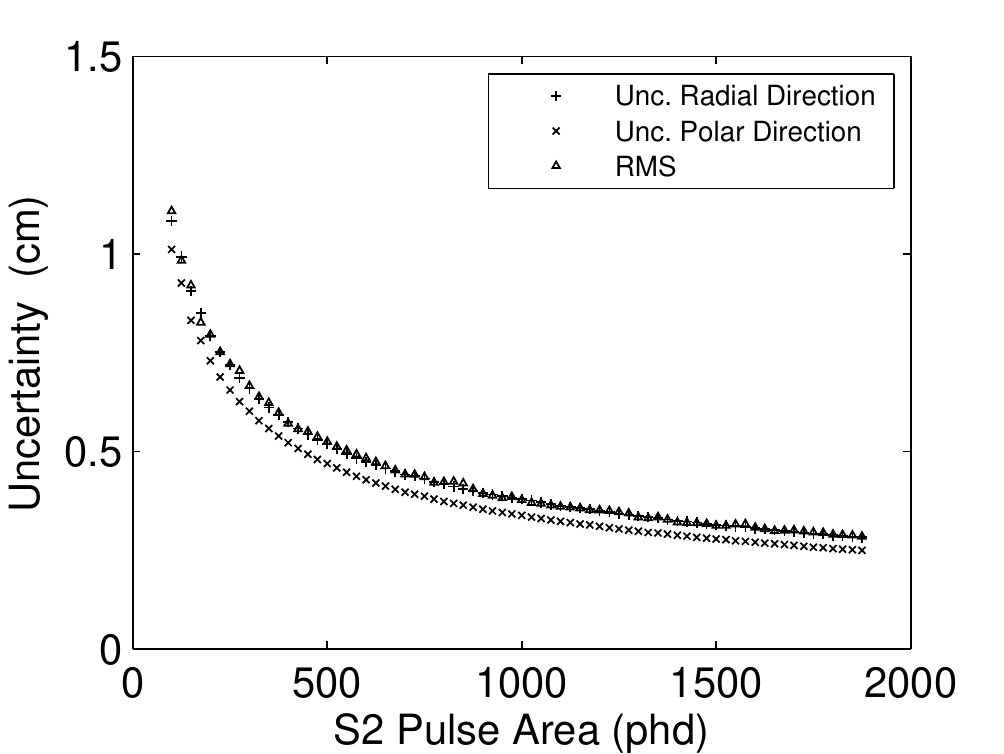}
\caption{Comparison of the centripetal and polar uncertainties in \dd{} simulations to the root mean square of the difference between the reconstructed position and the true position.}
\label{fig:Merc_Sim_Check}
\end{center}
\end{figure}

An advantage of comparing the uncertainty in simulation to that of data is that in a simulation the true event locations are known, allowing for a consistency check, as shown in Fig.~\ref{fig:Merc_Sim_Check}. The calculated statistical uncertainties are very close to the differences between the real and reconstructed positions.

\begin{figure}
\begin{center}
\includegraphics[width=0.48\textwidth,clip] {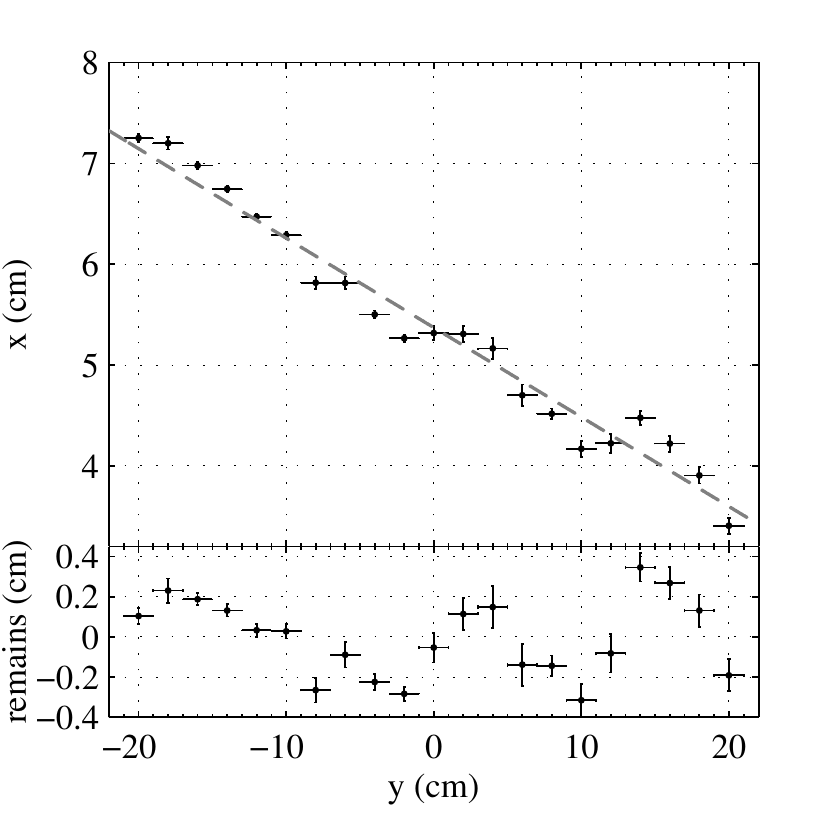}
\caption{\textit{Top}: The crosses show the average reconstructed $x$ position as a function of reconstructed $y$ from \dd{} calibration data. In both cases $x$ and $y$ are corrected for the effects of the non-uniform electric field. The dashed line shows a linear fit to the \dd{} beam direction \textit{Bottom}: Deviation from the linear fit in the $x$ direction.}
\label{fig:Mercury_DD_Beam}
\end{center}
\end{figure}

In addition to the statistical uncertainty in the $xy$- (R$\theta$) positions, there is a systematic uncertainty that may be calculated using the position of events in the beam from \dd{} calibrations. Because the \dd{} 
events arise from a well collimated beam, the events in the \dd{} calibrations appear as a line in the $xy$-position (R$\theta$) plane. The deviation from a straight line may therefore be used as a measure of the systematic uncertainty in the position reconstruction. Such a check does not depend on the actual angle of the beam, only that the beam is collimated with a small angular divergence. A linear fit to the \dd{} path and the reconstructed $xy$-position in the \dd{} data is shown in Fig.~\ref{fig:Mercury_DD_Beam}. The maximum deviation from the linear fit within the fiducial volume is 4~mm.

\subsubsection{\label{sec:field}Effects of non-uniform field}

As noted in Sec. \ref{sec:LUXSim}, the average drift field in LUX is 180 V/cm. Here we treat the position dependence in more detail for some relevant corrections where using the average field is not optimal.

Due to field leakage through the cathode, the electric field in the active xenon is not perfectly uniform, resulting in a small radial component. Figure 8 (left) in~\cite{Akerib:2017:ModelingElectricField} shows the reconstructed positions of events from $^{83\mathrm{m}}$Kr calibrations. The non-axial component to the field results in the electron cloud from an event being pushed inward, away from the wall of the detector. The position reconstruction is performed using the $S2$ signal, resulting in events being assigned an $xy$-position based on where in the liquid the electrons were extracted. 
Consequently, the calculated $xy$ position is closer to the center than the true $xy$-position, with a vertical dependence arising from the field's
depth-dependence.

To understand this phenomenon, the detector electric field was modeled in two dimensions using COMSOL version 4.3b. The model was assumed to be axisymmetric and included the grids and insulators with their proper voltages and dimensions. The predicted field strength is shown in Fig.8 (right) in~\cite{Akerib:2017:ModelingElectricField}. At high radius the field lines become non-parallel, leading to events being reconstructed at smaller radius than the true event position. Along the vertical axis of the detector ($x=y=0$) the field varies from~120~V/cm just above the cathode, to~220~V/cm just below the gate. This variation has a negligible effect on the light and/or charge yields from low-energy nuclear or electron recoils. Furthermore, any such variation in ER response over the fiducial volume is averaged out in the measurement of the tritium ER band.

\begin{figure}
\begin{center}
\includegraphics[width=0.45\textwidth,clip] {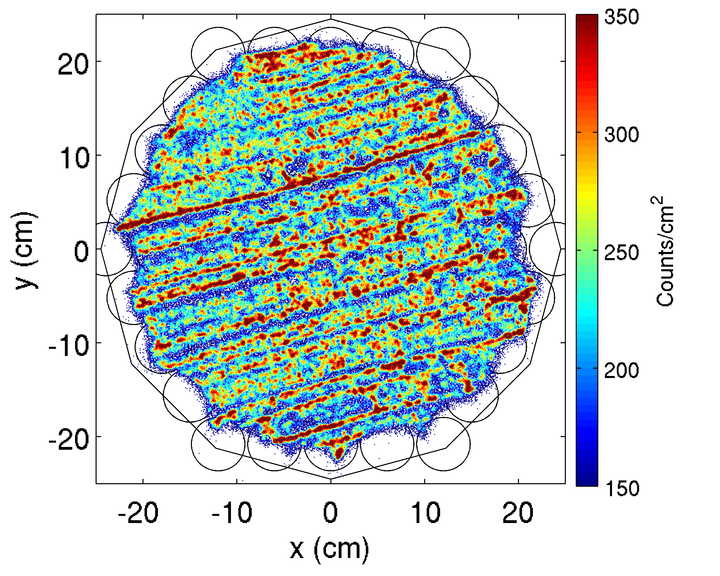}
\includegraphics[width=0.45\textwidth,clip] {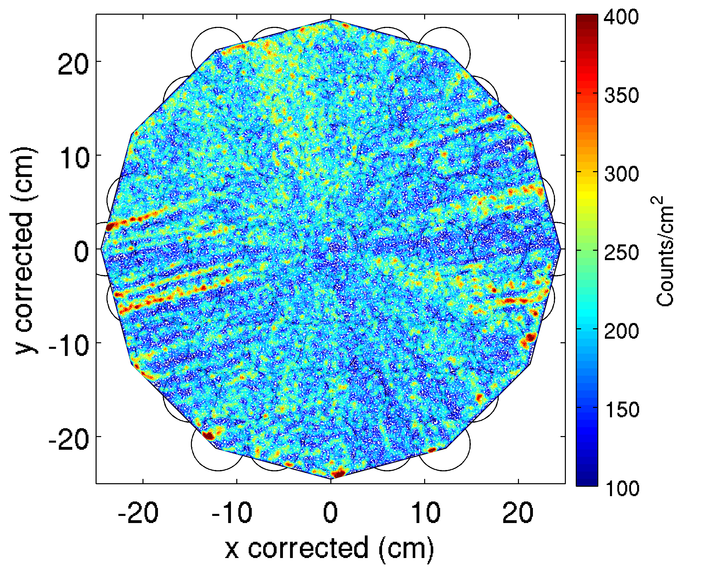}
\caption{\textit{Top}: uncorrected $xy$-positions for $^{83\mathrm{m}}$Kr data with drift times between 200 and 300~\textmu{}s. \textit{Bottom}: The same data, after the correction to $xy$-position has been applied.}
\label{fig:Position_Corrections}
\end{center}
\end{figure}

The distribution of $^{83\mathrm{m}}$Kr is distributed uniformly throughout the detector, allowing it to be used to produce a mapping of reconstructed $xy$-positions to real $xy$-positions. The detector is cut into 30~\textmu{}s slices in drift time and each slice is then further segmented into 60 sections in the polar directions. The first 30 minutes of any injection are ignored to ensure enough time for uniform mixing. The remaining events in each section are placed into 600 uniform radial bins and the average radius in each bin is used to calculate the radial correction. This scheme enforces the reconstructed positions to be radially uniform within each drift time slice. Figure~\ref{fig:Position_Corrections} shows the effects of this radial correction for data with drift times between 200 and 300~\textmu{}s. Effects of non-uniform fields on drift time are found to be negligible.

The field non-uniformity also has a small effect on the pulse area corrections. Because position-dependent pulse area corrections are applied based on $^{83\mathrm{m}}$Kr data, they are affected by the non-uniformity of the field. In principle, this would not be a problem if the light and charge yields for events in the WIMP search region had the same dependence with field as $^{83\mathrm{m}}$Kr events. Unfortunately however, at lower energies the light and charge yields are less sensitive to field. This results in a systematic uncertainty on the $S2$ corrections that can be calculated through comparison to tritium data, and stands at 4\% at the center of the detector. The effect on the $S1$ signal is less than 2\% at any location in the detector.

\subsection{\label{sec:ER}Electron-recoil response}
\subsubsection{\label{sec:Doke}Combined energy model for electron recoils.}
Single-scatter events in the TPC are interpreted with the combined energy model~\cite{platzman}:

\begin{equation}
\label{eq:etotal}
E_{total} = W\cdot(n_e + n_{\gamma}) = W\cdot\left(\frac{S1}{g_1} + \frac{S2}{g_2}\right),
\end{equation}

\noindent
where $g_1$ and $g_2$ represent gain factors that convert $S1$ and $S2$ signals to electron number ($n_e$) and photon number ($n_{\gamma}$), respectively. $W$ is the energy scale factor of LXe in units of eV/quantum, $g_1$ is the product of the average photon collection efficiency and the average QE of the PMTs, while $g_2$ is the product of the electron extraction efficiency at the liquid-gas surface ($\epsilon_{ee}$) and the single electron size. For ER events in LUX, a constant $W$ value of 13.7 eV/quantum is assumed~\cite{Dahl:2009}.

The gain factors $g_1$ and $g_2$ may be determined by observing two or more ER line sources of known energy in which the average light and charge yields differ. $g_1$ and $g_2$ are then fixed by requiring that $E_{total}$, computed with Eq.~\ref{eq:etotal}, reproduces the true energy of each source.  In ER events the average yields vary with energy and electric field due to changes in the average recombination of ionization electrons with Xe$^+$ ions.

\begin{center}
\begin{table}
\setlength{\extrarowheight}{3pt}
\caption{Table of sources used in the analysis of Fig.~\ref{fig:dokeplot}.}
\begin{tabularx}{\linewidth}{rScX}
\hline\hline
Source & \multicolumn{1}{c}{\quad E (keV)\quad} & Type & Origin \\ 
\hline
$^{127}$Xe & 5.3 & L-shell X-ray & Run~3 Data \\
$^{83\mathrm{m}}$Kr & 41.55 & IC & internal calibration source \\ 
$^{131\mathrm{m}}$Xe & 163.9 & IC & early Run~3 Data \\  
$^{127}$Xe & 208.3   & L-shell X-ray + $\gamma$ & Run~3 Data \\  
$^{127}$Xe & 236.1   & K-shell X-ray + $\gamma$ & Run~3 Data \\  
$^{129\mathrm{m}}$Xe & 236.1 & IC & early Run~3 Data  \\  
$^{127}$Xe & 409  & K-shell X-ray + $\gamma$ & Run~3 Data \\  
$^{214}$Bi & 609 & $\gamma$ & detector background \\ 
$^{137}$Cs & 661.6 & $\gamma$ & external calibration source \\
\hline\hline
\end{tabularx}
\begin{tablenotes}
\item[] All source data were collected at 180 V/cm. The $^{129m}$Xe decay and one of the $^{127}$Xe processes completely overlap at 236.1~keV. IC = internal conversion.
\end{tablenotes}
\label{tab:linesources}
\end{table}
\end{center}

\begin{figure}
\begin{center}
\includegraphics[width=0.48\textwidth,clip] {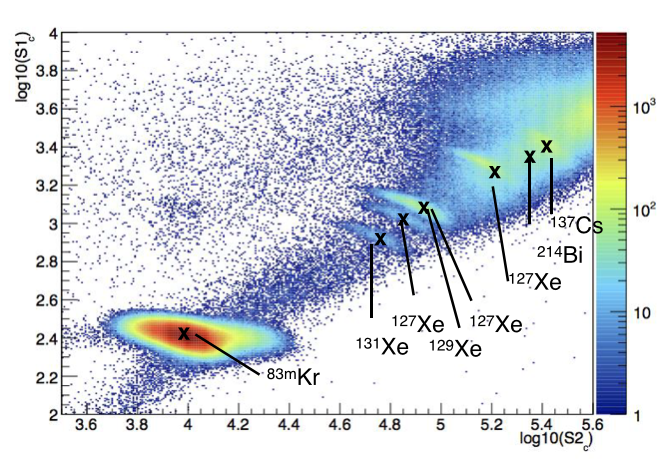}
\caption{$S1_c$ vs. $S2_c$ for a compilation of LUX line source data. $^{83\mathrm{m}}$Kr and $^{137}$Cs data were collected during dedicated calibration runs. All other lines were present in the low background WIMP search data. $^{127}$Xe~\cite{Akerib:xenon127}, $^{129}$Xe and $^{131}$Xe were only present early in Run~3 due to their cosmogenic origin. The anti-correlation between $S1$ and $S2$ is due to recombination.}
\label{fig:linesources}
\end{center}
\end{figure}

The nine sources listed in Table~\ref{tab:linesources} were used to extract values for $g_1$ and $g_2$ in LUX. A scatter plot of $S1_c$ vs $S2_c$ for these data is shown in Fig.~\ref{fig:linesources}. A strong anti-correlation between $S1$ and $S2$ is apparent in each line due to recombination fluctuations. The data are fitted with a rotated two-dimensional Gaussian to determine $\langle$S1$\rangle$ and $\langle$S2$\rangle$ for each line source. To reduce the dependence of the result on the data selection, each fit has data selected within two Gaussian widths of the mean, as determined by the initial fit. Variation in the $S2$ signal and $S1$ signal due to PMT saturation and single electron size variation are included as systematic errors in each fit.

\begin{figure}
\begin{center}
\includegraphics[width=0.48\textwidth,clip] {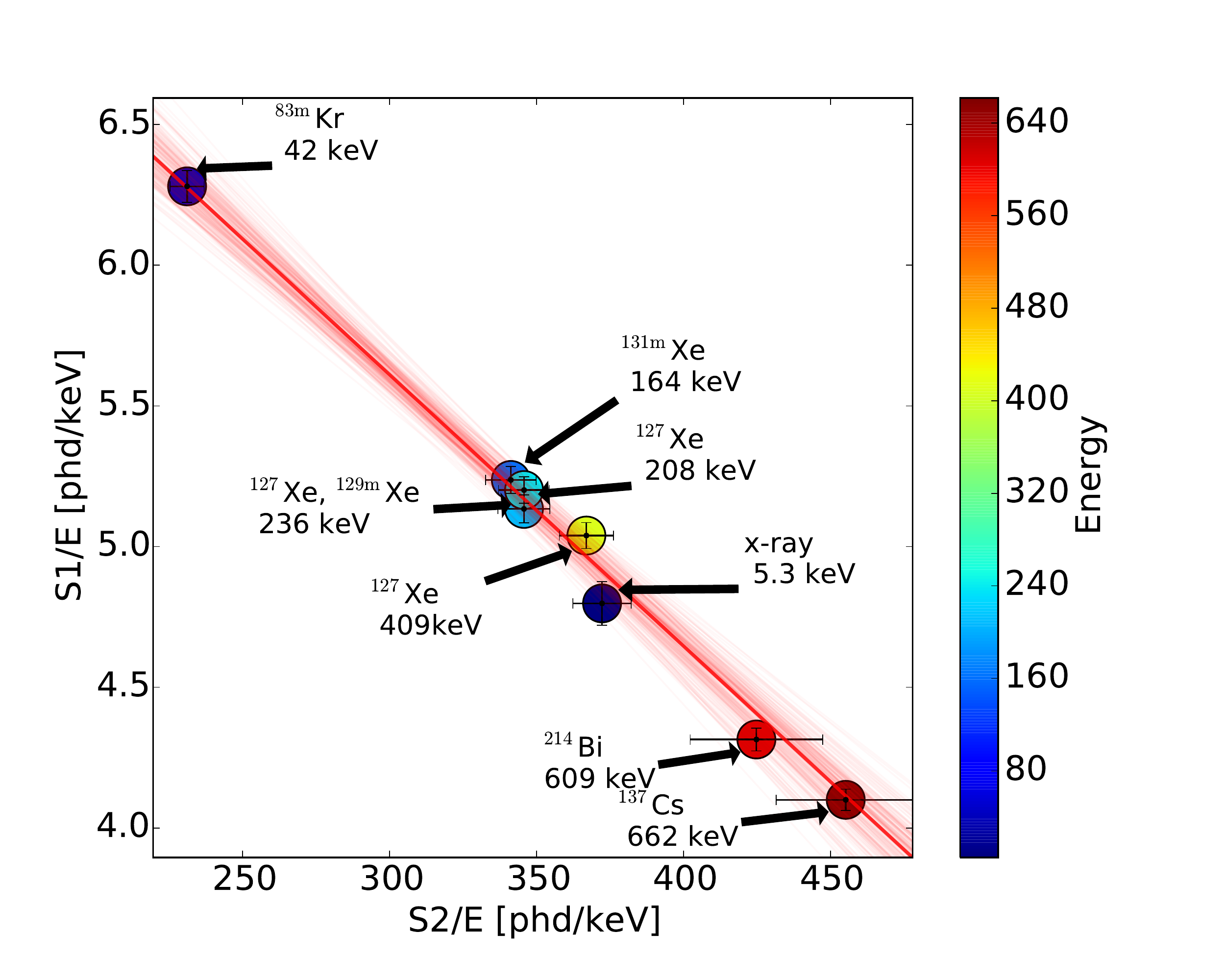}
\caption{Extraction of the gain factors $g_1$ and $g_2$, adhering to the method described in the text. Each point represents a single line source listed in Table~\ref{tab:linesources}.  }
\label{fig:dokeplot}
\end{center}
\end{figure}

The values of $g_1$ and $g_2$ were extracted by plotting $x = \rm{\langle S2 \rangle}/E$ vs $y = \langle $S1$ \rangle /E$ for each source and electric field value, as shown in Fig.~\ref{fig:dokeplot}. A linear fit, $y = mx+b$, was then performed to the eight points, where $m = -g_1/g_2$, and $b = g_1/W$. Values of $g_1 = 0.117 \pm 0.003$ and $g_2 = 12.1 \pm 0.8$ were found, implying $\epsilon_{ee} = 0.491 \pm 0.032$. The errors were determined by toy Monte Carlo where each of the eight points was varied within its error. The extracted value for $g_1$ is in good agreement with the results of the optical model described in Sec.~\ref{sec:OpticalModel}. 

The observed value of $\epsilon_{ee}$ may be compared to previous measurements if the electric field value above and below the LUX liquid surface is known. Uncertainty in the location of the liquid level between the gate grid and the anode leads to uncertainty in the electric field, such that a precise comparison is not possible. The COMSOL field model described in Sec. \ref{sec:field} indicates that the observed value of $\epsilon_{ee}$ is consistent with measurements from~\cite{Gushchin:1979,Gushchin:1982} for a liquid level 3.6~mm above the gate grid. Such a liquid level is consistent with expectations based upon the design of LUX and uncertainties in the thermal expansion coefficients of the TPC insulators.
These results have been further validated by fitting tritium data to the tritium beta spectrum, finding $g_1 = 0.115 \pm 0.005$ and $g_2 = 12.1 \pm 0.9$, implying $\epsilon_{ee} = 0.509 \pm 0.038$~\cite{Akerib:2015:tritium}.  

\subsection {\label{sec:NEST_ER_model}Electron recoils: Implications for NEST ER model}

\begin{figure}
\begin{center}
\includegraphics[width=0.45\textwidth,clip] {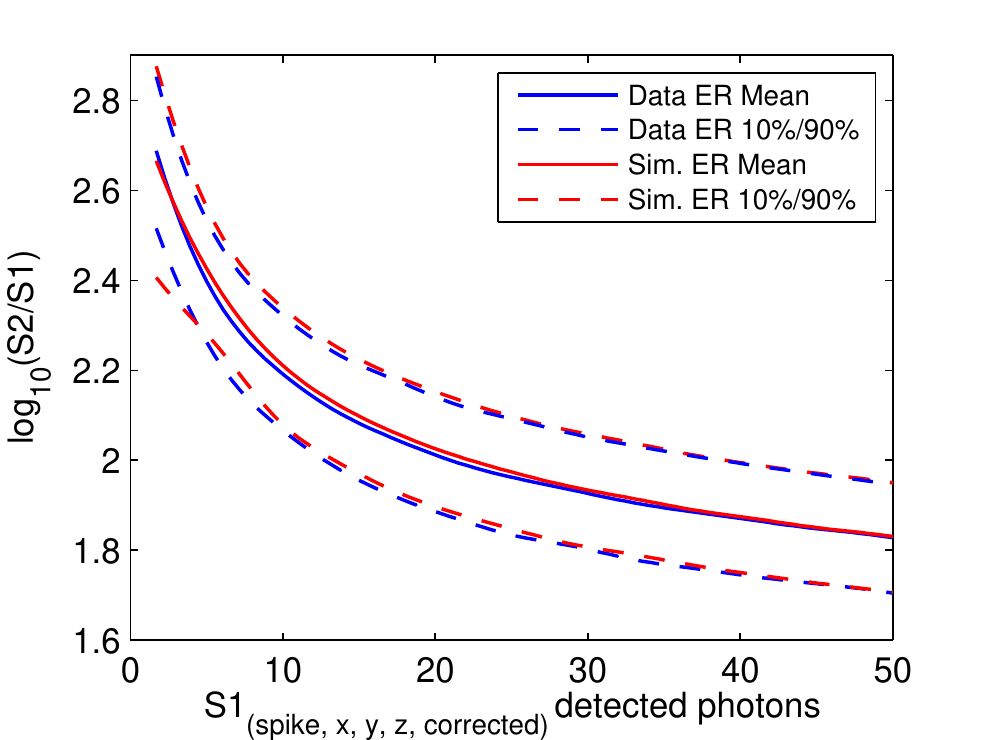}
\caption{The mean and width of the ER band as determined from Gaussian fits to the tritium population in $S1$ slices are depicted in blue, with a tuned MC based on the fundamental yields extracted using the known beta energy spectrum, at the same discrete points in $S1$, in red. The latter is based on a fit to the yields, but not to the band, and it is thus in a sense a prediction.}
\label{fig:erband}
\end{center}
\end{figure}

The electron recoil model is based upon the high-statistics tritium data sets described in~\cite{Akerib:2015:tritium}. The light and charge yields in simulations were tuned to reproduce the tritium data. A comparison between the tuned simulation and the tritium data is shown in Fig.~\ref{fig:erband}.

As shown in~\cite{Akerib:2015:tritium}, NEST version 0.98 for ER from 2013~\cite{Szydagis:2013} disagrees by $\sim$10\% at the lowest and highest energies of interest. As thoroughly discussed in~\cite{Szydagis:2011}, NEST uses the Thomas-Imel model of recombination, but here, an extra energy dependence was added to force agreement with the high statistics LUX tritium data for all $S1$ and $S2$ yields. This is done on a purely empirical basis, although a first-principles physics justification is to be explored in the future that will, by extension, also explore the ER $S2$/$S1$ band behavior. This was not required in previous analyses, and still is not required for NR analysis, as seen next. 

\subsection{\label{sec:NR}Nuclear-recoil response}
\subsubsection{\label{sec:DD}D-D neutron calibration}

Mono-energetic neutrons from a deuterium-deuterium (\dd{}) fusion source were used to measure the response of the LUX detector in $S_{1c}$ and $S_{2c}$ to nuclear recoils in the ranges 1.1-74 keV and 0.7-74 keV, respectively~\cite{Akerib:2015:dd}. 

The ratio of the ionization to scintillation signal is used to discriminate between nuclear and electron recoils in liquid xenon TPCs. Neutrons from the \dd{} source are used to calibrate the nuclear recoil band over the $S1$ range used for the WIMP search analysis. Double scatter nuclear recoil events are used to kinematically reconstruct the energy of the first recoil to determine the response versus true recoil energy (see~\cite{Verbus:2017:DDCalibrationTechnique,Akerib:2015:dd} for details). Subsequently, a simulated nuclear recoil band is compared to the \dd{} calibration data to demonstrate consistency of the nuclear recoil signal model used to generate $S1$ and $S2$ PDFs for the WIMP search profile likelihood ratio analysis, described in Sec.~\ref{sec:limit}. 

\begin{figure}
\begin{center}
\includegraphics[width=0.48\textwidth,clip] {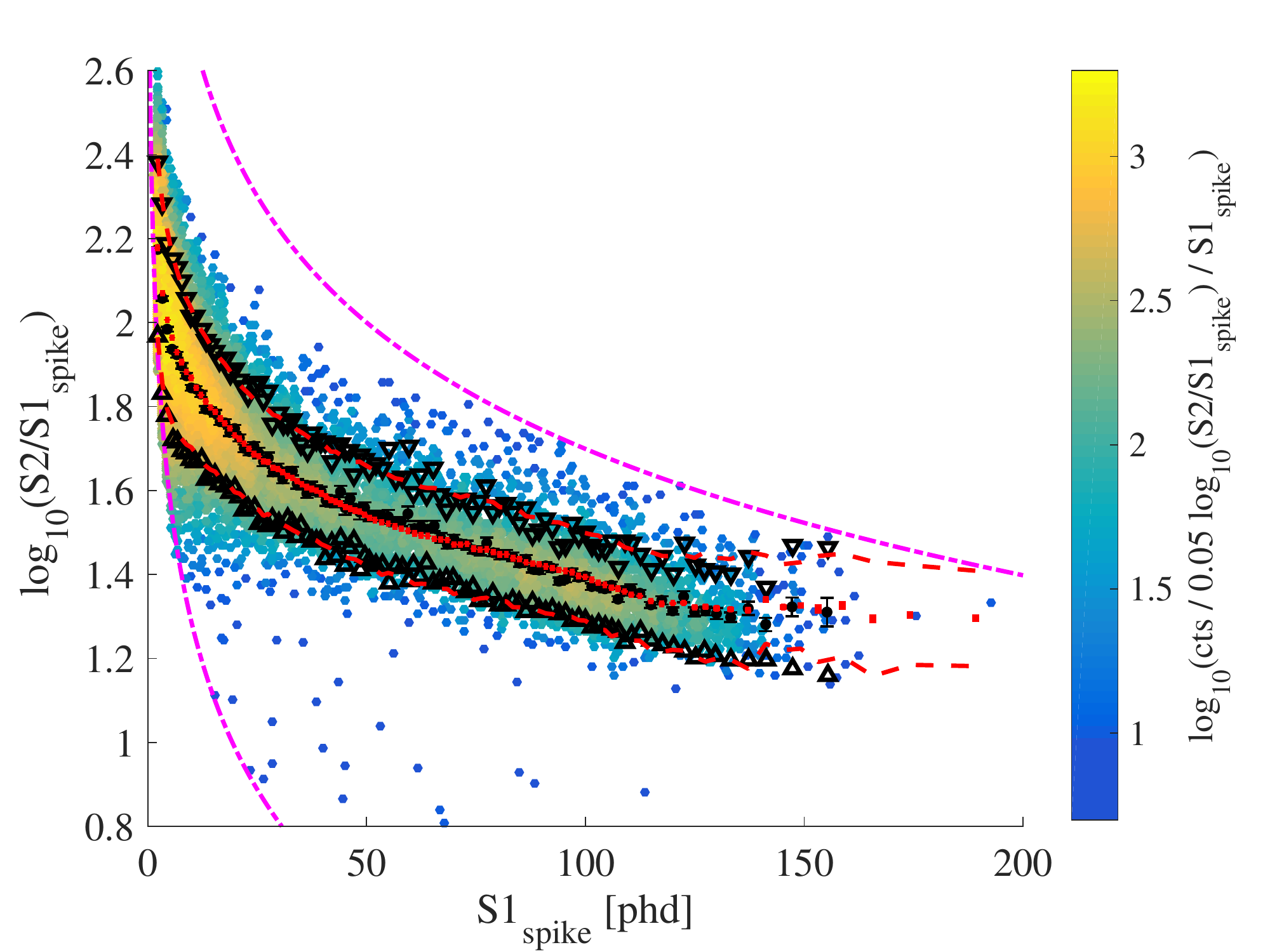}
\caption{The measured events from the D-D exposure defining the nuclear recoil. 
There are 19249 events
remaining after all cuts. The black data points are the Gaussian fit mean values for each $S1$ spike bin. The red data points are the corresponding Gaussian fit mean values for the simulated nuclear recoil band. The black triangles and red dashed line indicate the 90\% one-sided limits from data and simulation, respectively. The lower magenta dot-dashed line indicates the $S2$ threshold at 165~phd, and the upper magenta dot-dashed line shows the applied upper cut for $S2$ ($S2$$<$5000~phd). The lower density of points in the center reflects  the differential cross-section behavior for 2.45~MeV neutrons incident on a xenon target. Error bars on the black and red data points are statistical only, and are generally too small to see.}
\label{fig:NR_band_fullRange}
\end{center}
\end{figure}

The nuclear recoil band was measured using single-scatter events from the \dd{} calibration dataset. An $S2$ threshold at 165~phd was applied on the raw $S2$ area before position correction for consistency with the WIMP search (Sec.~\ref{sec:S2threshold}). The simulations were produced using LUXSim/Geant4 and the Lindhard-based NEST model described in Sec.~\ref{sec:Model}, and were passed through the same LUX data processing framework as the \dd{} calibration data. The same cuts and analysis used for nuclear recoil band data were applied to the resulting reduced simulation waveforms. Figure~\ref{fig:NR_band_fullRange} shows the nuclear recoil band for the full range of the \dd{} calibration data, after the analysis cuts. The cuts applied are described in~\cite{Akerib:2015:dd}, but the $S1$ range is extended from 50 phd to 200 phd. This shows the data up to the recoil energy spectrum endpoint at 74 $\rm keV_{nr}$. The simulated nuclear recoil band is consistent with \dd{} calibration data within the systematic uncertainty intrinsic to the simulation process.

\begin{figure}
\begin{center}
\includegraphics[width=0.48\textwidth,clip] {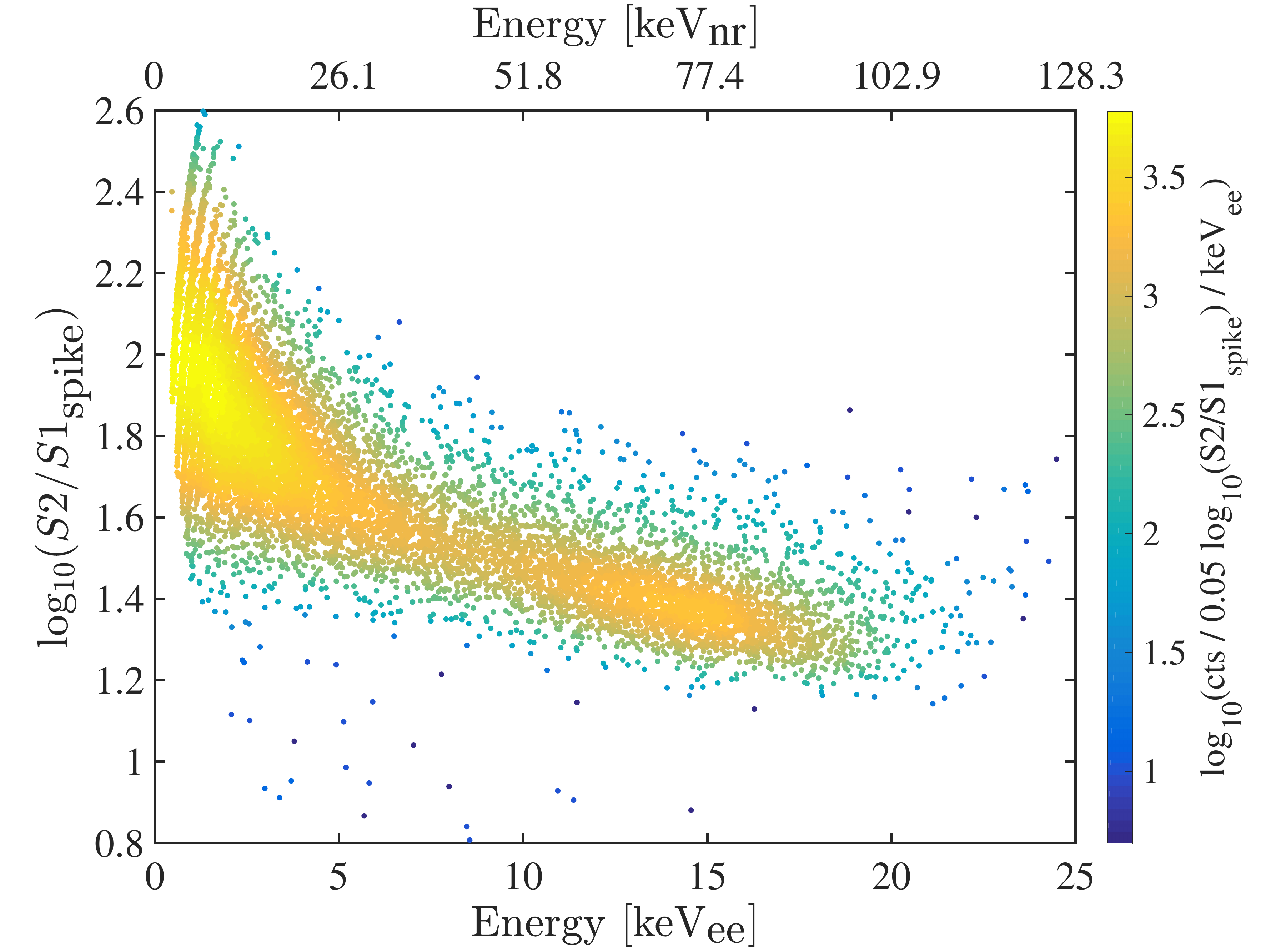}
\caption{The measured events defining the nuclear recoil band are shown in the scatter plot as a function of energy in $\rm keV_{ee}$ and $\rm keV_{nr}$. They are the same 19249 events, after analysis cuts, shown in Fig.~\ref{fig:NR_band_fullRange}. The density scale is different from the previous nuclear recoil band plot due to normalization by bin width.}
\label{fig:NR_band_keVee}
\end{center}
\end{figure}

Figure~\ref{fig:NR_band_keVee} shows the same events, applying the same cuts, but now as a function of combined energy in $\rm keV_{ee}$, and as a function of $\rm keV_{nr}$, where nr indicates the energy scale for nuclear recoils. Explicitly, the $\rm keV_{ee}$ energy scale is calculated through Eq. \ref{eq:etotal}, where the gains during the \dd{} calibration time period are $g_1 = 0.115 \pm 0.004$ and $g_2 = 11.5 \pm 0.9$~\cite{Akerib:2015:dd}. In turn, the $\rm keV_{nr}$ is calculated using

\begin{equation}
E[{\rm keV_{nr}}] = \frac{E{\rm[keV_{ee}]}}{L{\rm[keV_{nr}]}},
\end{equation}

\noindent 
where $L$ is Lindhard's factor \cite{Lindhard}, described in Sec.~\ref{sec:Model}. Again, this plot shows data up to the recoil energy spectrum endpoint at 74~$\rm keV_{nr}$. The non-zero width of the vertical bands of events at low energy is due to corrections for $S1$ spike overlap in the per-channel waveforms, as well as to 3D position-based detector corrections. 

\subsubsection{\label{sec:Model} Modeling nuclear recoils in liquid xenon}

A model of nuclear recoil response for use in simulations and WIMP searches has been built and constrained following the techniques used in \cite{Lenardo:2015}. An energy deposit $E_0$ in the liquid is distributed between the formation of excitons and of electron-ion pairs. Some energy is additionally lost as an unmeasurable dissipation of heat. The process is modeled with a modified Platzman equation for rare gases \cite{platzman}, to which an efficiency factor, $L$, is applied to account for the energy lost to atomic motion rather than the detectable electronic channels:
\begin{equation}
E_0 = \frac{( N_{ex} + N_i ) W}{L},
\end{equation}
where $W$ is the average energy required to produce a quantum (either an exciton or ion) in the liquid. Again, $W = 13.7~\text{eV}$ is assumed. 

Three different stages determine how the generated quanta are divided into the final light ($S1$) and charge ($S2$) signals. At the first stage, a fixed exciton-to-ion ratio $N_{ex} / N_i$ is defined, that determines how much of the energy initially goes to excitation as opposed to ionization. At the second stage, the ion-electron pairs 
recombine with probability $r$ to produce further excitons.  At the third stage, biexcitonic collisions cause some excitons to de-excite either through heat or through Penning ionization. This effect
is modeled by multiplying the resulting number of photons by a fraction $f_l$, and allowing some fraction $P$ of the quenched excitons to become ions. At the end of these three stages, the leftover excitons de-excite to produce scintillation photons, while the electrons that escape recombination contribute to the charge signal. The final equations for the number of photons ($n_{ph}$) and electrons ($n_{e}$) produced by an energy deposition $E_0$ are
\begin{equation}
n_{ph} = L \times f_l \times\frac{E_0}{W}
\left[ 1 -  A (1-r) \right],
\end{equation}
\begin{equation}
n_e = L \times \frac{E_0}{W} 
\left[ B(1 - A)(1-r) + A \right],
\end{equation}
where $A~=~1/(1 + N_{ex} / N_i)$ and $B~=~P(1 - f_l)$. The expressions for $L$, $f_l$, and $r$ are described in detail below.

\subparagraph{\label{sec:L_factor} The Lindhard factor}

A factor $L$  determines the fraction of energy in a nuclear recoil event that goes into scintillation and ionization, rather than atomic motion. In a single Xe-Xe collision, this fraction is given by ${s_e}/(s_e + s_n)$, where $s_e$ and $s_n$ are the electronic and nuclear stopping powers (energy loss per unit distance). $L$ represents this fraction for the entire cascade of collisions in an ionization event. The model given by Lindhard's theory \cite{Lindhard} is adopted, with

	\begin{equation}
	L = \frac{k\,\,g(\epsilon)}{1 + k\,\,g(\epsilon)},
	\label{eq:lindhard}
	\end{equation} 
\noindent
where $k$ is a proportionality constant between $s_e$ and the velocity of the recoiling nucleus, and $\epsilon$ is a dimensionless energy scale equal to $11.5~(E_{nr}/{\rm keV})~Z^{-7/3}$. The function $g$ models the ratio of electronic to nuclear stopping powers under the Thomas-Fermi approximation, and is given by

	\begin{equation}
	g(\epsilon) = 3\,\epsilon^{0.15} + 0.7\,\epsilon^{0.6} + \epsilon.
	\end{equation}

\noindent
The parameter $k$ is treated as a free parameter for fitting of this model to the nuclear recoil data. (For ER, $L = 1$ fits all measurements to date.)

In addition to Lindhard's theory, an alternative model has been investigated that gives more favorable signal strength at energies below 2~keV. For this, the Thomas-Fermi approximation is replaced with the Ziegler {\em et al.} parameterization of nuclear stopping power, as described in \cite{Bezrukov}. Ziegler's expression is calculated using the universal screening function. The reduced stopping powers (given in terms of the dimensionless energy) in this model are given by
\begin{equation}
s_e(\epsilon) = 0.166 \epsilon^{0.5},
\label{eq:s_e}
\end{equation}
\begin{equation}
s_n(\epsilon_\text{Z}) = \frac{\text{ln}(1 + 1.1383\,\epsilon_\text{Z})}{2[\epsilon_\text{Z} + 0.01321\,\epsilon_\text{Z}^{0.21226} + 0.19593\,\epsilon_\text{Z}^{0.5}]},
\end{equation}
where $\epsilon_\text{Z}=1.068\epsilon$. The slight difference in energy scales is due to different assumed screening lengths in the calculation of the dimensionless energy. Following \cite{Bezrukov}, an additional prefactor $\alpha$ is introduced, which multiplies the entire expression to account for the cascade of collisions generated by a single initial nuclear recoil, thus $\alpha > 1$. The final expression for the $L$-factor under this alternative model is 
\begin{equation}
L = \alpha \frac{s_e}{s_e + s_n}.
\label{eq:bez}
\end{equation} 

\noindent
The Lindhard model (Eq. \ref{eq:lindhard}) is used in the signal model to set the WIMP-nucleon cross section limit, while the alternative (Eq. \ref{eq:bez}) is used only to demonstrate the effect of more optimistic assumptions in yields at low energies.

\subparagraph{\label{sec:Recombination} Recombination}

The probability of recombination, $r$, is calculated using 
the Thomas-Imel box model \cite{ThomasImel}, which gives 
\begin{equation} 
r = 1 - \frac{\text{ln}(1 + N_i \, \varsigma)}{N_i \, \varsigma}.
\end{equation}
The energy dependence in this equation is contained in the number of ions $N_i$.  The quantity $\varsigma$ is dependent on the applied electric field. However, since LUX is operated and calibrated at a constant 180~V/cm, data provide no constraint on this property; consequently the field dependence in this work is ignored and $\varsigma$ is treated as a constant parameter.

\subparagraph{\label{sec:Biexcitonic_collisions} Biexcitonic collisions}

A final quenching and ionization is applied to the light signal to account for Penning effects, in which 
two excitons can interact to produce one exciton and one photon \cite{MeiHime}, or one photon and one electron. Both processes remove quanta from the photon signal. Following the analysis in~\cite{Bezrukov}, this quenching is parameterized as the fraction derived from Birks' saturation law \cite{BirksTheoryAndPracticeOfScintillationCounting}
\begin{equation}
f_l = \frac{1}{1 + \eta\, s_e},
\end{equation}
where $\eta$ is a strength parameter and $s_e$ is given by Eq. \ref{eq:s_e}. Here, $f_l$ represents the proportion of excitons that remain; the fraction of quenched quanta is given by $(1 - f_l)$. 
This expression exhibits an increased quenching effect with increasing energy, due to higher excitation density along
the track of the recoiling xenon atom. Penning ionization manifests as  some fraction of the collisions resulting in the release of electrons.  The parameter $P$ is introduced to model the unknown ratio of energy lost to ionization vs. heat in biexcitonic processes.

\subparagraph{\label{sec:Constrain} Constraining the model}
To fit the model, a global likelihood is constructed that is simultaneously constrained by measurements of the light yield, charge yield, and nuclear recoil band mean. The yields are constrained by an analytical model, while the nuclear recoil band requires a full MC simulation to generate the $S1$ and $S2$ signals from which the mean $\text{log}_{10}(\text{S}2/\text{S}1)$ is calculated in bins of $S1$. The global likelihood can be separated into the product
	\begin{equation}
	\mathcal{L}_{global} = \mathcal{L}_{yields} \times \mathcal{L}_{band}.
	\end{equation}

The likelihood function for light and charge yields is constructed by assuming each point is a 2D Gaussian distribution in energy and yield, with the width given by the $x$ and $y$ uncertainties. A joint likelihood is calculated using Eq. 4.1 in \cite{Hsiao}:
	\begin{equation}
		\begin{split}
		\mathcal{L}_{yields} (\vec{\theta} | E_i, Y_i) = 
		\prod_{i=0}^n \,\,\int_0^\infty  A_i \, \text{exp}\left( \frac{- ( E_i - E)^2}{2 \sigma_{E_i}^2} \right) \times \\
		\text{exp} \left( \frac{- (Y_i - \mu_{Y})}{2 \sigma_{Y_i}^2} \right) dE,
		\end{split}
	\end{equation}
where $A_i$ is a normalization constant, $E_i$ is the energy of the point $i$, $Y_i$ is either charge or light yield ($L_y$ or $\mathcal{Q}_y$) of point $i$, and $\mu_{Y}$ is the model prediction at the energy $E$. The vector $\vec{\theta}$ is the vector of free parameters in the model.

The likelihood for the band mean is constrained by the mean values of $\text{log}_{10}(\text{S}2/\text{S}1)$ in the nuclear recoil data, binned by $S1$. A full MC simulation of the NR band is performed using the model defined by $\vec{\theta}$ calculating mean values in the same $S1$ bins to compare to data. The likelihood equation is given by

	\begin{equation}
		\begin{split}
		\mathcal{L}_{band} & (\vec{\theta} \,|\, \overline{\text{log}_{10}(\text{S}2/\text{S}1)}_j ) = \\ 
		\\
		 &\prod_{j=0}^m B_j \, \text{exp}\left( \frac{ - (\,\,  \overline{\text{log}_{10}(\text{S}2/\text{S}1)}_j - \mu_{sim} )^2 }{2 (\sigma_{j}^2 + \sigma_{sim}^2)} \right).
		\end{split}
	\end{equation}
In the above, $B_j$ is a normalization constant, 
$\overline{\text{log}_{10}(\text{S}2/\text{S}1)}_j$ is the measured band mean in the $j$th $S1$ bin, and $\sigma_j$ is the measured uncertainty. The values for $\mu_{sim}$ and $\sigma_{sim}$ are the band mean and uncertainty calculated from the simulation. Due to the need to run the simulation repeatedly while constraining the model, the simulation is run with statistics comparable to the measured values.

To optimize this model, a Metropolis-Hastings Markov Chain Monte Carlo algorithm (MCMC) was used to produce 10,000 samples of $\mathcal{L}_{global} (\vec{\theta}) $ \cite{HastingsMCMC}.  The advantage of this method is that it produces a map of the likelihood distribution across all parameters, naturally incorporating correlations.  To extract an optimal model and uncertainties, the MCMC sample was used to calculate the means and standard deviations of each of the five parameters. The results are shown in Table~\ref{tab:fits}. The MCMC was also used to construct 68\% and 95\% confidence bands on the model.  The calculation of $\mathcal{Q}_y$ and $L_y$ from the best fit model with confidence intervals is shown in Fig.~1 of \cite{Akerib:2015:run3}, and the best-fit simulation of the nuclear recoil band is shown in Fig.~13 of \cite{Akerib:2015:dd}.

\begin{table}
\setlength{\extrarowheight}{3pt}
\caption{Means and standard deviations of the free parameters using both the Lindhard and the alternative model, calculated from the MCMC sample set.}
\label{tab:fits}
\centering
\begin{tabularx}{\linewidth}{X D{,}{\pm}{4.4} D{,}{\pm}{4.4}}
\hline\hline
Parameter & \multicolumn{1}{c}{\qquad Lindhard model \qquad \qquad} & \multicolumn{1}{c}{Alternative model} \\
\hline
$k$         & 0.1735,0.0060 & \multicolumn{1}{c}{N/A}   \\
$\alpha$    &  \multicolumn{1}{c}{N/A}                & 2.212,0.081\\
$\varsigma$ & 0.0168,0.0020 & 0.0145,0.0018 \\
$N_{ex}/N_i$& 0.482,0.069   & 0.487,0.058 \\
$\eta$      & 13.2,2.3      & 6.7,3.4 \\
$P$         & 0.111  ,0.060 & 0.18,0.10 \\
\hline\hline
\end{tabularx}
\end{table}

\subsection{\label{sec:discrimination}ER and NR discrimination}

\begin{figure}[htbp]
\centering
\includegraphics[width=0.48\textwidth,clip]{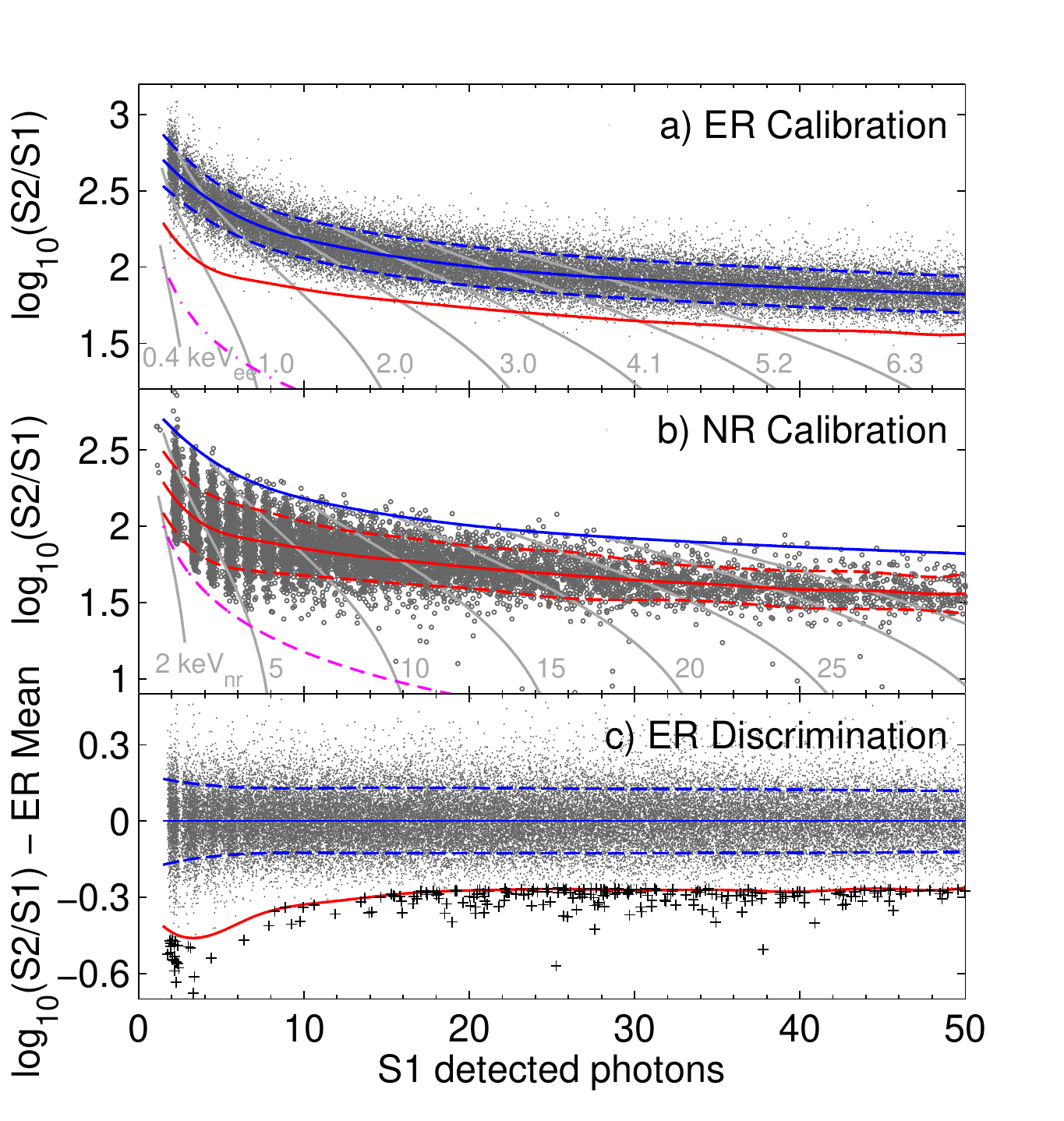}
\caption{Calibrations of the detector response in the fiducial volume, and leakage of ER events below the NR mean. Panel $(a)$ shows the ER (tritium) calibration, while panel $(b)$ shows the NR calibration (obtained with mono-energetic neutrons from a \dd{} generator). Solid lines show band means, while dashed lines indicate the $\pm1.28\sigma$ contours (blue for ER and red for NR). Also shown is the $S2$ threshold applied in the analysis (dot-dashed magenta line). Panel $(c)$ shows $\text{log}_{10}(\text{S}2/\text{S}1)$ normalized to the mean of the ER band as a function of the $S1$ for the tritium data. The 248 events that fall below the NR mean are plotted using a different marker (vertical cross) for clarity.}
\label{fig_Dis01}
\end{figure}

The tritium data used are shown in Fig.~\ref{fig_Dis01}~(a) and (c). In panel (c) the 248 events below the NR mean are represented by a different marker for clarity. The ER band width is represented by the dashed lines, and is mostly flat for $S1$ signals larger than 5~phd with a small broadening below this value. The difference in 
$\text{log}_{10}(\text{S}2/\text{S}1)$ between the NR and ER means, shown in solid lines in the different panels of Fig.~\ref{fig_Dis01}, has a significant variation below 20~phd resulting in a large dependency of the discrimination on $S1$. 

\begin{figure}[htbp]
\centering
\includegraphics[width=0.48\textwidth,clip]{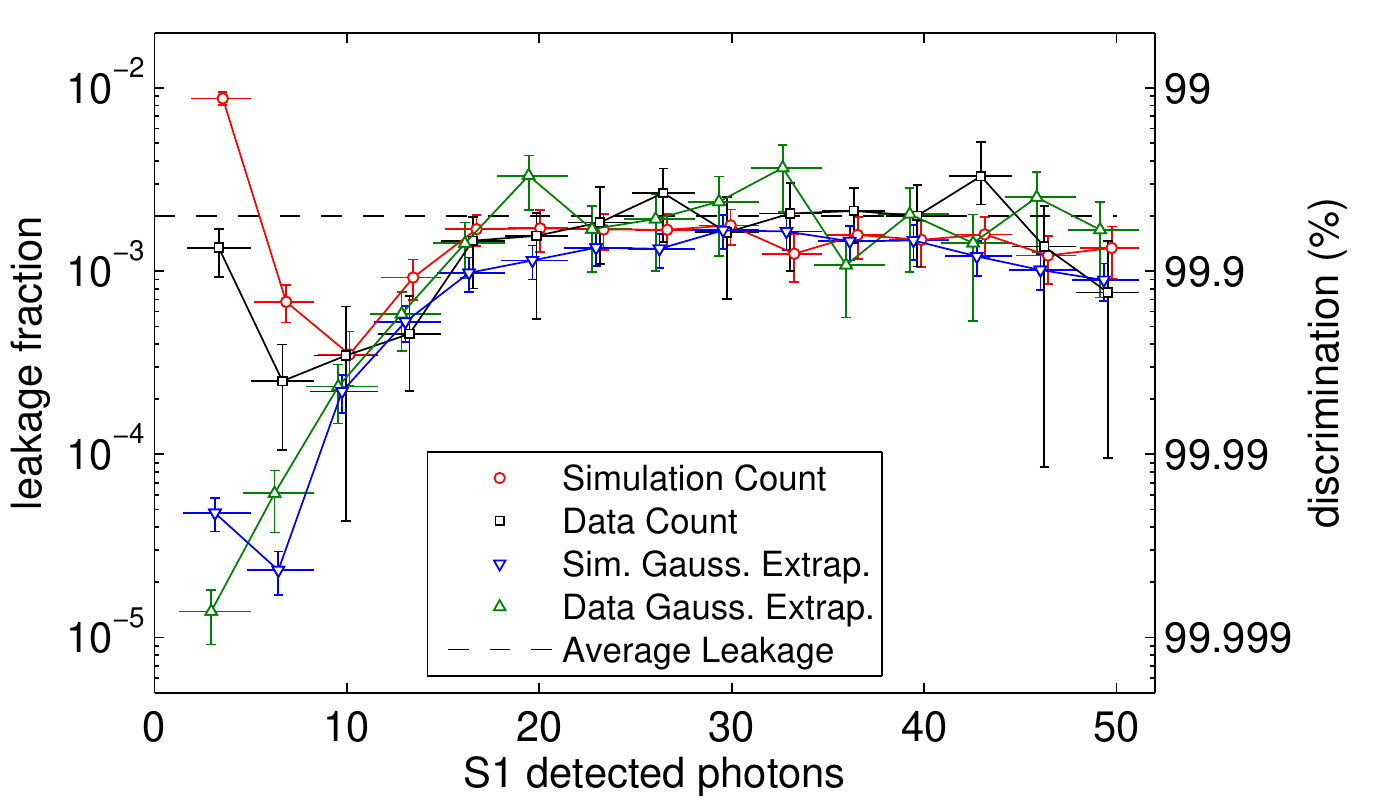}
\caption{Leakage (left) and discrimination (1$-$leakage, right) vs. $S1$. Values obtained from a Gaussian extrapolation (blue and green triangles) are shown alongside those obtained from counting tritium events. The dashed line corresponds to the average discrimination of 99.80\% in data count. To achieve good agreement on observed leakage (data count) the $g_1$ and $g_2$ used in simulations had to be adjusted by 5\%, well within uncertainty. The Monte Carlo neither underestimates nor overestimates the leakage fraction systematically, alternating between the two within error, so, for both total leakage and Gaussian-only, MC (first principles) and data agree fairly well.}
\label{fig_Dis02}
\end{figure}

To estimate the discrimination as a function of $S1$, the leaking events were sliced in 3.3~phd wide bins with the first slice centered at 3.35~phd and the last at 49.55~phd. This ensured enough events in each bin to estimate the leakage fraction --defined as the ratio between the number of leaking events and the total number of tritium events observed in each bin-- while maintaining consistency with the slicing used in both the NR and ER calibrations. The results are shown by black squares in Fig.~\ref{fig_Dis02}. One can see that the leakage fraction decreases in the first few bins but then increases with $S1$ up to $\sim$20~phd, remaining constant thereafter. This behavior could not be observed in past detectors, which saw only the smooth increase, because it requires a lower threshold (which is driven by a high $g_1$). The green triangles in the same figure represent the leakage fraction obtained from a pure Gaussian extrapolation: in this case Gaussian fits to the ER band slices are used to obtain the band width as a function of the $S1$, which is then used to estimate the number of events that are below the NR band centroid. For the first bins, this Gaussian estimate under-predicts observed leakage due to contributions from upward $S1$ fluctuations of events with mean energy below threshold: these events contain relatively smaller $S2$s, and will thus end up well below the ER band. The greatest consistency with Gaussianity occurs at high $S1$. The non-Gaussian or ``anomalous'' leakage may be explained as above, and was expected based on simulation with LUXSim and NEST in Geant4 (red circles).

The dashed line in Fig.~\ref{fig_Dis02} is the weighted average of the leakage, in which the weights are the number of events in the WIMP search data for each $S1$ slice. The statistical uncertainty takes into account the uncertainty from the NR calibration and the binomial uncertainty from the number of the leaking events, while the systematic uncertainty is dominated by the field variations along the drift time --resulting in changes in the discrimination between different $z$ positions in the chamber-- with a smaller contribution coming from the variation of the single electron size between the NR and ER calibrations.

Considering the number of ER events in the WIMP search run after quality and fiducial cuts are applied, the expected leakage below the NR mean is 0.7$\pm$0.1(stat)$\pm$0.3(syst) events.

\section{\label{sec:data}Data Analysis}
\subsection{\label{sec:EventSelec}Event selection: analysis cuts}

Data acquired between April 21st and September 1st 2013 have been analyzed, a total of 132 calendar days, of which 101 days correspond to the data sample for the WIMP search. The remaining data samples correspond to ER and NR calibrations, xenon circulation outages and two periods in early August following tritiated methane injections (due to a higher electron recoil background following tritium injections, a conservative decision was taken to exclude these data from the final analysis). The final live-time for the actual dark matter search corresponds to 95 days, which accounts for exclusions due to periods of detector instability (2.5\%), the trigger hold-off (1.8\%) and other minor contributions (lost/corrupted files and DAQ dead-time, of about 1.6\%). 

Single scatter events (dubbed golden events), potentially indicative of WIMP elastic scattering off nuclei, are selected for further analysis (see Sec.~\ref{sec:EventClassification}). Monitoring of the detector stability was performed using slow control (SC) parameters that could influence its response: liquid level, HV grid voltages and currents, outer vacuum vessel pressure, circulation flow rate and detector pressure. All these parameters were scanned for the entire duration of the run looking for out-of-bounds periods, with excursions occurring less than 5 minutes apart being merged into a single unified period. Zero-length outages (a single data point outside bounds) were extended to the past and the future by the average of the update period for the corresponding SC sensor in the 10 min interval containing the excursion. Zero length outages were not considered for the liquid level and the outer vacuum pressure, as they would likely be caused by sensor fluctuations; all other outages resulted in data being excluded from the analysis. An outage was considered to be over once the sensor returned to within nominal bounds, with the exception of circulation failures: in these cases, the outage period was extended until the next $^{83\mathrm{m}}$Kr calibration, which provided updated information on the electron lifetime. Additionally, the trigger rate of each dataset was visually inspected and periods of 30~s around the rate excursions were excluded from the final analysis.

A radial fiducial cut was placed at 20~cm, driven by the leakage of decay products from $^{214}$Pb (a $^{222}$Rn daughter) implanted on the detector walls as discussed in Sec.~\ref{sec:NR_BG}. Additionally, the height of the fiducial volume was defined to extend from 38 to 305~\textmu{}s in drift time (corresponding to a total liquid height of 40.3$\pm$0.2~cm) to reduce backgrounds from the PMT arrays and electrodes. Finally, the $S1$ and $S2$ thresholds (1~phd and 165~phd respectively, described in detail in Sec.~\ref{sec:eff}) were applied to the remaining events, as well as an $S1$ upper limit of 50~phd.

The cleanliness of the data acquired allowed for an analysis with only a single data-quality cut, to exclude periods of high rate of single electron background. Small $S2$ pulses due to single electrons leaving the liquid surface are prevalent in the extended tails of genuine high energy events (for up to several ms), and may exceed the $S2$ threshold either due to a fluctuation in a single electron signal size, or more commonly due to multiple electrons leaving the liquid within a few \textmu{}s. These $S2$s may be associated with a random isolated $S1$ or a ``fake'' $S1$ resulting from the pulse finder algorithm over-splitting a prior single electron and producing a false event. Excluding events with a significant area outside the ($S1$ and $S2$) signal pulses effectively removes these fake events while ensuring a large acceptance for WIMP-like events.

\begin{figure}
\begin{center}
\includegraphics{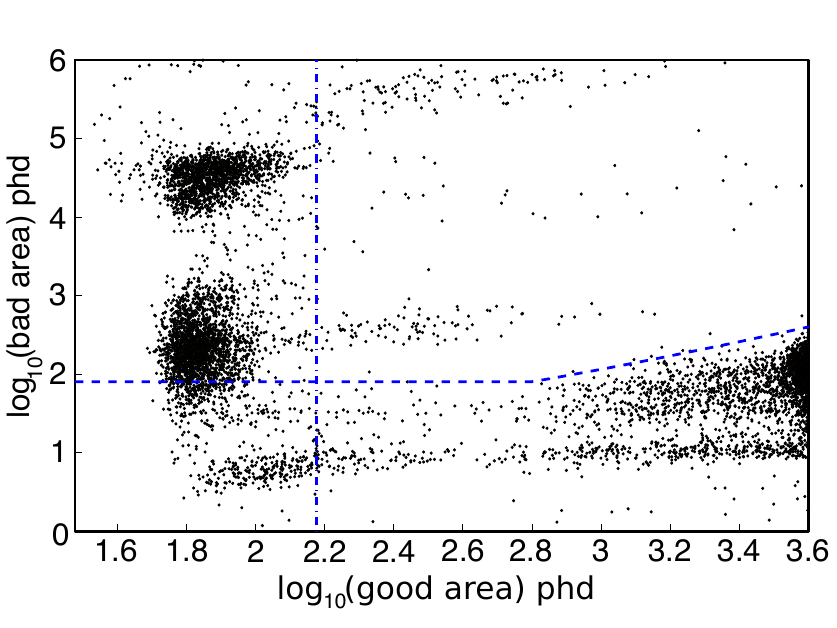}
\caption{Cleanliness cut applied to golden events, shown by the dashed lines, which removes periods of high-rate single electron background, is defined as a function of extra non-signal area contained within the 1~ms window of each event. The 165~phd $S2$ threshold used in the analysis is also shown (vertical dot-dashed line).}
\label{fig:badarea}
\end{center}
\end{figure}

\begin{figure}
\begin{center}
\hspace*{-0.5cm}
\includegraphics[width=0.5\textwidth,clip]{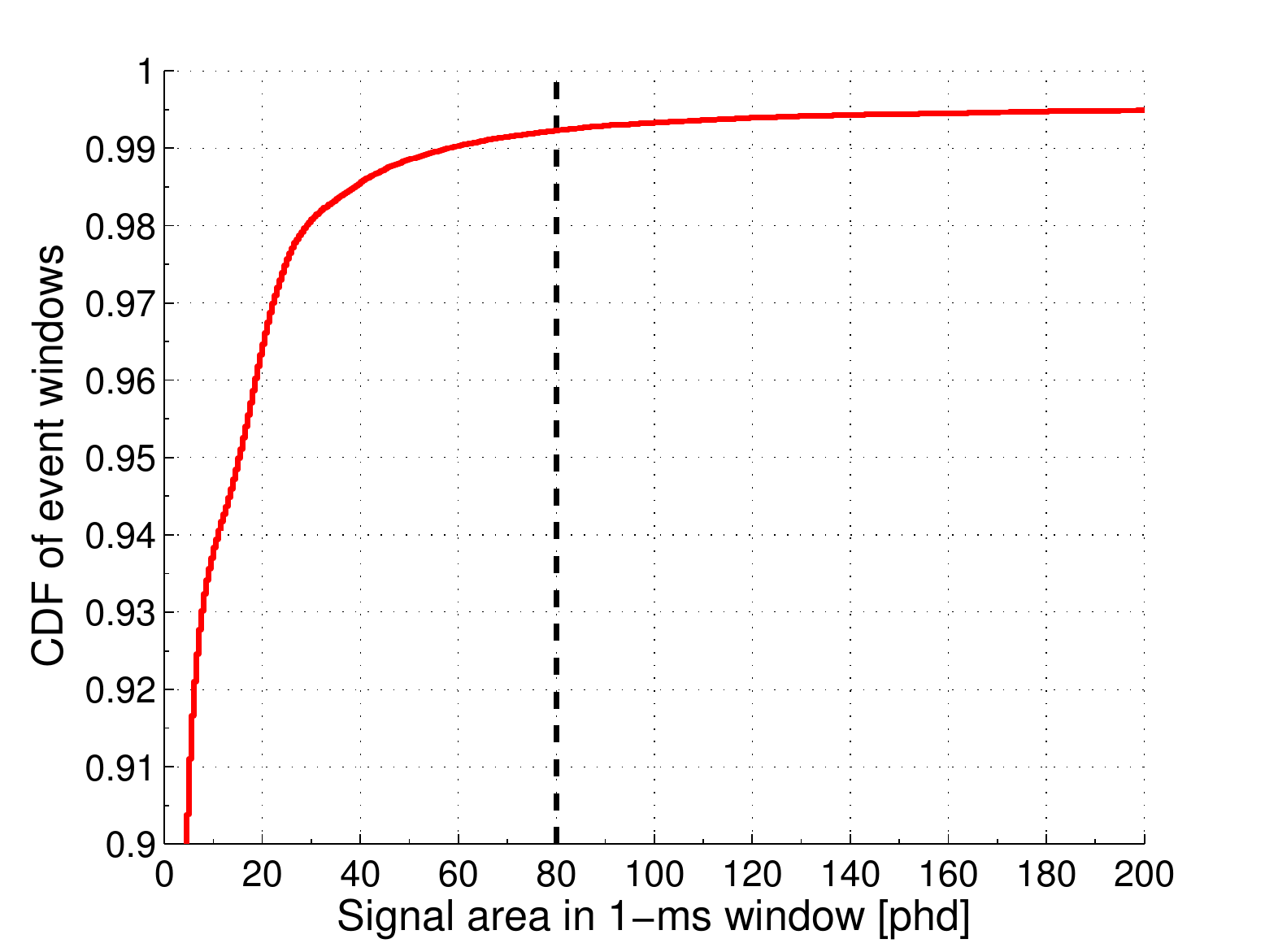}
\hspace*{-0.8cm}
\includegraphics{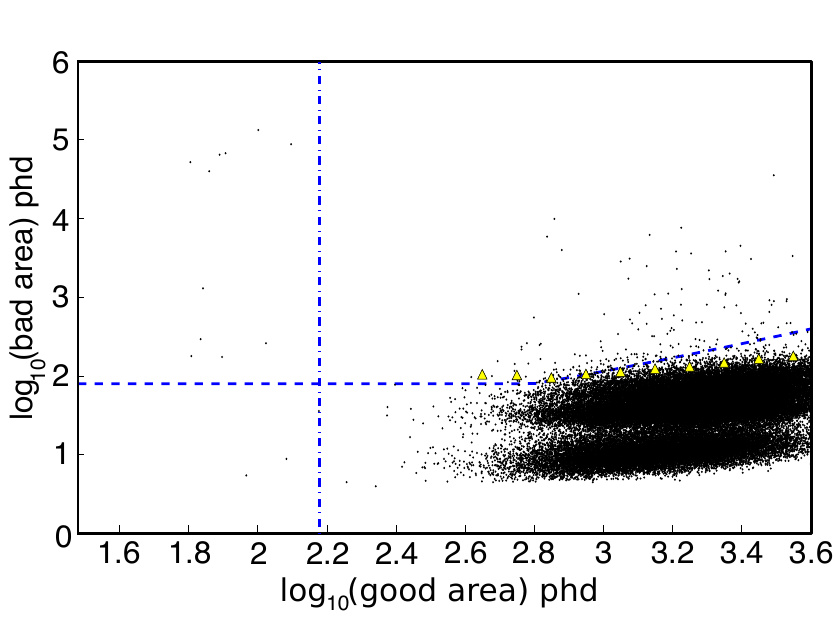}
\caption{Top: Cumulative number of 1~ms time window slices from a randomly picked WIMP search dataset as a function of the event area contained within, showing 99\% acceptance at 80~phd. Bottom: Cleanliness cuts (dashed lines) applied to a tritium calibration dataset, confirming an acceptance of $\gtrsim$99\% for good areas above 400~phd. Yellow triangles indicate the 99\% percentile in each good area bin; the vertical dot-dashed line shows the 165~phd $S2$ threshold.}
\label{fig:badareaacceptance}
\end{center}
\end{figure}

Figure~\ref{fig:badarea} shows golden events from the WIMP search data plotted as a function of the total raw signal area ($S1$$+$$S2$ areas) and the extra (non-signal) area contained within the same event. The dashed line indicates the threshold for the maximum allowed non-signal area in an event, starting at 80~phd (which corresponds to $\sim$3 extracted electrons) per 1~ms event window for small energy deposits and slowly increasing from this value for events with a total signal area above 630~phd. The associated acceptance was determined by slicing randomly chosen background datasets in 1-ms time windows and evaluating the total area contained within. The top plot in Fig.~\ref{fig:badareaacceptance} shows the distribution of cumulative areas inside these 1~ms intervals in one of the tested datasets, with 99\% of the windows containing less area than the 80~phd value used in the cut. This value was confirmed using tritium calibration data for signal areas above 400~phd (see bottom plot in Fig.~\ref{fig:badareaacceptance}), with the cut excluding less than 1\% of the events in all signal area bins.

Table~\ref{table:event-level_cuts} summarizes the effect of each of the analysis cuts and thresholds applied to the WIMP search data in the number of events acquired during Run~3.

\begin{table}
\setlength{\extrarowheight}{3pt}
\caption{Effect of each of the event level cuts on the total number of acquired WIMP search triggers. Cuts have been applied sequentially.}
\begin{center}
\begin{tabularx}{\linewidth}{Xr}
\hline\hline
Cut  &  Events remaining after cut\\
\hline
All triggers & 93\,126\,362  \\
Single scatters & 5\,375\,716\\
Stability cuts &  5\,221\,580 \\
Fiducial volume & 1\,923\,367\\
$S1$ range (1 -- 50~phd) & 6\,898\\
$S2$ threshold ($>$165~phd) & 840\\
Cleanliness cut & 591\\
\hline\hline
\end{tabularx}
\end{center}
\label{table:event-level_cuts}
\end{table}

\subsection{\label{sec:eff}Thresholds and efficiencies}

\subsubsection{\label{sec:S1threshold}$S1$ threshold}
The $S1$ threshold is applied on the fully $xyz$-position corrected number of spikes, whose minimum is restricted by the 2-fold coincidence requirement in the $S1$ definition implemented in the pulse classification module (see Sec.~\ref{sec:PulseFinderClassifier}). As a result of position corrections, $S1$s smaller than 2 detected spikes are possible despite the two-fold coincidence, but in the original analysis~\cite{Akerib:2015:run3} an additional 2~phe threshold was imposed. Closer investigation of the region below this threshold did not reveal any additional noise or unexpected populations, and it was therefore decided to lower the $S1$ threshold to 1~phd for this analysis. Doing so increases the efficiency for finding a valid $S1$ by 15\% at 3~keVnr for events with a valid $S2$.

The fully integrated background model (described in Sec.~\ref{sec:background}), including contributions from the cosmogenically activated $^{127}$Xe that limited the upper bound of the $S1$ scale to 30~phe in the original analysis, allows the extension of the upper $S1$ limit to 50~phd. As described in Sec.~\ref{sec:NR_BG}, $S1$s above 50~phd (outside the region of interest for the WIMP search) were used to estimate the leakage of wall events into the fiducial volume. 

\subsubsection{\label{sec:S2threshold}$S2$ threshold}
The threshold on the $S2$ signal is applied to the uncorrected raw pulse areas, and is driven by the definition of the fiducial volume and the power of the cleanliness cut to exclude events during periods of high rate single electron background following large $S2$ signals. 

Over a broad range of energies, events occurring close to the detector walls were found to occasionally be reconstructed to lie inside the fiducial volume due to incomplete charge collection, and to therefore be mistakenly identified as WIMP-like signals. Since, for small sized $S2$ signals, these events behave in a similar manner for different $S1$ regions, their expected leakage into a given fiducial volume in the $S1$ WIMP search range of interest ($S1<$50~phd) may be estimated for different $S2$ thresholds using events from the WIMP search data with an $S1$ greater than 50~phd: see Sec.~\ref{sec:NR_BG} for more details on this study. 

Ultimately, the $S2$ threshold was set at 165~phd, a compromise between the capability of the cleanliness cut to remove events in periods of high rate single electron background (in which false $S2$s may be paired with coincident sphes or $S1$s from no-charge-collection regions of the detector) and the low efficiency for events with smaller $S2$s to have a valid $S1$.

\subsubsection{\label{sec:Efficiencies}Efficiencies}

The efficiency to detect single scatters, which is used in the final PLR analysis, is shown by the black line in the lower panel of Fig.~1 in \cite{Akerib:2015:run3}. It was estimated using simulated NR event waveforms, generated with LUXSim using the Lindhard parametrization to the light and charge yields obtained from the \emph{in situ} nuclear recoil calibrations \cite{Akerib:2015:dd}. These simulated events have been analyzed with the full data processing framework and using the same cuts and thresholds that have been applied to the WIMP search data, and show an efficiency for detecting golden events of 0.3\% at the signal cut-off energy of 1.1~keV, rising to 50\% at 3.3~keV and reaching 100\% at around 6~keV. The drop in efficiency above 30~keV is determined by the $S1$ upper limit of 50~phd.

\begin{figure}
\begin{center}
\includegraphics[width=0.48\textwidth,clip]{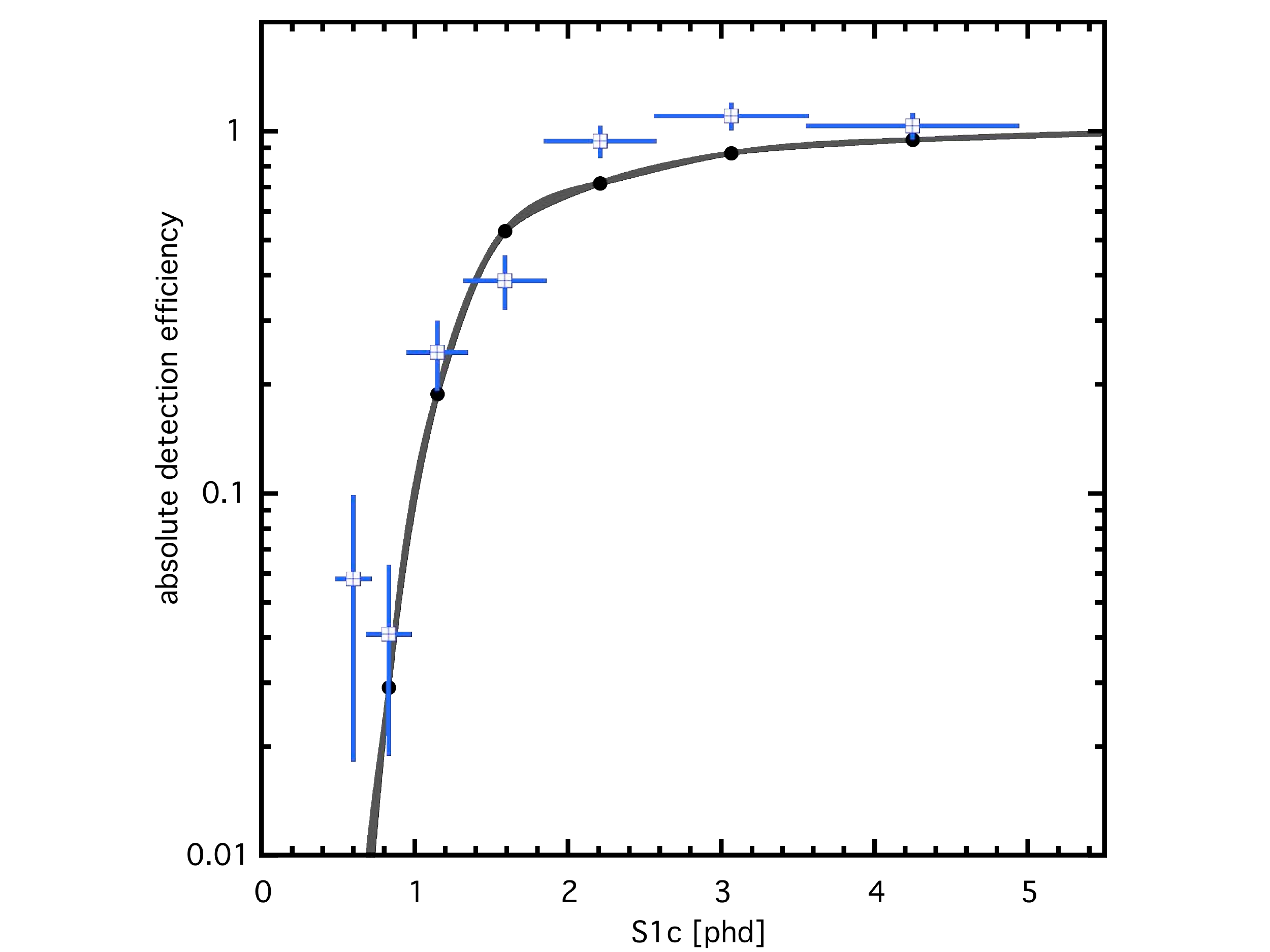}
\caption{Comparison of the single scatter efficiency curve obtained from the simulation of NR events (black dots, the line is added just to guide the eye) with the efficiency obtained from analysis of \dd{} data (blue empty squares), in the relevant region of small $S1$ pulses. For larger signals ($S1>$5~phd) the efficiency stays flat at 100\% up to the $S1$ 50~phd upper limit.}
\label{fig:efficiency_in_s1}
\end{center}
\end{figure}

The efficiency estimated from this simulated NR dataset was compared with the efficiency obtained from \dd{} calibration data (see Fig.~\ref{fig:efficiency_in_s1}), showing a good agreement in the relevant region of small $S1$ pulses. To estimate the efficiency from nuclear recoils generated by \dd{} neutrons, the $S1$ spectrum obtained from the analysis of the data, which used similar cuts and thresholds to the ones used for the WIMP search data except for a tighter fiducial cut around the beam line projection inside the detector, was compared with that obtained from a simulation of the \dd{} neutron beam using LUXSim for the same number of generated neutrons. This simulation was then used to generate the energy deposits inside the xenon volume created by the nuclear recoils resulting from neutron elastic scatterings, with the $S1$ and $S2$ spectra being estimated using the same NEST version used for the \dd{} analysis and the PLR analysis. 

To ensure that all pulse topologies occurring in the data, as well as potential losses due to the restriction to 10 recorded pulses in the DPF, are covered in the efficiency estimation, an absolute DPF efficiency has been determined from 4000 NR calibration hand-scanned events, obtained with an AmBe source (see Sec.~\ref{sec:EventClassification}). A list of golden single scatter events that were identified by manual scanning was compared to the output of the data processing framework. Applying all analysis cuts, it was found that 97.5$\pm$1.7\% of all golden events had been successfully captured and identified by the LUX DPF. This absolute efficiency is applied as a scaling throughout the entire energy range.

\subsection{\label{sec:background}Backgrounds}
\subsubsection{\label{sec:coincidence} Random coincidence background}
Apparent golden events can be generated by the random coincidence of an $S1$-only event and an $S2$-only event. The two unrelated pulses must be separated in time by less than the maximum physical drift time, with the $S1$ preceding the $S2$. $S1$-only events, {\it i.e.}, a minimum of two PMTs recording a photoelectron within 100~ns of one another, without an accompanying $S2$, may be caused by energy deposition in the sub-cathode dead region, random coincidence of dark-count photoelectrons, or Cherenkov light (emitted, for example, in PTFE components or the window of a PMT). $S2$-only events are caused by low-energy tracks in an $S2$-live volume, {\it i.e.}, the drift or extraction regions, which generate fewer than two detected $S1$ photons. The $z$-position inferred from apparent drift time in these events is a uniform random variable, so they constitute a background even in large xenon TPCs where conventional single scatters from peripheral radioactivity are excluded by the fiducial volume.

To contribute to the coincidence background, an $S2$ must have raw area above the 165~phd analysis threshold, be reconstructed within the fiducial radius, and occur in an event window which passes the cleanliness cut and contains no preceding $S1$. The spectrum of $S2$s meeting these criteria was obtained from a representative sample of WIMP search datasets. The isolated $S2$ rate from threshold to 4500~phd raw area --covering the entire NR signal region-- is 5~mHz. Beta particles and alpha-decay daughter nuclei originating on the wires of the gate electrode are a common cause of $S2$-only events due to the high field and the obscuration of $S1$ light: the characteristic short duration of gate $S2$s, for which the electron cloud diffuses very little in $z$ on the way to surface, could be used to reduce this background at some penalty in efficiency, though this was not required for the LUX $S1$-$S2$ WIMP search.

\begin{figure}
\begin{center}
\includegraphics[width=0.48\textwidth,clip]{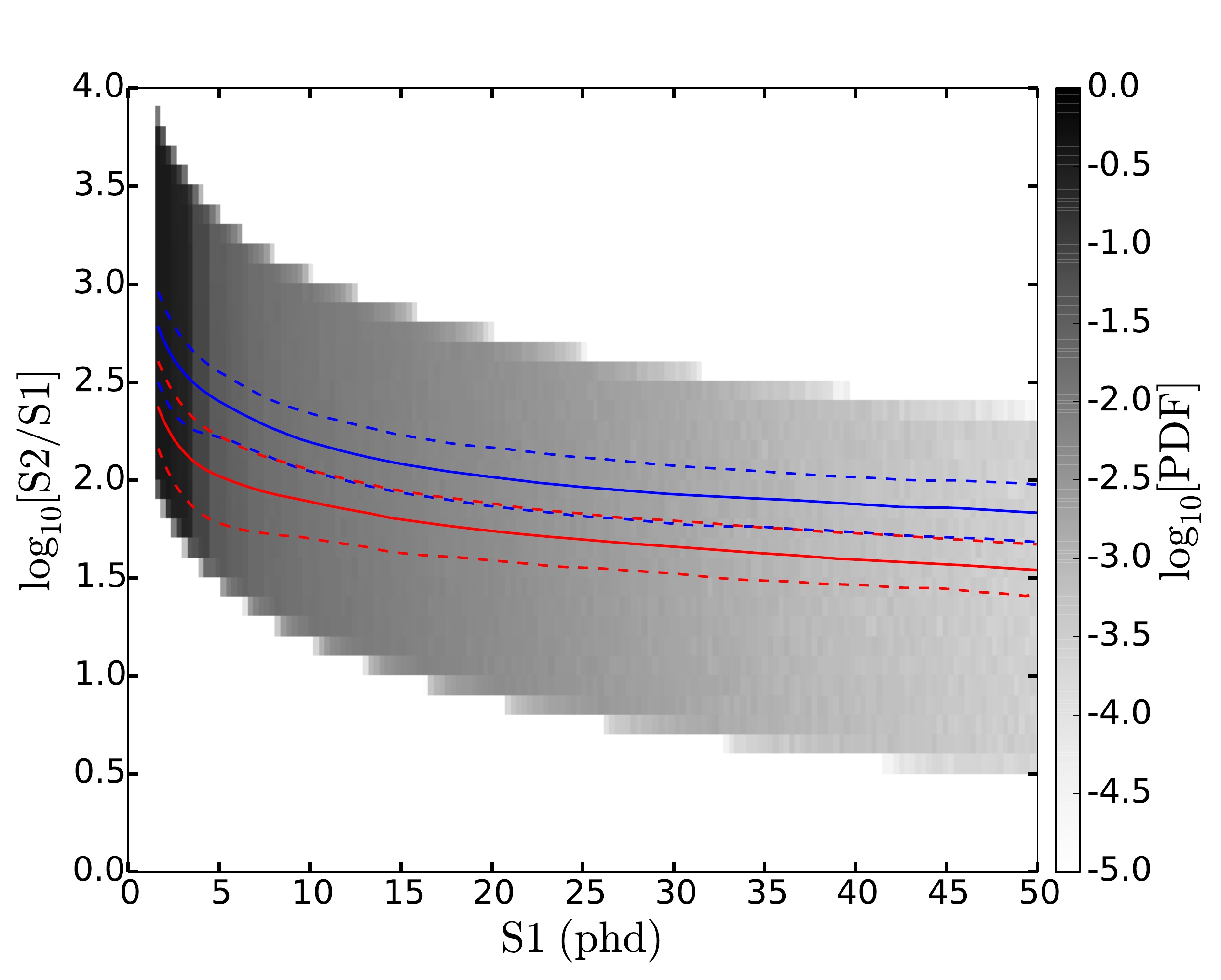}
\caption{Distribution in $\text{log}_{10}(\text{S}2/\text{S}1)$ vs. $S1$ of events caused by random coincidence of $S1$-only and $S2$-only events. The blue and red lines have the same meaning as in Fig.~\ref{fig_Dis01}.}
\label{fig:coincidence_pdf}
\end{center}
\end{figure}

The trigger efficiency for $S1$s rolls off to zero in the search range (below 50~phd) but, because all samples passing zero-suppression are written to disk by the DAQ, the rate of isolated $S1$s can still be measured by applying the pulse-finder algorithm to representative WIMP search data, irrespective of the trigger. The isolated $S1$ rate from threshold to 50~phd raw area is 1~Hz, with a falling spectrum. Figure \ref{fig:coincidence_pdf} shows the resulting distribution of background events due to coincidence $S1$-only and $S2$-only events, which leads to a prediction of 1.1 background events in the 95.0-day Run~3 reanalysis sample.

\subsubsection{\label{sec:NR_BG} Nuclear-recoil and surface backgrounds}

Nuclear recoils contributing to the WIMP background in LUX can be caused by elastic scattering of fast neutrons or by surface alpha decays in which the daughter nucleus recoils into the liquid xenon target. Neutron background rates are estimated in detail in~\cite{Akerib:2014:bg} and contribute a negligible expectation of 0.08 events in the Run~3 reanalysis.

\begin{figure}
\begin{center}
\includegraphics[width=0.45\textwidth,clip]{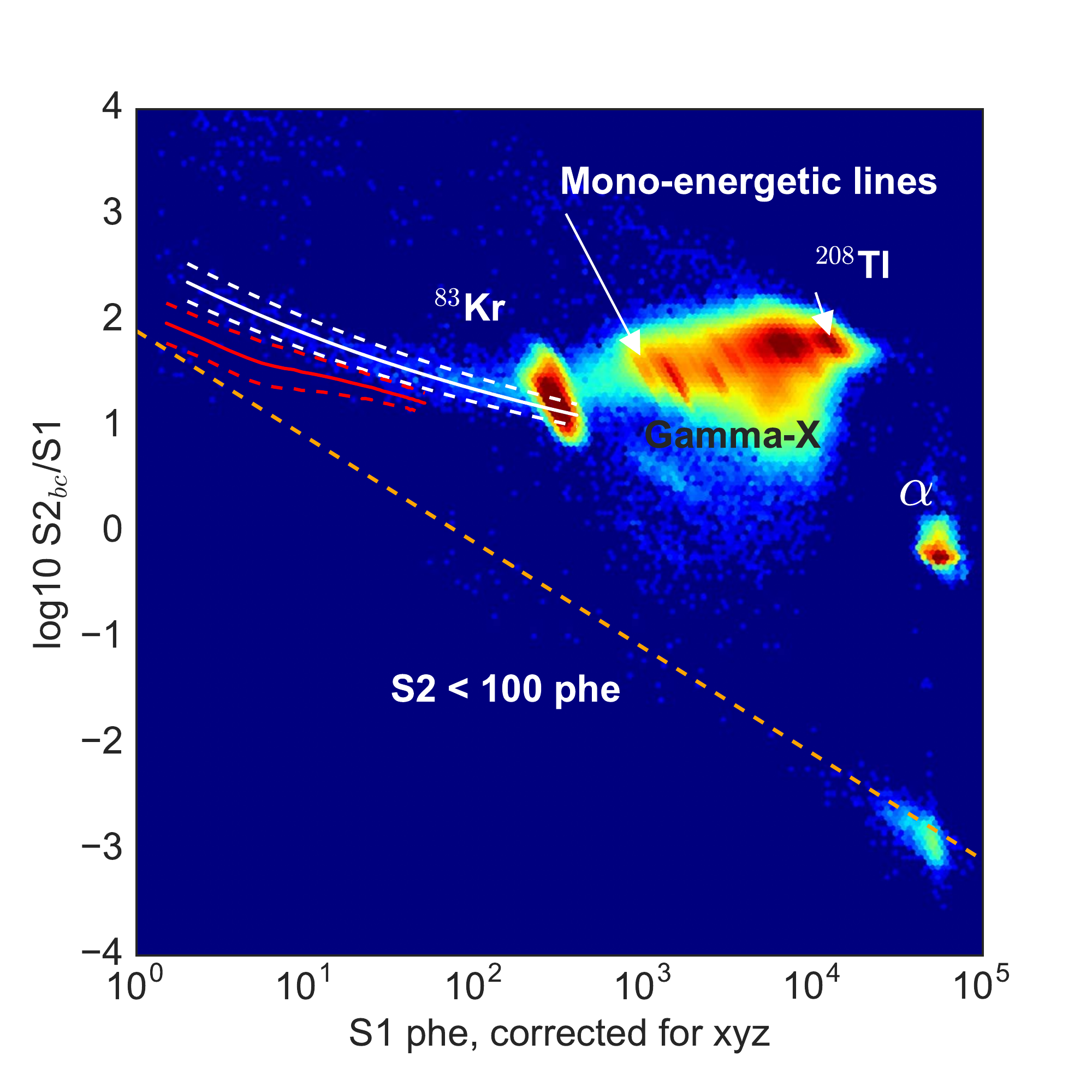}
\includegraphics[width=0.45\textwidth,clip]{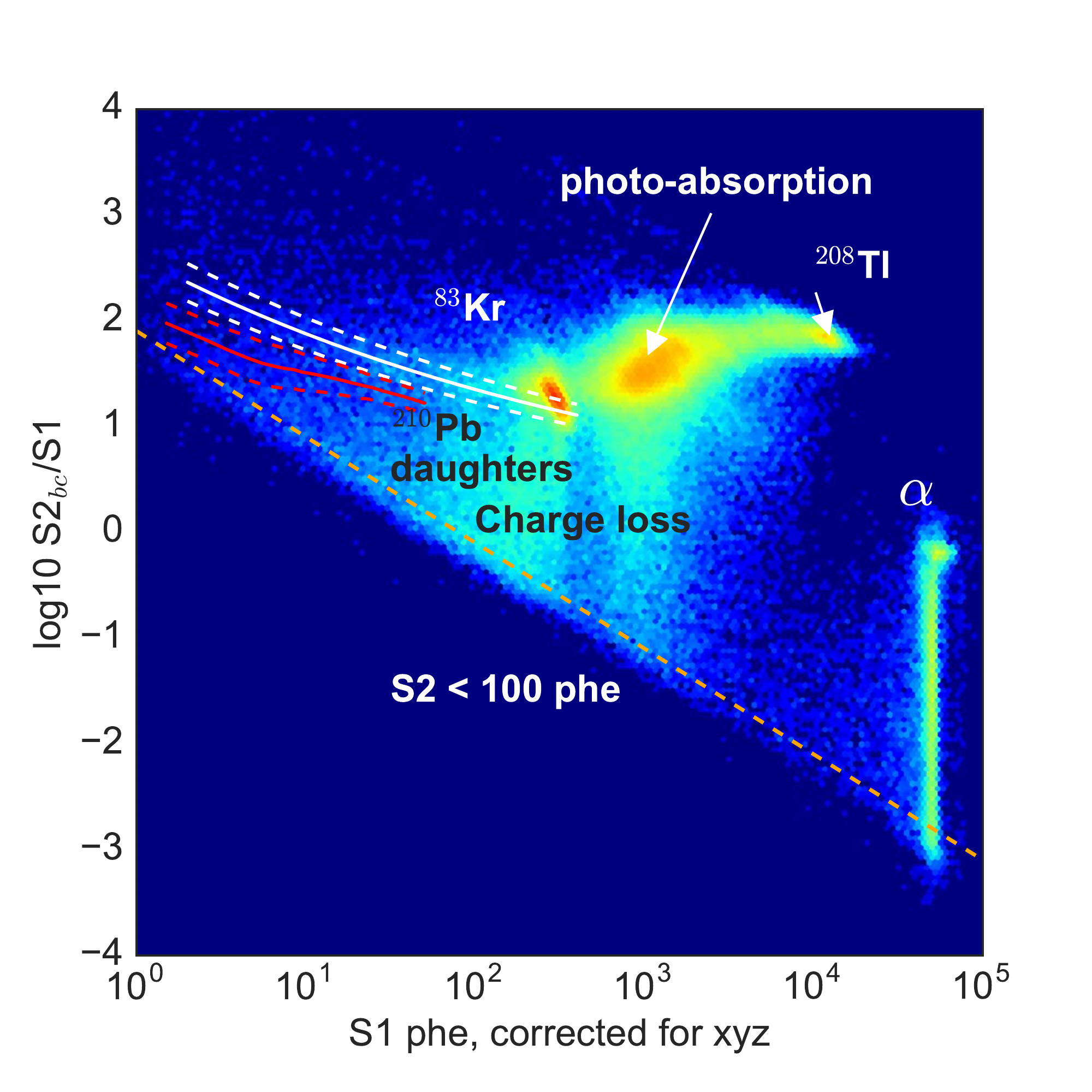}
\caption{Distribution in $S1$ vs. log10($S2_{b}/S1$) of background events from 86.3 live-days. Top: Events within the liquid xenon bulk defined by a radius of less than 15~cm and a drift time between 10 to 320~$\mu$s. Bottom: events near the PTFE wall, defined by the distribution of $^{210}$Pb daughters. More details on the wall reconstruction can be found in~\cite{Lee:2015}. The spread of events below the main peaks illustrates the incomplete charge collection for events near the wall.}
\label{fig:wall_s1_s2}
\end{center}
\end{figure}

\begin{figure}
\begin{center}
\includegraphics[width=0.48\textwidth,clip]{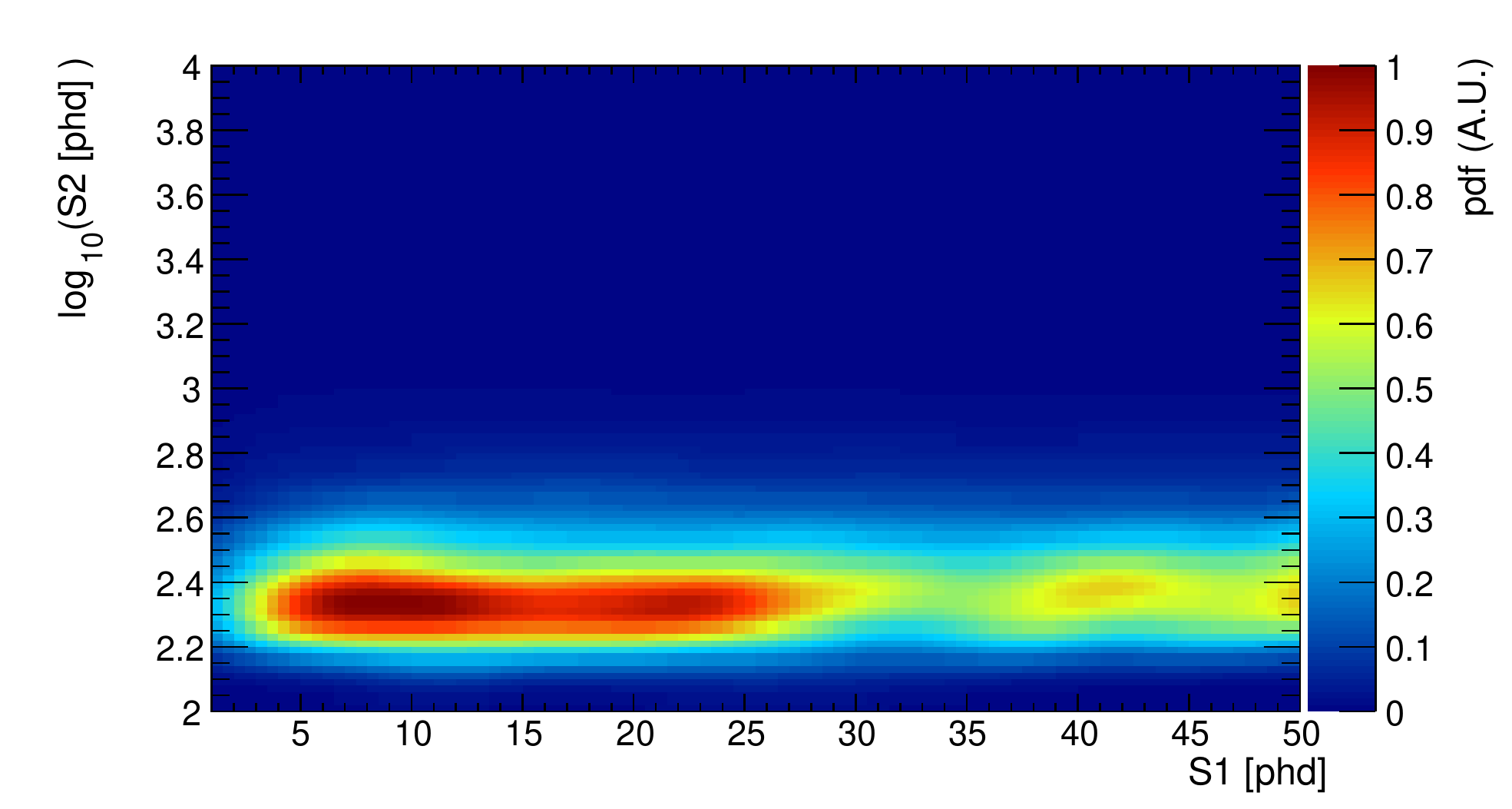}
\caption{Kernel-density estimated distribution, in $S1$ and $S2$, of wall background events reconstructed within the fiducial volume. The predicted count is 24 events within the 145-kg fiducial volume over 95.0 live-days.}
\label{fig:wall_pdf}
\end{center}
\end{figure}

Wall events constitute a background to the WIMP search only when reconstructed inside the fiducial radius. The suppressed $S2/S1$ ratio described above tags wall events without reference to their reconstructed radius, allowing one to model the resolution of the wall in $r$ empirically with a high-$S1$, low-$S2/S1$ sideband away from the WIMP signal region. The low-$r$ side of the resolved wall is well described by an $S2$-dependent Gaussian distribution with standard deviation 
\begin{equation}
\sigma[\mathrm{cm}]=\frac{1}{0.61\log_{10}\left(S_{2c}[\mathrm{phd}]\right)}.
\end{equation}
The rate and $S1$-$S2$ distribution of the wall background were derived from the data themselves, non-parametrically. X-rays, Compton scatters and $^{206}$Pb daughter nuclei recoiling against undetected alpha particles all contribute to the wall population, but their signal yields are not readily simulated. Instead, an empirical wall radius was calculated, for each bin of polar angle, from the median radius of the extreme low-$S2/S1$ population, and kernel density estimation was applied to all events beyond this radius to construct a wall probability density function (PDF) in $S1$-$S2$ space. Multiplying by the Gaussian conditional distribution for $r$, given $S2$, results in the complete model for wall events in the fiducial volume, shown in Fig.~\ref{fig:wall_pdf}, and a prediction of $24\pm7$ counts in the 95.0-liveday sample.

\subsubsection{\label{sec:ER_BG} Electron-recoil backgrounds}
Single-site electron recoils within the LUX fiducial volume are caused by $\upbeta$ and electron-capture decays within the liquid target and by $\upgamma$ decays in the surrounding material. To account for systematically uncertain rates, the background model is subdivided into 6 populations (Table I of \cite{Akerib:2015:run3}), which are described here in descending order of low-energy event rate.

Uranium-chain, thorium-chain, $^{40}$K, and $^{60}$Co nuclei in the detector construction materials emit high-energy $\upgamma$-rays that may, with low probability, Compton scatter once in the fiducial volume and then leave the active region. A fit of isotope activities to data in the range 0.5--3~MeV electron-equivalent energy ({\it i.e.}, well above the dark matter region of interest) gives generally good agreement \cite{Akerib:2014:bg}. Two anomalies in the fit are still under investigation, though neither has significant implications for WIMP searches: the spectrum around the 969~keV line is higher than other Th-chain activity implies, and there is an elevated continuum background in the lowest few~cm of the active region. To account for the latter, the PDF for Compton scattering events was split into two source groups, with rates allowed to vary independently: one consisting of decays in the bottom PMT arrays, the surrounding support structure, and the bottom $\upgamma$-ray shield; and one consisting of decays elsewhere, principally in the top-array and the sidewalls.

A spatially uniform population of beta-decays, unaccompanied by other self-vetoing energy depositions in the active region, arises from two isotopes that are present in the liquid xenon target: $^{85}$Kr (which is present in the atmosphere primarily due to nuclear fuel reprocessing) and $^{214}$Pb (a radon daughter). A process of gas-phase chromatography was developed to reduce the former background \cite{Akerib:2016:kryptonRemoval}: the concentration by mass (g/g) of natural Kr in the LXe was reduced from \num{130e-9} to \num{3.5e-12}. Together with independent measurements of the ratio $^{85}$Kr/$^{\mathrm{nat}}$Kr, this concentration implies (0.13$\pm$0.07)\num{e-3} events/keV/kg/day in the region of interest with a flat energy spectrum. The rate of $^{214}$Pb decays also has a spectrum that is effectively flat in the region of interest and was constrained to \num{0.11e-3} -- \num{0.22e-3} at 90\%~CL. This range is constrained from below by the rate of alpha decays from daughter $^{214}$Po nuclei, and from above by the total measured event rate at 300~--~350~$\rm keV_{ee}$~\cite{Akerib:2014:bg}.

Electron-capture decays by $^{127}$Xe and $^{37}$Ar dispersed in the liquid contribute to the background of golden ER events. $^{127}$Xe is a cosmogenic isotope, produced by neutron capture on $^{126}$Xe. Roughly 800,000 $^{127}$Xe nuclei decayed in the LUX active volume during data-taking in 2013, predominantly early in the run (having a half-life of 36~days). Subsequent atomic X-rays lose their energy to electrons in the bulk xenon, with the total energy determined by the shell from which the initial capture took place. Of the resulting energy lines, those at 1.1 and 5.2~keV fall in the region of interest for WIMP searches. The daughter $^{127}$I nucleus emits a $\upgamma$-ray at 203~keV or 375~keV with 43\% and 57\% probability, respectively. The X- and $\upgamma$-ray emission are practically simultaneous, given the sub-nanosecond transitions and the tens-of-nanosecond scale of detector response and DAQ sampling. Only when the $\upgamma$-ray escapes undetected is the X-ray energy measured as a golden event, and thus a potential background. Simulation of this $\upgamma$-ray inefficiency was used to predict the rate and spatial distribution of golden $^{127}$Xe X-ray events, given the initial concentration of \mbox{ $490\pm95\;\upmu\mathrm{ Bq\,kg}^{-1}$ } measured via the $\upgamma$-ray lines.

There is some evidence for an additional line in the low-energy spectrum, consistent with a source that is uniform in position. This is thought to be due to $^{37}$Ar. The fit to search data is $12\pm8$ such events in $1.4\times10^{4}\;\mathrm{kg\,days}$ of exposure.

\begin{figure}
\begin{center}
\includegraphics[width=0.48\textwidth,clip]{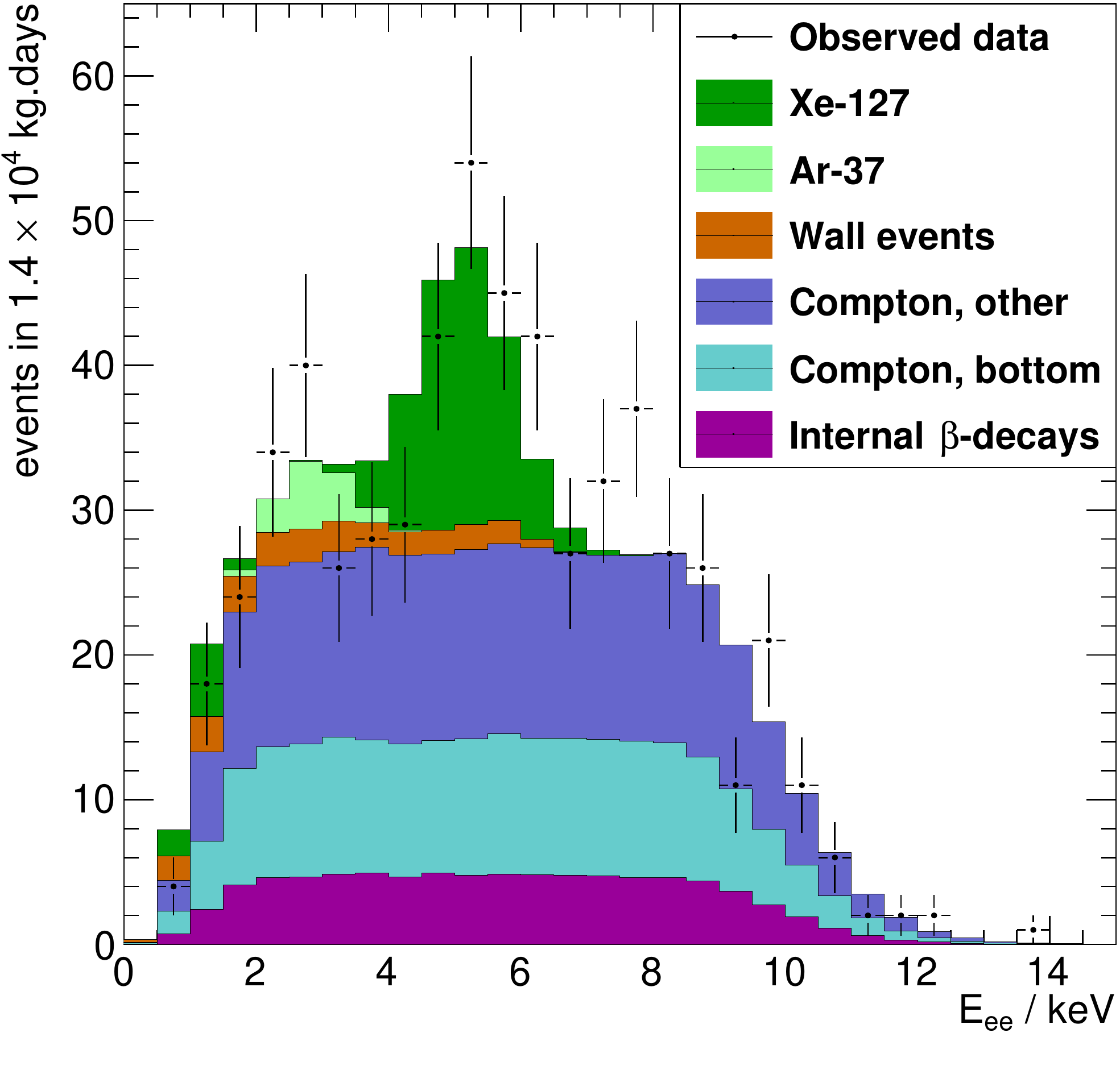}
\caption{Electron-equivalent-energy histogram of the LUX Run~3 search events with $S1<$50~phd (points) together with the best-fit model (stacked histogram), which has zero signal. The $x$-axis is the linear, combined light-and-charge energy estimator, \mbox{$E_{\mathrm{ee}}=W\cdot \left(S1/g_{1}+S2/g_{2} \right)$}, with $W$=13.7~eV. The roll-off in efficiency either side of 2~--~9~keV is primarily due to cuts on $S1$: search events are required to have at least 2-PMTs in coincidence and a maximum size of 50~phd.}
\label{fig:Eee_noSignal}
\end{center}
\end{figure}

For all electron-recoil backgrounds, the NEST model (Sec.~\ref{sec:NEST_ER_model}) and LUXSim simulation of the detector and its response were used to generate MC waveforms. The same event-reconstruction software and selection cuts used for search data were applied to MC data, yielding events sampled from a simulation model that included efficiency and resolution effects. Smoothed histograms of the MC events were used to estimate the background PDFs in $S1$-$S2$ and also, for the non-spatially-uniform populations of $^{127}$Xe and Compton-scatter events, those in $r$-$z$. Figure~\ref{fig:Eee_noSignal} shows the background model and LUX Run~3 data as a function of recoil energy.  

\subsection{\label{sec:signal}Signal model}

\begin{figure}
\begin{center}
\includegraphics[width=0.46\textwidth,clip]{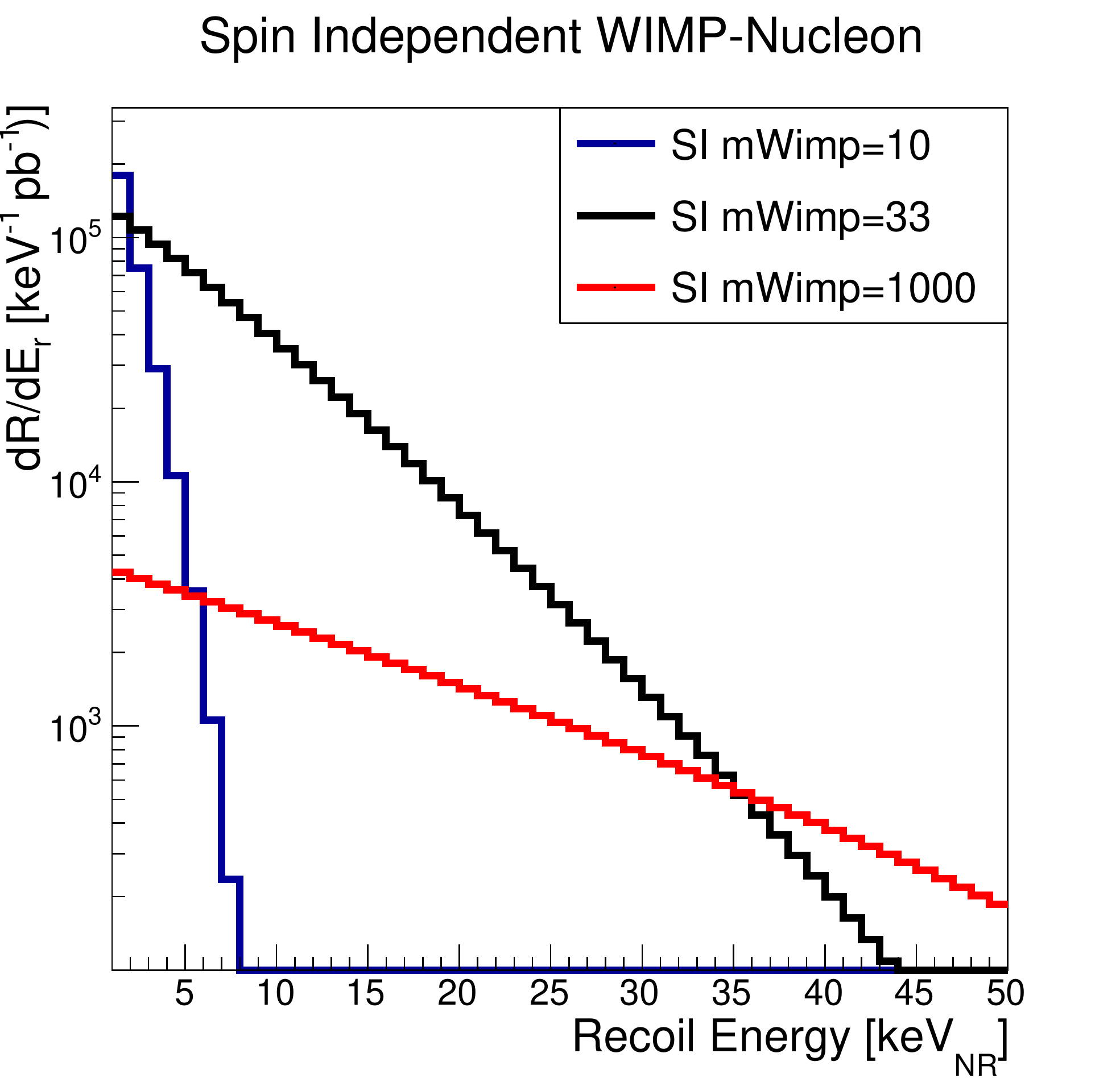}
\includegraphics[width=0.46\textwidth,clip]{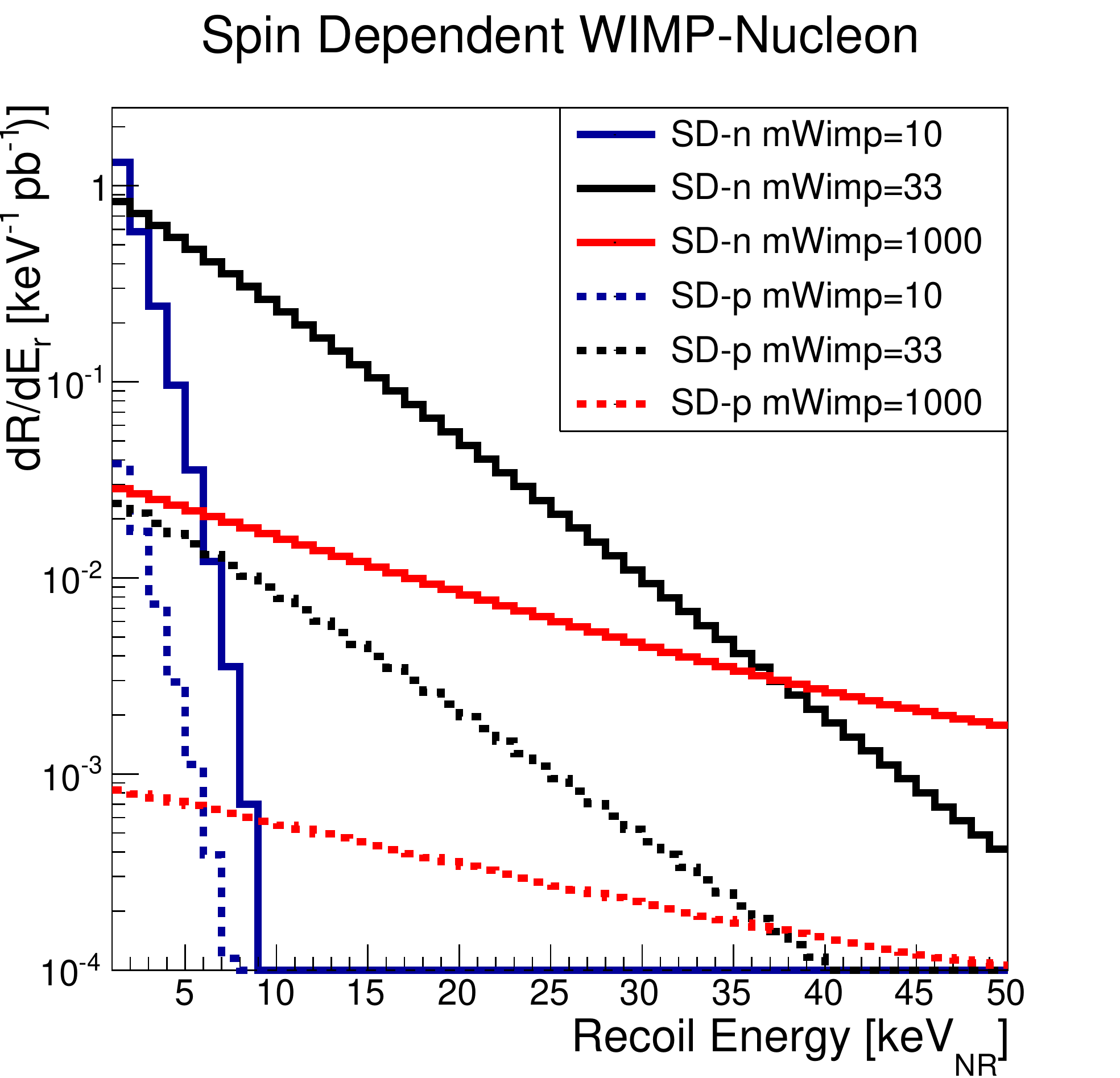}
  \caption{Top: SI WIMP-nucleon interaction differential rate functions for WIMP mass = 10~GeV (blue), 33~GeV (black) and 1~TeV (red). Bottom: SD WIMP-neutron (solid) and WIMP-proton (dashed) interaction rate for WIMP mass = 10~GeV (blue), 33~GeV (black) and 1~TeV (red).}
  \label{fig:WIMPSpectrum}
  \end{center}
\end{figure}

Both spin-independent (SI) and spin-dependent (SD) WIMP-nucleon scattering models are considered in this paper. The WIMP velocity profile is derived from a standard Maxwellian velocity distribution with $v_{0}~=~220$~km/s and $v_{esc}~=~544$~km/s as presented in~\cite{McCabe1005.0579}. The WIMP mass density, $\rho_{0}$, is taken to be $0.3$~GeV/cm$^{3}$. For the SI case, the scattering rate is computed under the assumption that the xenon nucleons interact coherently in amplitude, thus enhancing the overall differential rate in recoil energy. In the SD case, the scalar WIMP-nucleon interaction is assumed to be suppressed and only the axial coupling contributes to the interaction. The coherent enhancement is not present in the SD case. Typically, the axial-vector coupling constants are chosen so that the WIMP couples only to protons, or only to neutrons. Xenon has an even number of protons and two naturally occurring isotopes (53\% abundance) with an odd number of neutrons. Therefore, in xenon the SD proton-only interaction is suppressed relative to neutron-only. However, in the proton-only case, scattering may also occur through two-body currents (three-body interactions) where both the protons and the unpaired neutron contribute, regaining some proton-only sensitivity. The SD model is detailed in \cite{Akerib:2016:SD} using calculations from~\cite{Klos}. In both the SI and the SD models, differential interaction rates are calculated as the functions of the recoil energies, as shown in Fig.~\ref{fig:WIMPSpectrum} for some examples of WIMP masses.

Using the differential rate function for each interaction model and WIMP mass, $S1$ and $S2$ observables are generated using NEST (Sec.~\ref{sec:NEST}). NEST was calibrated using \emph{in situ} nuclear recoil data that allowed simulation of the detector response to WIMP-nucleon interactions. Using the $g_1$ and $g_2$ obtained from the energy calibration in Sec.~\ref{sec:ER}, a global fit to both the charge and light yields was applied, using the Lindhard model \cite{Lindhard} for the energy dependence response in nuclear recoil interactions. These parameters are shown in Sec.~\ref{sec:DD} with their best fit values and uncertainties shown in Table~\ref{tab:fits}. The uncertainties in these fit parameters are treated as nuisance parameter in the probability likelihood ratio (PLR) method. For the signal models, the spatial observables, $r$ and $z$, are taken to be uniformly distributed by volume. The four signal observables are combined into a 4-dimensional PDF with a normalization derived from the signal efficiency (see Fig.~1 in~\cite{Akerib:2015:run3}) and the total rate expected from a particular WIMP mass and interaction coupling. The signal model is defined by:

\begin{equation}
M_{sig} = N_{0} \times \epsilon(\kappa_L) \times P(S1,\log_{10}(S2),r,z),
\label{EQ:SignalModel}
\end{equation}
where $N_{0}$ is the total number of events expected for a particular WIMP model, $\epsilon$ is the signal acceptance efficiency which depends on the nuisance parameter $\kappa_L$, and $P(S1,\log_{10}(S2),r,z)$. $\kappa_L$ is the Lindhard factor and $P(S1,\log_{10}(S2),r,z)$ is 4-dimensional PDF for the signal model.

\begin{figure}
\begin{center}
\includegraphics[width=0.46\textwidth,clip]{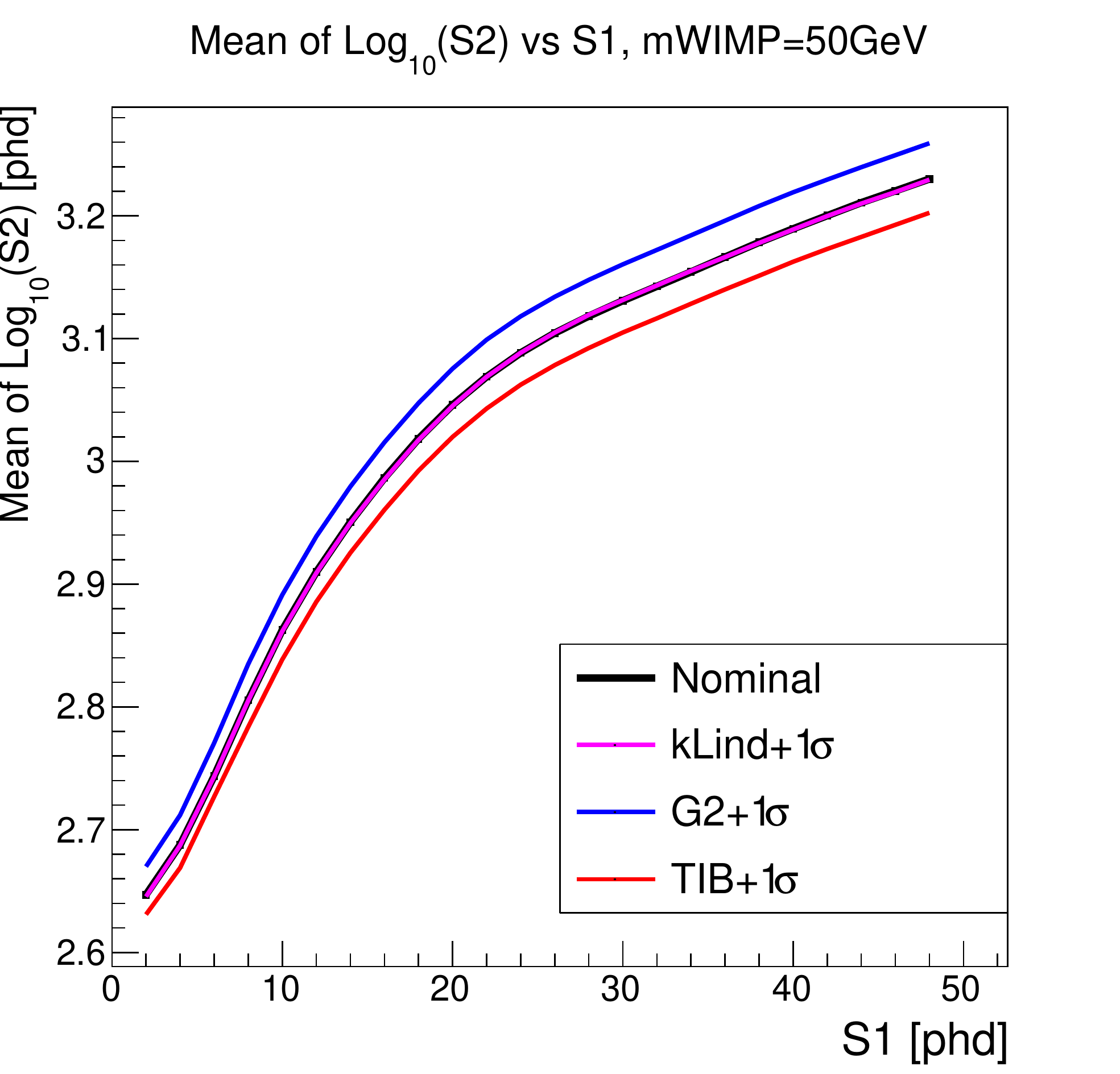}
\includegraphics[width=0.46\textwidth,clip]{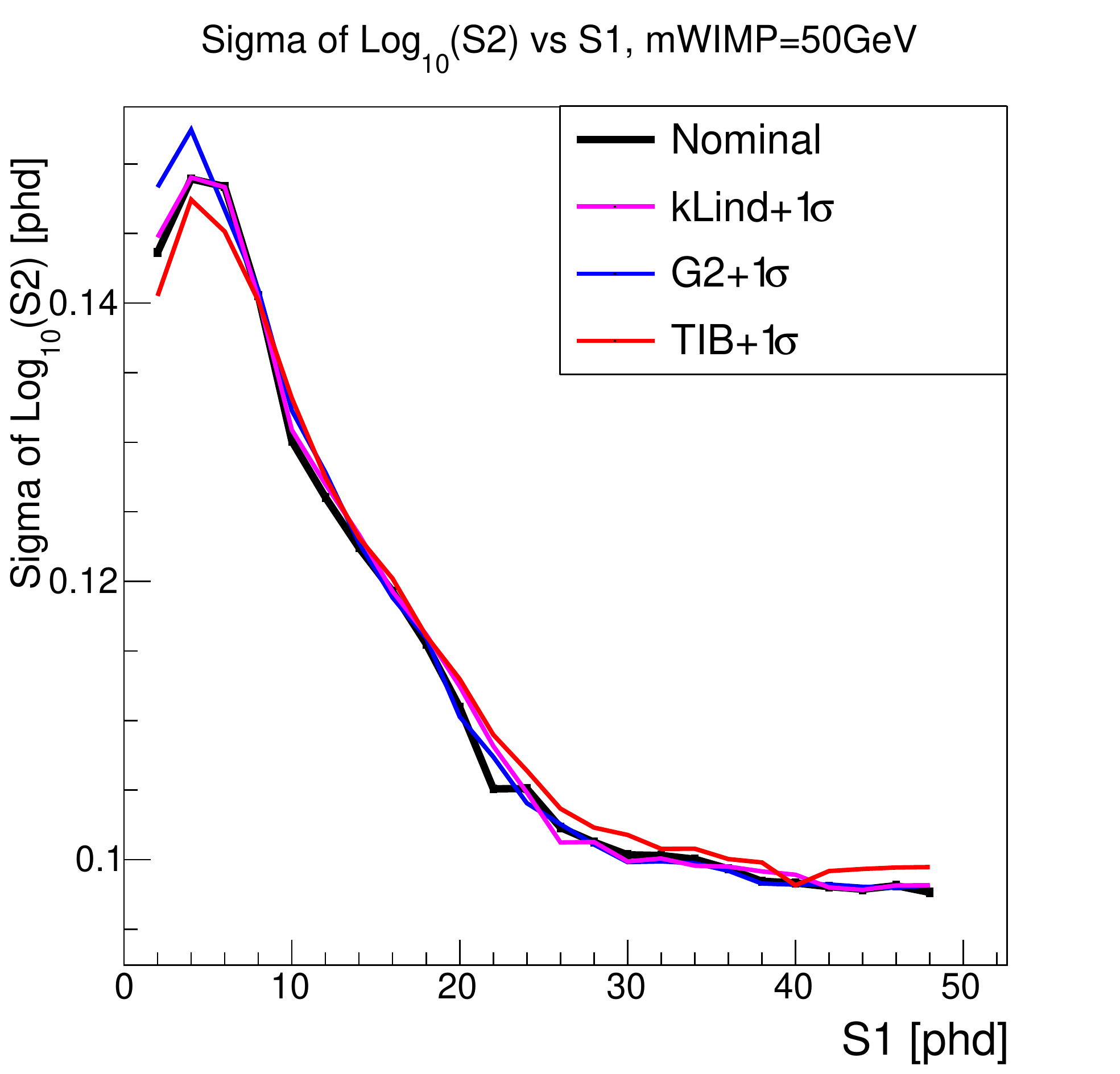}
\caption{Top: Mean of the $\log_{10}(S2)$ band as a function of $S1$. Bottom: Gaussian width of the $\log_{10}(S2)$ band as a function of $S1$. (Black) Nominal value of all nuisance parameters, (red) TIB varied by $+1\sigma$,  (blue) $g_2$ varied by $+1\sigma$,and (magenta) $\kappa_L$ varied by $+1\sigma$. $g_1$, $N_{ex}/N_i$, $\eta$,and $P$ variations are not plotted but they fall within the $g_2$ and TIB variations.}
\label{fig:NuisParVar}
\end{center}
\end{figure}

\begin{figure}[!htp]
\begin{center}
\includegraphics[width=0.5\textwidth,clip]{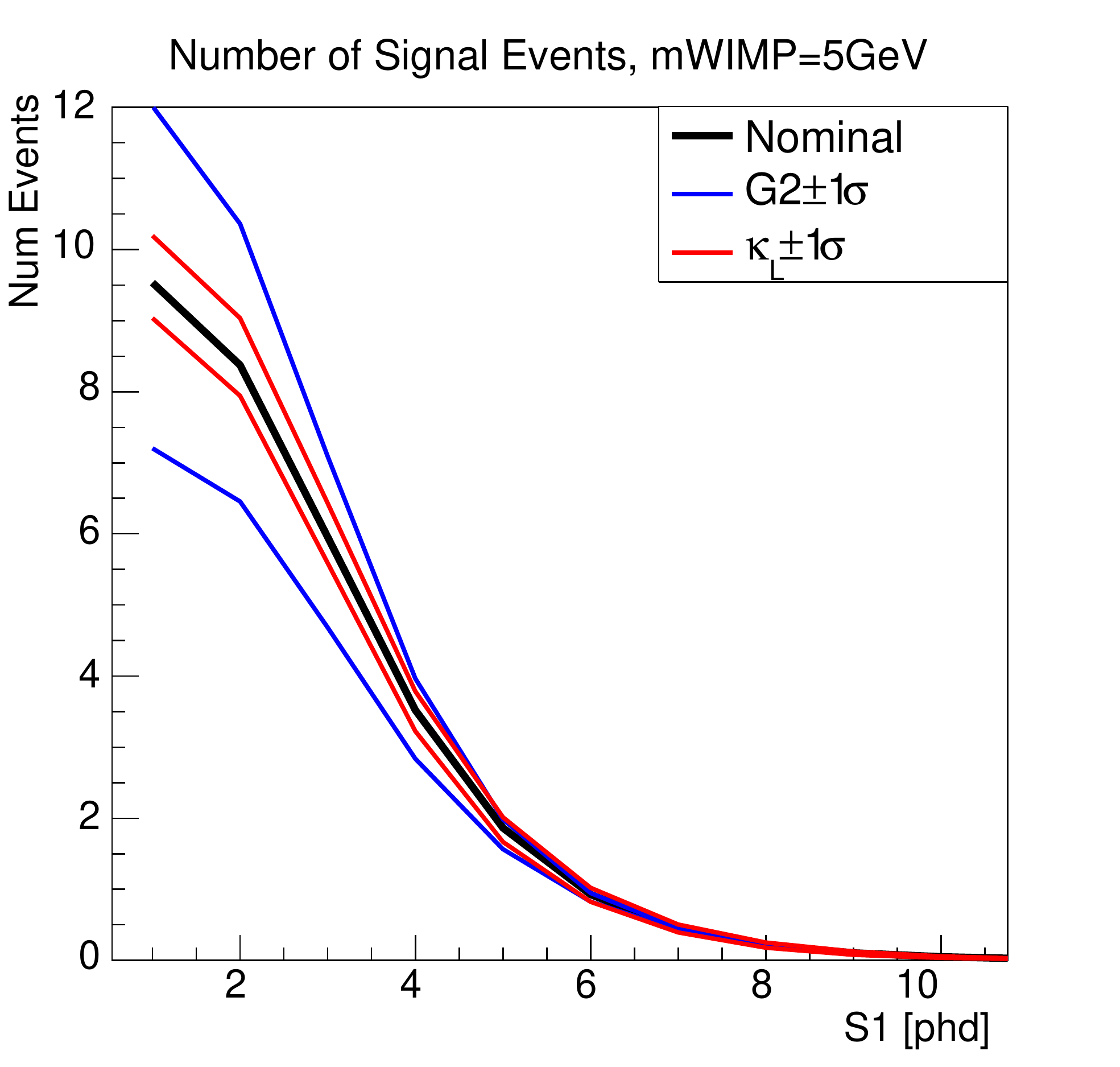}
\includegraphics[width=0.5\textwidth,clip]{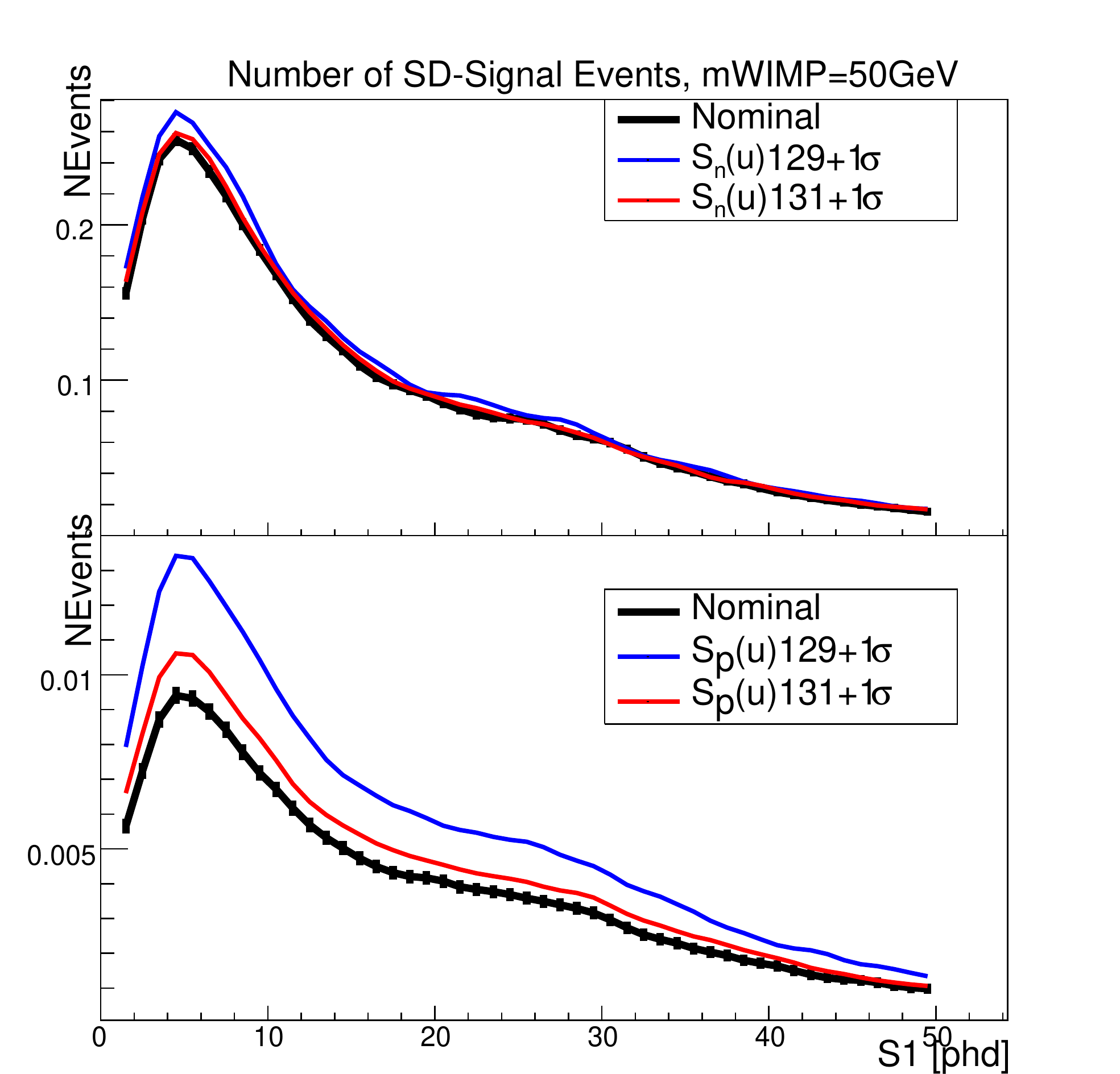}
\caption{Top: Number of SI WIMP-nucleon interaction events passing the analysis cuts scaled to the Run 3 exposure for a WIMP mass of 5~GeV. The black line corresponds to nominal values of all nuisance parameters, blue shows $g_2$ varied by $\pm1\sigma$, and red shows $\kappa_L$ varied by $\pm1\sigma$. Bottom: The number of SD WIMP-neutron (upper) and SD WIMP-proton (lower) interaction events passing the analysis cuts scaled to the Run 3 exposure for a WIMP mass of 50 GeV. In these plots, black corresponds to nominal values of all nuisance parameters, blue shows the structure factors $S_n(u)129$ and $S_p(u)129$, as defined in Section C of~\cite{Klos}, varied by $+1\sigma$, and red shows $S_n(u)131$ and $S_p(u)131$ varied by $+1\sigma$.}
\label{fig:NuisParVar2}
\end{center}
\end{figure}

To determine the effects of these nuisance parameters, the parameters were varied by $\pm~1\sigma$ from their best fit values. Figure~\ref{fig:NuisParVar} compares the mean (top) and width (bottom) of $\log_{10}$($S2$) as a function of $S1$ for a 50~GeV WIMP. Figure~\ref{fig:NuisParVar2} shows the change the number of events expected with our exposure for both spin independent and dependent WIMP models. In these plots, only the $g_2$, $\kappa_L$ and Thomas-Imel box model parameter (TIB) \cite{ThomasImel} variations are shown. These distributions are different for each WIMP mass so $\chi^{2}$, as defined by Eq.~\ref{EQ:Residual}, is computed for each mass and parameter variation, 

\begin{equation}
\chi^{2} = \sum_{S1=1}^{S1_{max}} \frac{(Y_{var}-Y_{nom})^2 N_{S1}}{Y_{nom}},
\label{EQ:Residual}
\end{equation}

\noindent where $Y_{nom}$ is the nominal value of the means or widths of the $\log_{10}(S2)$ band, $Y_{var}$ are the $\pm1\sigma$ variation, and $N_{S1}$ is the number of events in that particular $S1$ bin to account for the weights. 
The $\chi^2$ values are shown in Table~\ref{Table:NuisPars} for a few representative WIMP masses. The $\chi^{2}$ in the means and the widths of the bands are dominated by the variation in $g_2$ followed by the TIB parameter. The third set of values in Table~\ref{Table:NuisPars}, $\Delta(\epsilon)$, measures the change in the signal acceptance efficiency due to variation in the nuisance parameters. In the case of signal acceptance efficiency, $g_2$ and $\kappa_L$ are the dominating nuisance parameters. Since the signal model must be regenerated during the variation of the signal model profiling, each nuisance parameter variation increases the computing time of the limits calculation by a factor of 8. We computed the exclusion limits with a few representative points with additional nuisance parameters allowed to vary. The result only fluctuate about the original by about 5\%. Consequently, only $\kappa_L$ and $g_2$ are the two nuisance parameters allowed to vary during the profile likelihood ratio analysis, where $\kappa_L$ only affects the signal acceptance. For the SD analysis, the effects of 
structure factors were also explored. In particular, the factors $S_p(u)$ and $S_n(u)$ as defined in Section C of~\cite{Klos} for $^{129}$Xe and $^{131}$Xe were allowed to vary by $\pm1\sigma$ in their theoretical uncertainties. For the WIMP-neutron interactions, $S_n(u)$, the signal strength changed by less than $7\%$ for all masses. For the WIMP-proton interactions, $S_p(u)$, the signal strength changed by less than $40\%$ for all masses. For this result, $S_p(u)$ and $S_n(u)$ are fixed at their central values and do not vary in the PLR. 

\newcolumntype{Y}{>{\centering\arraybackslash}X}
\begin{table}
\setlength{\extrarowheight}{3pt}
\caption{Effects of Spin-Independent nuisance parameter variations for various WIMP masses. Each parameter is varied by their respective $\pm~1\sigma$ uncertainties, $\Delta_{par}$. Four representative WIMP masses and all 7 possible nuisance parameter variations  are tabulated here (g1, g2 and signal model parameters from Table \ref{tab:fits}).}
    \begin{tabularx} {\linewidth}{@{}lYYYYYYY@{}}
	\hline\hline
	Mass$\,\,\,$  & \multicolumn{1}{c}{$g_1$} & \multicolumn{1}{c}{$g_2$} & \multicolumn{1}{c}{$\kappa_L$} & \multicolumn{1}{c}{TIB} & \multicolumn{1}{c}{\footnotesize ${N_{ex}}/{N_i}$} & \multicolumn{1}{c}{\footnotesize $\eta$} & \multicolumn{1}{c}{\footnotesize $P$}\\
	\hline
    &\multicolumn{7}{c}{ $\Delta_{par}$  (\%)}\\
    \hline
	All & {$\pm$2} & {$\pm$7} & {$\pm$4} & {$\pm$12} & {$\pm$14} & {$\pm$18} & {$\pm$54}\\
	\hline
    &\multicolumn{7}{c}{ $\chi^2_{\mathrm{mean}}$}\\
	\hline
    5  & {\textless0.01}& {\textless0.01}& {\textless0.01}& {\textless0.01}& {\textless0.01}& {\textless0.01} & {\textless0.01}\\
    10 & 0.03 & 1.5 & 0.1 & 0.3 & 0.5 & 0.1 & 0.1 \\
    50 &  3.2 & 87 & 1.7 & 43 & 26 & 14 & 19 \\
    1000  & 0.3 & 9.4 & 0.1 & 6.7 & 1.8 & 2.0 & 2.9 \\
	\hline
    &\multicolumn{7}{c}{ $\chi^2_{\sigma}$}\\
    \hline
    5  & {\textless0.01}& {\textless0.01}& {\textless0.01}& {\textless0.01}& {\textless0.01}& {\textless0.01}& {\textless0.01} \\
    10  & 1.3 & 1.3 & 0.1 & 0.8 & 0.6 & 0.8 & 0.3\\
    50 & 6.9 & 8.1 & 13 & 29 & 12 & 14 & 13\\
    1000 & 1.0 & 2.1 & 1.7 & 2.2 & 2.2 & 2.1 & 1.7\\
    \hline
    &\multicolumn{7}{c}{$\Delta(\epsilon)$ (\%)}\\
    \hline   
    5 & 2.6 & 27& 15 & 0.84 & 1.9 & 1.3 & 1.2\\
    10  & 0.6& 4.5 & 2.9 & 1.2 & 1.5 & 1.9 & 0.1\\
    50 & 0.2 & 0.3 & 0.1 & 0.2 & 0.03 & 0.6 & 0.1\\
    1000 & 0.78 & 0.19& 1.0& 0.58 & 0.18 & 2.0 & 0.04\\
	\hline\hline
	\end{tabularx}

\label{Table:NuisPars}
\end{table}

\subsection{\label{sec:limit}Profile likelihood calculations 
of interaction cross section limits}

A model for the LUX data and auxiliary measurements ({\it i.e.}, \dd{} calibration and background rate estimation) is specified by a WIMP mass ($m$), a WIMP-nucleon cross section ({\it e.g.}, $\sigma_0$ in the spin-independent analysis), and a vector of nuisance parameters ($\boldmath{\theta}$). The latter consists of rate parameters for each background population, together with the two signal-model parameters that have significant uncertainty: the ratio of $g_2$ in WIMP-search versus \dd{} data-taking, and the Lindhard $k$ parameter defined in Sec.~\ref{sec:NR}. The total likelihood is a product of the likelihood of the observed data given the full model and an independent Gaussian constraint term for each element in $\boldmath{\theta}$, as specified in Table~I of \cite{Akerib:2015:run3}.

\begin{figure}[!htp]
\begin{center}
\includegraphics[width=0.48\textwidth,clip]{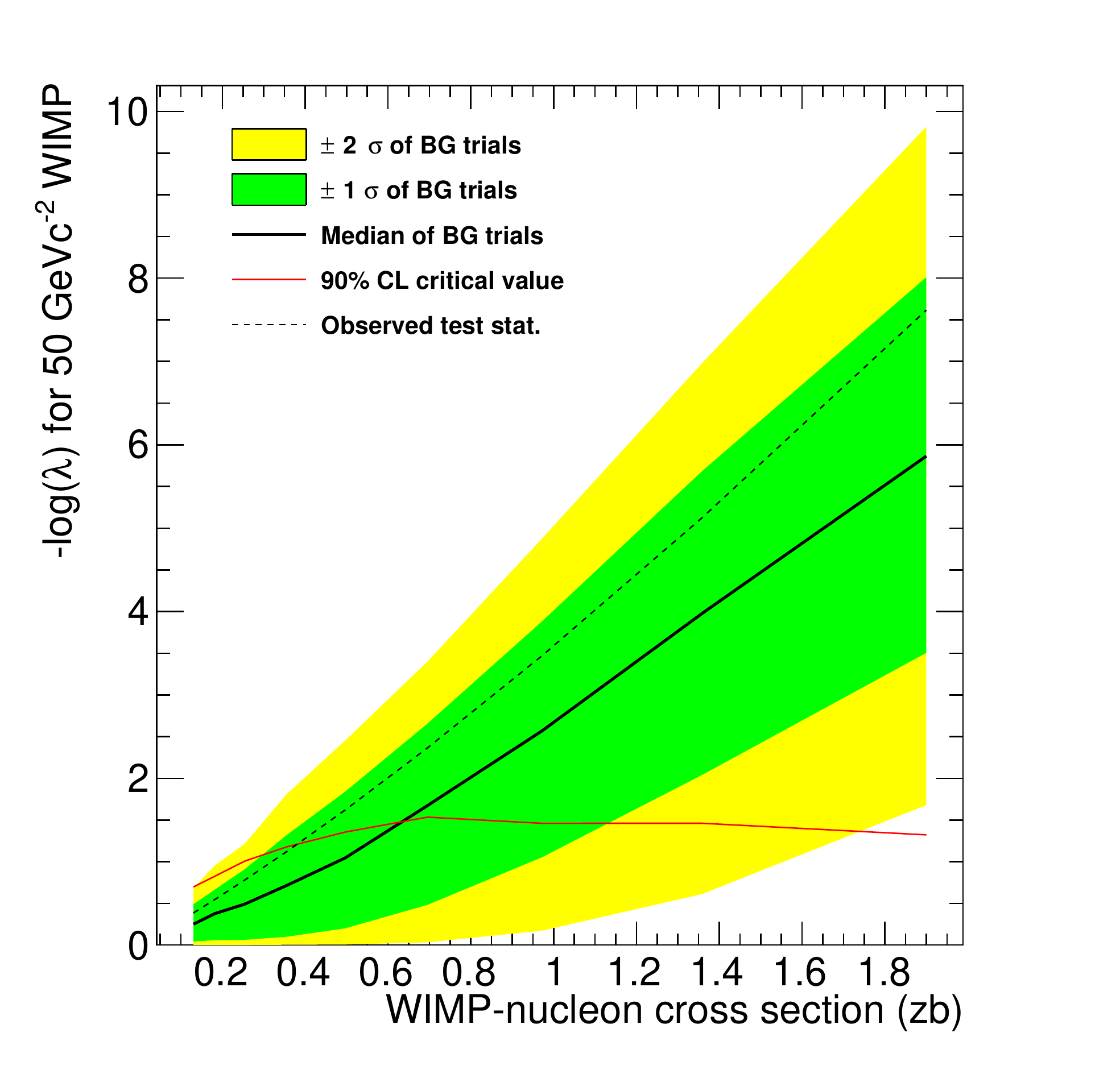}
\caption{Negative logarithm of the profile likelihood ratio versus spin-independent WIMP-nucleon cross section at a benchmark mass of 50~\GeVmass. The critical value to exclude a cross section at 90\% CL is in red. The raw observed test statistic is the dashed line. Given the best-fit BG-only model, the median and the central 68\% and 95\% ranges are shown in black, green and yellow, respectively.}
\label{fig:profile_likelihood_50GeV}
\end{center}
\end{figure}

WIMP-nucleus cross section limits are computed at 31 test WIMP masses, spanning 3.5 to 20 000~GeV/c$^{2}$, via a profile likelihood ratio (PLR) test. The test statistic  $\lambda\left(\sigma\right)$ is the ratio of the conditional maximum likelihood at a given cross section (maximizing over the nuisance parameters with $\sigma$ fixed) to the unconditional maximum likelihood. The unconditional maximum-likelihood fit to the observed data has $\sigma=0$. Figure~\ref{fig:profile_likelihood_50GeV} shows the function $-\log\left(\lambda\left(\sigma\right)\right)$ at an example WIMP mass. For each mass the \textsc{RooStats} \cite{root} hypothesis test inverter tool is used to compute a $p$-value for a range of cross sections and to interpolate for the observed limit, {\it i.e.}, the cross section at which \mbox{$(1-p_{\mathrm{obs}})=\mathrm{CL}=90\%$}. The same procedure is applied to MC samples from the BG-only model to obtain the distribution of limits expected in the absence of signal. The raw result of the PLR test using the observed data would be a limit on the low side of its expected distribution, as seen in Fig.~\ref{fig:profile_likelihood_50GeV} where $-\log\left(\lambda\left(\sigma\right)\right)$ lies above the median of the BG-only trials. It can be understood intuitively as the realized events fluctuating, by chance, away from the region of high signal-to-background density in the space of measured light, charge, radius and height.
This is analogous to the case of a counting experiment with background, in which the observed count happens to fall below the background-only expectation. So as not to exclude cross sections for which sensitivity is poor,  a power constraint is applied that restricts the reported limit at the median from background-only trials.

The cross section limits correspond to an absolute count between $2.2-4.8$ signal events in the dataset, increasing monotonically with WIMP mass. In the low-mass regime the energy spectrum of nuclear recoils is steeply falling and largely below the detection threshold in $S1$; the signal events that would be detected are those with upward binomial fluctuations in the number of detected $S1$ photons. Consequently, for a given measured $S1$ size, low-mass WIMP PDFs are better separated from the ER population than are high-mass ones.

\section{\label{sec:conclusion} Conclusions \& outlook}

The LUX experiment has led the SI hunt for WIMP scattering until Summer 2017, and it is still currently leading the SD search for neutron-coupling, thanks to progress across instrument performance, calibration sources, and data-analysis techniques. A search based on novel \emph{in situ} calibration of PMT response and of the efficiencies and PDFs for signal and background event populations has been described. Light and charge calibrations reach down to 1 keV for both electron and nuclear recoils, based on absolute energy scales from X-ray lines and scattering kinematics, respectively. The linear, combined $S1$-and-$S2$ energy scale is shown to resolve ER lines from 5.3 to 662 keV. These in turn are used to calibrate an absolute scale of initial quanta, both for scintillation and ionization. The fiducial target mass has been estimated in two ways, one by a direct measurement of the fiducial volume and the other based on a dispersed source in the energy region of interest. After processing with an updated suite of event-reduction algorithms, 99.8\% rejection of electron-recoil backgrounds was achieved based on $S1$ and $S2$ alone. Cross-section limits are derived from a profile likelihood ratio test which, in addition, exploits background discrimination in spatial coordinates. The statistical derivations and motivations are detailed in Cowan et. al. in \cite{Cowan:likelidhoodtest}. These advances from the first LUX underground run may be applied to its subsequent, year-long run and also to future searches with liquid-xenon TPCs~\cite{Elena:2016:XENON1T,Akerib:2015:cdr,Baudis:2016:DARWIN}.

\begin{acknowledgments}

This work was partially supported by the U.S. Department of Energy (DOE) under award numbers DE-FG02-08ER41549, DE-FG02-91ER40688, DE-FG02-95ER40917, DE-FG02-91ER40674, DE-NA0000979, DE-FG02-11ER41738, DE-SC0006605, DE-AC02-05CH11231, DE-AC52-07NA27344, and DE-FG01-91ER40618; the U.S. National Science Foundation under award numbers PHYS-0750671, PHY-0801536, PHY-1004661, PHY-1102470, PHY-1003660, PHY-1312561, PHY-1347449; the Research Corporation grant RA0350; the Center for Ultra-low Background Experiments in the Dakotas (CUBED); and the South Dakota School of Mines and Technology (SDSMT). LIP-Coimbra acknowledges funding from Funda\c{c}\~{a}o para a Ci\^{e}ncia e a Tecnologia (FCT) through the project-grant PTDC/FIS-NUC/1525/2014. Imperial College and Brown University thank the UK Royal Society for travel funds under the International Exchange Scheme (IE120804). The UK groups acknowledge institutional support from Imperial College London, University College London and Edinburgh University, and from the Science \& Technology Facilities Council for PhD studentships ST/K502042/1 (AB), ST/K502406/1 (SS) and ST/M503538/1 (KY). The University of Edinburgh is a charitable body, registered in Scotland, with registration number SC005336.

The $^{83}$Rb isotope used in this research were supplied by the United States Department of Energy Office of Science by the Isotope Program in the Office of Nuclear Physics.

This research was conducted using computational resources and services at the Center for Computation and Visualization, Brown University.

The collaboration gratefully acknowledge the logistical and technical support and the access to laboratory infrastructure provided to us by the Sanford Underground Research Facility (SURF) and its personnel at Lead, South Dakota. SURF was developed by the South Dakota Science and Technology Authority, with an important philanthropic donation from T. Denny Sanford, and is operated by Lawrence Berkeley National Laboratory for the Department of Energy, Office of High Energy Physics.

\end{acknowledgments}

\bibliography{main}

\end{document}